\documentclass[aps,prb,longbibliography,showpacs,floatfix,superscriptaddress,twocolumn]{revtex4-1}

\usepackage{graphicx,color}
\usepackage{amsfonts, amsbsy}
\usepackage[figuresright]{rotating}
\usepackage{amssymb}
\usepackage{amsmath}
\usepackage{bm}
\usepackage{blkarray}
\usepackage{mathtools}
\usepackage{psfrag}
\usepackage{caption}
\usepackage{subcaption}
\usepackage{multirow}
\usepackage{tabularx}
\usepackage{textcomp}
\usepackage{units}
\usepackage{lipsum}
\usepackage{soul}
\usepackage{titlesec}
\usepackage{times}
\usepackage{hyperref}
\usepackage{enumitem}

\DeclareMathAlphabet\mathbfcal{OMS}{cmsy}{b}{n}

\titlespacing*{\section}
{0pt}{1ex plus .25ex}{1ex plus 1ex}
\titlespacing*{\subsection}
{0pt}{1ex plus .25ex}{1ex plus 1ex}
\titlespacing*{\subsubsection}
{0pt}{1ex plus .25ex}{1ex plus 1ex}

\titleformat{\section}
 {\fontsize{10}{15}\bfseries }
  {\thesection}
  {1em}
  {}

\titleformat{\subsection}
  {\bfseries }
  {\thesubsection}
  {2em}
  {}

\titleformat{\subsubsection}
  {\itshape}
  {\thesubsubsection}
  {2em}
  {}

\def\beq{\begin{eqnarray}}
\def\eeq{\end{eqnarray}}

\newcommand{\ket}[1]{\left| #1 \right>} 
\newcommand{\bra}[1]{\left< #1 \right|} 
\let\baraccent=\= 
\renewcommand{\=}[1]{\stackrel{#1}{=}} 
\newcommand{\mc}[1]{\mathcal{ #1}} 

\newcommand{\red}[1]{\textcolor{red}{#1}} 

\makeatletter

\titleclass{\subsubsubsection}{straight}[\subsection]

\newcounter{subsubsubsection}[subsubsection]
\renewcommand\thesubsubsubsection{\thesubsubsection.\arabic{subsubsubsection}}

\titleformat{\subsubsubsection}
   {\normalfont\normalsize\itshape \centering}{\thesubsubsubsection}{1em}{}
\titlespacing*{\subsubsubsection}
{0pt}{1ex plus .3ex}{1ex plus .3ex}

\makeatletter
\renewcommand\paragraph{\@startsection{paragraph}{5}{\z@}%
  {3.25ex \@plus1ex \@minus.2ex}%
  {-1em}%
  {\normalfont\normalsize}}
\renewcommand\subparagraph{\@startsection{subparagraph}{6}{\parindent}%
  {3.25ex \@plus1ex \@minus .2ex}%
  {-1em}%
  {\normalfont\normalsize}}
\def\toclevel@subsubsubsection{4}
\def\toclevel@paragraph{5}
\def\toclevel@paragraph{6}
\def\l@subsubsubsection{\@dottedtocline{4}{7em}{4em}}
\def\l@paragraph{\@dottedtocline{5}{10em}{5em}}
\def\l@subparagraph{\@dottedtocline{6}{14em}{6em}}
\makeatother

\begin{document}
\title{Multiplicative Majorana zero-modes}
\author{Adipta Pal}
\affiliation{Max Planck Institute for Chemical Physics of Solids, Nöthnitzer Strasse 40, 01187 Dresden, Germany}
\affiliation{Max Planck Institute for the Physics of Complex Systems, Nöthnitzer Strasse 38, 01187 Dresden, Germany}

\author{Joe H. Winter}
\affiliation{Max Planck Institute for Chemical Physics of Solids, Nöthnitzer Strasse 40, 01187 Dresden, Germany}
\affiliation{Max Planck Institute for the Physics of Complex Systems, Nöthnitzer Strasse 38, 01187 Dresden, Germany}
\affiliation{SUPA, School of Physics and Astronomy, University of St.\ Andrews, North Haugh, St.\ Andrews KY16 9SS, UK}

\author{Ashley M. Cook}
\affiliation{Max Planck Institute for Chemical Physics of Solids, Nöthnitzer Strasse 40, 01187 Dresden, Germany}
\affiliation{Max Planck Institute for the Physics of Complex Systems, Nöthnitzer Strasse 38, 01187 Dresden, Germany}

\begin{abstract}
Topological qubits composed of unpaired Majorana zero-modes are under intense experimental and theoretical scrutiny in efforts to realize practical quantum computation schemes. In this work, we show the minimum four \textit{unpaired} Majorana zero-modes required for a topological qubit according to braiding schemes and control of entanglement for gate operations are inherent to multiplicative topological phases, which realize symmetry-protected tensor products---and maximally-entangled Bell states---of unpaired Majorana zero-modes known as multiplicative Majorana zero-modes.  We introduce multiplicative Majorana zero-modes as topologically-protected boundary states of both one and two-dimensional multiplicative topological phases, using methods reliant on multiplicative topology to construct relevant Hamiltonians from the Kitaev chain model. We furthermore characterize topology in the bulk and on the boundary with established methods while also introducing techniques to overcome challenges in characterizing multiplicative topology. In the process, we explore the potential of these multiplicative topological phases for an alternative to braiding-based topological quantum computation schemes, in which gate operations are performed through topological phase transitions.
 \end{abstract}
\maketitle

Topological quantum computation schemes are central to study of topological condensed matter and viewed as one of their most important and practical applications. In particular, they hold great promise for overcoming challenges of decoherence associated with scalable quantum computation schemes~\cite{kitaev2003fault}. These schemes rely upon realization of topological qubits consisting of quasiparticles with non-Abelian exchange statistics, with the simplest and most widely-studied of these quasiparticles being the unpaired Majorana zero-mode (MZM) \cite{aasen2016milestones}. This area of research has expanded rapidly in the last two decades, with many recent experimental works reporting signatures associated with unpaired Majorana zero-modes\cite{strubi2011interferometric,jack2019observation}, along with a tremendous number of theoretical proposals for experimental realization and practical application \cite{karzig2017scalable,lian2018topological}.

 In order to construct a topological qubit from unpaired Majorana zero-modes, \textit{two pairs} of unpaired Majorana zero-modes are required at minimum by proposals based on braiding ~\cite{plugge2017majorana,leijnse2012parity}, and some gate operations required for topological quantum computation utilize controlled entanglement~\cite{calzona_2020}. The recently introduced multiplicative topological phases (MTPs)~\cite{cook2022mult}---topological phases of matter corresponding to a symmetry-protected tensor product structure in which multiple parent topological phases may be combined in a multiplicative fashion to realize novel topology---present an opportunity to elegantly meet these requirements. If two parent topological phases, each realizing unpaired Majorana zero-modes, are combined in this manner, states consisting of tensor products of unpaired Majorana zero-modes are possible. As shown in work introducing MTPs~\cite{cook2022mult}, it is furthermore possible to selectively entangle topologically-protected boundary modes while respecting symmetries protecting the multiplicative topological phase in the bulk, which could potentially be used to introduce entanglement in a controlled manner for the purpose of gate operations.

For these reasons, we introduce  multiplicative topological phases constructed from parent phases realizing unpaired Majorana zero-modes in this work, and introduce the concept of a multiplicative Majorana zero-mode (MMZM), a single quasiparticle composed of two or more MZM states in a tensor product---or maximally-entangled--- at the simplest level. We choose parent Hamiltonians to be instances of the canonical Kitaev chain model~\cite{Kitaev_2001}.  We find that, for the models considered, MMZMs realize a variety of two-qubit states in different regions of the phase diagram. This indicates MMZMs have the potential to serve as an alternative platform for topological quantum computation to braiding schemes, in which each parent of the multiplicative phase provides a qubit, and the minimum number of MZMs for a qubit is instead effectively two.

We also explore the potential of multiplicative topology to realize novel physics in this work of interest beyond quantum computation schemes: while the Kitaev chain realizes a one-dimensional topological phase, a multiplicative topological phase constructed from two parent Kitaev chains can actually be one-dimensional or two-dimensional. We consider both constructions in this work using Kitaev chain parent phases, realizing one-dimensional and two-dimensional multiplicative Kitaev chain (MKC) constructions, and studying the multiplicative Majorana zero-modes resulting in each case. To characterize the arising multiplicative phases, we study the Wannier center spectrum of the MKC and find that its eigenvalues are sums of the eigenvalues of the parent Wannier center spectra. As a result, Wilson loops can fail to characterize multiplicative topology in certain cases. We show, however, that the MKC can be decomposed into parts, and winding numbers for these components used to characterize topological phases realized by the MKC.

We begin by first reviewing the Kitaev chain and its topological classification in section \ref{sec:parent_model_hamiltonians}. In section \ref{sec:child_model_hamiltonian_parallel} we introduce a one-dimensional MKC and present its spectrum and bound states. Finally, in section \ref{sec:child_model_hamiltonian_perpendicular} we introduce a two-dimensional MKC, also characterizing its spectral properties and bulk-boundary correspondence.

\section{Parent Hamiltonians} \label{sec:parent_model_hamiltonians}

To realize topologically-protected states analogous to unpaired Majorana zero-modes in multiplicative topological phases, we construct them from two parent Hamiltonians. The latter are described by a Hamiltonian core to many leading experimental proposals~\cite{2deg_halperin, 2deg_karsten, Lutchyn2018, science_majoranas, ali_proposal, jason_alicea_proposal, Alicea2011} for realization of unpaired Majorana zero-modes and topological qubits known as the Kitaev chain~\cite{Alicea_2012, Beenakker_review}. Given the foundational nature of the Kitaev chain in topological quantum computation~\cite{Kitaev_2001, sankar_das_sarma_review}, our results are broadly-relevant to study of quasiparticles in multiplicative phases relevant to topological quantum computation. We further show that phases in which multiplicative Majorana zero-modes are realized exhibit a number of unique features of considerable fundamental interest in study of topological phases of matter and promising for topological quantum computation schemes.

First, we review the Kitaev chain model and its significance to platforms for topological quantum computation. The one-dimensional Kitaev chain model is a foundational tight-binding model describing spinless complex fermions hopping between nearest-neighbor sites, with additional $p+ip$ superconducting pairing~\cite{Alicea_2012}. More specifically, the real space Hamiltonian for the Kitaev chain (KC) takes the form~\cite{Kitaev_2001},
\begin{equation}
\begin{aligned}
H_{KC} =& \sum_{j=1}^N-\mu c^\dagger_jc_j -t(c^\dagger_jc_{j+1}+h.c.) \\
&+\Delta(c_jc_{j+1}+h.c.),
\end{aligned}
\end{equation}
where here $c^\dagger_j$ creates an electron at site $j$, $\mu$ is the chemical potential, $t$ is the nearest-neighbor hopping integral, and $\Delta$ is the superconducting pairing strength.

The fermion number parity conservation of the superconductor yields two sectors of the Hilbert space, one with even ground state parity and one with odd ground state parity~\cite{Kitaev_2001}. For odd parity and open boundary conditions (OBC) for the chain, the ground state manifold is degenerate and composed of states strongly-localized at its ends. Within the ground state manifold, furthermore, states may be constructed with wavefunctions strongly-localized at only one end of the chain or the other, which are of Majorana character~\cite{Kitaev_2001}. These two Majorana bound states constitute a physical fermion that allows information to be encoded non-locally, providing a robust platform for quantum computing.

The single-particle sector of the model also displays the desired unpaired Majorana zero-modes at the ends of the chain for open boundary conditions and it is widely-studied and experimentally relevant~\cite{j_sau_2010}. This version is sufficient for the purpose of introducing multiplicative topological phases based upon the Kitaev chain and we restrict ourselves to this case for the remainder of the manuscript.

We first consider the infinitely-long chain in the single-particle regime with periodic boundary conditions. Fourier-transforming the Hamiltonian and imposing particle-hole symmetry (PHS) through a redundancy, we express the model in terms of a Bogoliubov de Gennes Hamiltonian $H_{BdG}(k)$,
\begin{align}
H_{KC} =& \frac{1}{2}\sum_k\Psi^\dagger_kH_{BdG}(k)\Psi_k, \\
H_{BdG}(k) =& -(2t\cos k+\mu)\tau^z+2\Delta\sin k\tau^y. \label{eq:Hbdg_kitaev}
\end{align}

Here, $\Psi_k=(c^{}_k, c^\dagger_{-k})^T$ with $c^{}_k$ annihilating a complex, spinless fermion with momentum $k$, reflecting the particle-hole degree of freedom incorporated explicitly into the Hamiltonian, and $\tau^j$ with $j \in \{x,y,z\}$ is a Pauli matrix.

For this effectively mean-field description of a superconductor, the Bloch Hamiltonian may be diagonalized to compute the bulk spectrum as $\epsilon_\pm(k)=\pm\sqrt{(2t\cos k+\mu)^2+4\Delta^2\sin^2k}$. From this expression, we see that the Bloch Hamiltonian is gapped for $|\mu| < 2t$ and $|\mu| > 2t$, with gap closings occurring at $k=\pi$ ($k=0$) for $\mu=2t$ ($\mu=-2t$). A topologically non-trivial phase is realized in the former regime, which may be characterized in the bulk by various methods~\cite{chiu2016classification} as well as explicit verification of unpaired Majorana zero-modes. The latter is facilitated by considering the Majorana representation of the finite Kitaev chain~\cite{Kitaev_2001}. For now, we consider the latter and express the BdG Hamiltonian in terms of Majorana operators with the convention $c_j=\frac{1}{2}(\gamma_{+,j}+i\gamma_{-,j})$, where $\{\gamma_\alpha,\gamma_\beta\}=2\delta_{\alpha\beta}$ and $\gamma^\dagger_\alpha=\gamma^{}_\alpha$, yielding the following expression for the Hamiltonian:
\begin{align}
H_{KC} =& \sum_{j=1}^N-\frac{\mu}{2}(1+i\gamma_{j,+}\gamma_{j,-}) \nonumber \\
&-\frac{t}{2}(i\gamma_{j,+}\gamma_{j+1,-}+i\gamma_{j+1,+}\gamma_{j,-}) \nonumber\\
&+\frac{\Delta}{2}(i\gamma_{j,+}\gamma_{j+1,-}-i\gamma_{j+1,+}\gamma_{j,-}).
\end{align}

Notice that for $t=\Delta$ and $\mu=0$, we have $[H_{KC},\gamma_{1,-}]=[H_{KC},\gamma_{N,+}]=0$, which implies we have two Majorana zero-modes, each with zero energy and localized at one end of the chain.


We will now construct multiplicative topological phases (MTP) with two parent Kitaev chain Hamiltonians $H_{p,1}(k_i)$ and $H_{p,2}(k_j)$, where $k_i$ and $k_j$ are momenta in directions $i$ and $j$. We take $i$ and $j$ to either be parallel ($i$ and $j$ are each taken to be $x$ and $k_i = k_j = k_x$ corresponds to momentum in the $x$ direction, for instance) or perpendicular ($i$ and $j$ are taken to be $x$ and $y$, for instance, with $H_{p,1}(k_x)$ describing a Kitaev chain parallel to the $x$-axis and $H_{p,2}(k_y)$ describing a Kitaev chain parallel to the $y$-axis in the $x$-$y$ plane). In this way, we may realize multiplicative topological phases that are either one-dimensional ($i=j$) or two-dimensional ($i \neq j$). We express parent Hamiltonian $\alpha$ (with $\alpha \in \{1,2\}$) using a vector of momentum-dependent parameters $\boldsymbol{d}(k)_{\alpha} = \left(d(k)_{1\alpha},d(k)_{2\alpha},d(k)_{3\alpha} \right)$ dotted into a vector of Pauli matrices $\boldsymbol{\tau} = \left(\tau_x, \tau_y, \tau_z \right)$ for parent $1$ and $\boldsymbol{\sigma} = \left(\sigma_x, \sigma_y, \sigma_z \right)$ for parent $2$,
\begin{subequations}
\begin{equation}
H_{p,1}(k_i) = \boldsymbol{d}(k_i)_{1}\cdot \boldsymbol{\tau},
\end{equation}
\begin{equation}
H_{p,2}(k_j) = \boldsymbol{d}(k_j)_{2}\cdot \boldsymbol{\sigma},
\end{equation}
\begin{equation}
H^c_{12}(\boldsymbol{k}) = \boldsymbol{d}(k_i)_{1}\cdot\boldsymbol{\tau}\otimes (-d(k_j)_{12},d(k_j)_{22},-d(k_j)_{32})\cdot \boldsymbol{\sigma},
\end{equation}
\end{subequations}
and the momentum vector $\boldsymbol{k} = k_i \hat{i} + k_j \hat{j}$ being simply $\boldsymbol{k} = k_i \hat{i}$ for $i=j$. The tensor product structure is protected by a combination of symmetries enforced on the child Hamiltonian and symmetries enforced on the parent Hamiltonians as discussed in Cook and Moore~\cite{cook2022mult}. This results in the child Hamiltonian possessing the following symmetries according to standard analysis purely at the level of child~\cite{Groth_2014, Varjas_2018}:

\begin{subequations}
\begin{equation}
    \mathcal{T}= \mathcal{K},
\end{equation}
\begin{equation}
    \mathcal{P}_1= \mathcal{I} \otimes \sigma^x \mathcal{K},
\end{equation}
\begin{equation}
    \mathcal{C}_1= \mathcal{I} \otimes \sigma^x,
\end{equation}
\begin{equation}
    \mathcal{P}_2=  \tau^x \mathcal{K} \otimes \mc{I},
\end{equation}
\begin{equation}
    \mathcal{C}_2=  \tau^x \otimes \mc{I},
\end{equation}
\end{subequations}
where $\mathcal{T}$, $\mathcal{P}$ and $\mathcal{C}$ correspond to time-reversal, particle-hole and chiral symmetry, respectively. Besides the discrete symmetries, the child Multiplicative Kitaev Chain has a \textit{unitary symmetry}, given by $\mc{U}=\tau^x\sigma^x$. Such a unitary symmetry naturally emerges in the child Hamiltonian by its tensor product form in terms of the parent Hamiltonians and each parent possessing chiral symmetry. This permits block diagonalization of the child Hamiltonian, as we show in this work. This motivates further development of methods for symmetry analysis, as such analysis at the level of the child and parents in combination rather than strictly at the level of the child reveals different information about the system useful in characterizing topological systems given the possibility of multiplicative topology.

We comment briefly on the interpretation of the basis for the child Hamiltonian, as there are two possible options. One possibility is to interpret the resultant Hamiltonian as quadratic, describing an effectively non-interacting system. However, we may also interpret the bases of the child Hamiltonians discussed here as tensor products of single-particle bases of the parents, corresponding to a basis for the child that is purely quartic. In this second interpretation, therefore, the Hamiltonian characterizes a strongly-correlated system. We focus on the first interpretation in this work, and will explore the second interpretation in greater detail in later work.

\section{Child Hamiltonian for parallel parent chains}
\label{sec:child_model_hamiltonian_parallel}

We first consider the MKC for two parallel parent Kitaev chains, corresponding to $i=j$ above. We therefore take $k_i = k_j = k$ to simplify notation. The parent and child Hamiltonians then take the following forms,
\begin{equation}
\begin{split}
H_{KC,1}(k) =& -(2t_1\cos k+\mu_1)\tau^z+2\Delta_1\sin k\tau^y,\\
H_{KC,2}(k) =& -(2t_2\cos k+\mu_2)\sigma^z+2\Delta_2\sin k\sigma^y,\\
H^c_{MKC,||}(k) =& [-(2t_1\cos k+\mu_1)\tau^z+2\Delta_1\sin k\tau^y]\\
&\otimes [(2t_2\cos k+\mu_2)\sigma^z+2\Delta_2\sin k\sigma^y].
\end{split}
\label{KCparallel}
\end{equation}

We characterize the MKC in this case first by studying the bulk spectrum and then by studying bulk boundary correspondence analytically and numerically.

\subsection{Bulk spectrum of the multiplicative Kitaev chain}

The spectrum of the child Hamiltonian $H^c_{MKC,||}(k)$ consists of doubly-degenerate eigenvalues given by
\begin{equation}
\begin{split}
E(k) &= \pm \sqrt{(2 t_1 \cos(k) + \mu_1)^2 + (2\Delta_1\sin(k))^2 } \\
&\sqrt{(2 t_2 \cos(k) + \mu_2)^2 + (2\Delta_2\sin(k))^2 }.
\end{split}
\label{MKCplldispersion}
\end{equation}
This corresponds to the bulk gap closing under the following conditions:
\begin{align}
\mu_{1,2} =
\begin{cases}
 - 2t_{1,2},    &\text{if } k=0 \\
 + 2t_{1,2},    &\text{if } k=\pi \\
 -2\cos(k)t_{1,2},   &\text{ if } \Delta_{1,2}=0.
\end{cases}
\label{eq:E(k)solutions}
\end{align}

\begin{figure}[h]
\includegraphics[width=8cm]{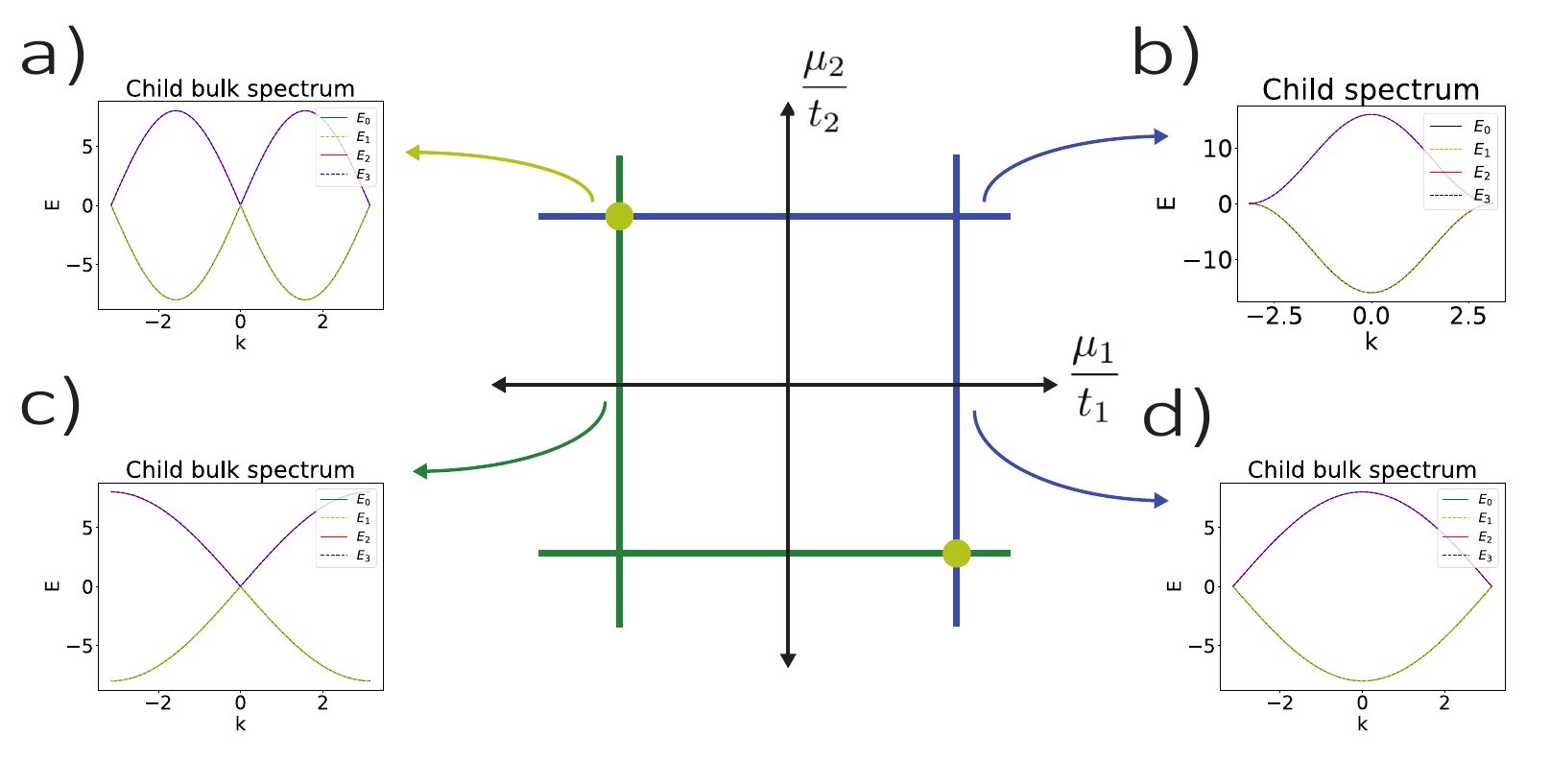}
\caption{Dependence of the bulk dispersion of the child Hamiltonian $H^c_{MKC,||}(k)$ in Eq.~\ref{KCparallel} on parameters of its parent Hamiltonians $H_{KC,1}(k)$ and $H_{KC,2}(k)$. Example bulk dispersions for $H^c_{MKC,||}(k)$ are shown in (a),(b),(c), and (d) for parameter values $\left( {\mu_1 \over t_1},  {\mu_2 \over t_2}\right) = \left( -2,  2\right)$, $\left( 2,  2\right)$, $\left( -2,  0\right)$, and $\left( 2,  0\right)$, respectively. The child bulk gap closes at $k=0$ along the green line, at $k=\pi$ along the blue line, and at $k=0,\pi$ on the yellow dots in agreement with Eq.~\ref{eq:E(k)solutions} for $\Delta_1 \neq 0, \Delta_2 \neq 0$.}
\label{fig1_MKCpar_bulk}
\end{figure}

We illustrate these in Fig.~\ref{fig1_MKCpar_bulk}, where we show the MKC spectrum as a function of $k$ for a set of representative points in a phase diagram generated by fixing $t_1=t_2=1$ and varying $\mu_1$ and $\mu_2$. Figs.~\ref{fig1_MKCpar_bulk} (a-d) show that the bulk gap closing points are inherited from the parents, as the eigenvalues of the MKC correspond to the product of two Kitaev chains eigenvalues for different configurations.

While the computation of bulk topological invariants for the parent Kitaev chains is known, this is not the case for the topological invariants of the MKC. Although the topology of multiplicative phases can be understood in terms of their parents' topological invariants, the methods for characterizing these Hamiltonians, without knowledge of their decomposition into parent Hamiltonians, have not been established.

One of the more robust methods for characterizing topology is the analysis of the Wilson loop spectrum. The Wilson loop~\cite{wilson_loop_paper} is a unitary operator defined over a closed path as:
\begin{equation}
    \begin{aligned}
    \mathcal{W} = \overline{\exp}{\Big [ i \int_{BZ} d \bm{k} \cdot \bm{A}(\bm{k}) \Big ]},
    \end{aligned}
    \label{WLdefA}
\end{equation}
where $\bm A$ is the non-Abelian Berry connection:
\begin{equation}
    \begin{aligned}
    \bm{A}_{mn}(\bm k) = i\bra{u_m(\bm k)} \nabla_{\bm k} \ket{u_n(\bm k)}.
    \end{aligned}
\end{equation}
Here $\ket{u_n(\bm k)}$ are Bloch states in the occupied subspace and $\bm A$ is defined a Hermitian operator. Consequently, $\mathcal W$ is a unitary operator whose eigenvalues are $e^{i2\pi \nu_j}$, where $\nu_j$ are the Wannier centers of charge.

We compute the Wannier centers for topologically-distinct regions of the phase diagram determined by the topological invariants of the parents. Each parent Hamiltonian has a $\mathbb{Z}_2$ topological classification, so the child Hamiltonian has a $\mathbb{Z}_2 \times \mathbb{Z}_2$ classification, with its invariant $\nu_C$ expressed in terms of the parent invariants $\nu^{(1)}$ and $\nu^{(2)}$ as $\nu_C = (\nu^{(1)}, \nu^{(2)})$, where $\nu^{(1,2)} \in \{0, 0.5\} \mod 1$ indicates the topological phase of each parent.

 In the trivial phase the Wannier center of the occupied band is located at the center ($\nu=0$) of the unit cell, and at the edge ($\nu=0.5$) in the topological phase. Therefore, we calculate for the MKC parallel case where we also have a single momentum component, and find that the two eigenvalues show a shift of the Wannier centers to the edge when only one of the parent phases is topological but not both. This inability of the Wilson loop method to detect some multiplicative phases results from the multiplicative dependence of the child Wilson loop on the Wilson loops of the parents.

\begin{figure}[h]
\includegraphics[width=8cm]{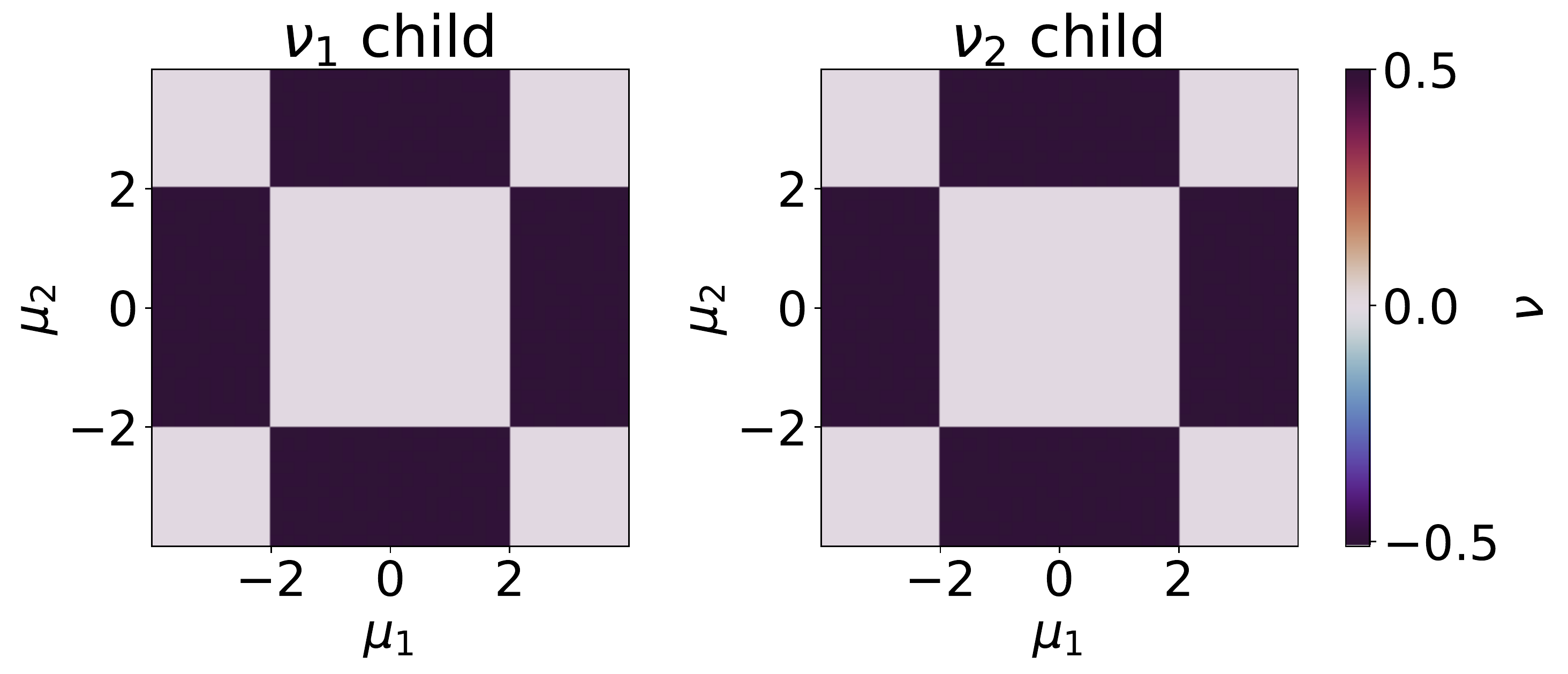}
\caption{Wannier centers for a child Hamiltonian at half-filling with parents with parameters $t_1=t_2$ and $\Delta_1=\Delta_2$. These correspond to the Wilson loop eigenvalues after integrating along $k_x$.}
\label{fig:Wannier-centers-parallel}
\end{figure}

The Wannier centers of charge of the MKC at half-filling ($M=2$ is the number of occupied orbitals) are shown in Fig.~\ref{fig:Wannier-centers-parallel}. The doubly-degenerate occupied states correspond to two equivalent Wannier centers $\nu_1$ and $\nu_2$, as shown by Fig.~\ref{fig:Wannier-centers-parallel} (a) and (b). Unexpectedly, both a MKC with $\lvert \mu_1 \rvert = \lvert \mu_2 \rvert < 2$ and a MKC with $\lvert \mu_1 \rvert = \lvert \mu_2 \rvert > 2$ have the Wannier centers localized at the center of the unit cell ($\nu_1=\nu_2=0$), despite the fact that finite chains with the former set of parameters have bound states, while finite chains with the latter set of parameters do not. It is only for parent Hamiltonians of different topology that the Wannier centers of a half-filled MKC localize at the edge ($\nu_1=\nu_2=0.5$), showing that the MKC Wilson loop eigenvalues correspond to the ones given by the parents' Wannier centers. This is analytically shown in Appendix~\ref{app:wilson_loop}.
\subsection{Quasiparticle velocities near critical points}

For the case of the parallel MKC, we examine the Dirac Hamiltonians near each of the gapless points. For both $\mu_1\sim -2t_1$ and $\mu_2\sim -2t_2$, the gap closes for $k=0$, so that we get the following Dirac Hamiltonian in its vicinity,
\begin{equation}
\begin{aligned}
H^c_{Dirac}(k) =& -(2t_1+\mu_1)(2t_2+\mu_2)\Gamma^{zz}\\
&+2\Delta_1(2t_2+\mu_2)k\Gamma^{yz}-2\Delta_2(2t_1+\mu_1)k\Gamma^{zy}
\end{aligned}
\end{equation}
where $\Gamma^{ij} = \tau^i\sigma^j$. Denote, $m_j = 2t_j+\mu_j$, $(j=1,2)$. The energies are double degenerate and given as,
\begin{equation}
E(k) = \pm\sqrt{4(\Delta_2m_1-\Delta_1m_2)^2k^2+m_1^2m_2^2}.
\end{equation}
Notice, for the gapless point, $\mu_1=-2t_1$, or $m_1=0$, we get $E(k)=\pm 2\Delta_1m_2k$, and for the gapless point, $\mu_2=-2t_1$, or $m_2=0$, one has $E(k)=\pm 2\Delta_2m_1k$. One must notice that if $m_1=0=m_2$ at once, one must expand till the quadratic order to get a $k$-dependence,

\begin{equation}
\begin{aligned}
H^c_2(k) =& -[m_1m_2-(t_1m_2+t_2m_1)k^2]\Gamma^{zz}+2\Delta_1m_2k\Gamma^{yz}\\
&-2\Delta_2m_1k\Gamma^{zy}+\Delta_1\Delta_2k^2\Gamma^{yy}.
\end{aligned}
\end{equation}

Denote $t_1m_2+t_2m_1 = M$. The spectrum is given by the 4 eigenvalues,
\begin{equation}
\begin{split}
E(k) = \pm\big{[}[(M\mp \Delta_1\Delta_2)k^2-m_1m_2]^2\\
+4k^2(\Delta_2m_1\mp\Delta_1m_2)^2k^2\big{]}^\frac{1}{2}.
\end{split}
\end{equation}
Then, for the case, $m_1=m_2=0$, we get quadratic dispersion relations,$E=\pm \Delta_1\Delta_2k^2$. Note the spectrum is doubly degenerate.

\subsection{Finite MKC parallel system with open boundary conditions}

Having presented the bulk spectrum of the multiplicative Kitaev chain, we now study bulk-boundary correspondence for this system. To do so, we characterize the bound states realized in topologically non-trivial regions of the phase diagram analytically for the case of parallel parent Kitaev chains. We then numerically study the low-energy spectrum as a function of chain length $L$ and chemical potentials $\mu_1$ and $\mu_2$ as well as localization of bound states.

\subsubsection{Analytical form of boundary modes}\label{Edge_MKCpll}
The procedure to get zero energy modes for the MKC parallel case is similar to the KC where we need to find the null-vectors of the Hamiltonian after the localization $k\rightarrow iq$. The details are worked out in the Supplementary section~\ref{Sup_sec_A} which yields the identity,
\begin{equation}
\begin{aligned}
&[(2t_1\cosh q+\mu_1)^2-4\Delta_1^2\sinh^2 q]\\
&\times[(2t_2\cosh q+\mu_2)^2-4\Delta_2^2\sinh^2 q] = 0
\label{eq:condition_boundary_modes_1}
\end{aligned}
\end{equation}
Here, one must use caution, since the boundary mode expressions depend on the parametric regime of $\mu_1$ vs. $\mu_2$, especially if $t_1\neq t_2$. This fact is related to which solutions(s) we choose for Eq~\ref{eq:condition_boundary_modes_1}. For semi-infinite boundary conditions $\Psi(0)=\Psi(x\rightarrow\infty)=0$, the general edge mode expression if just one of the parents were topological, and the other trivial is of the form,
\begin{equation}
\begin{aligned}
\Psi(x)\sim
\bigg{[}\bigg{(}\frac{\mp\mu_i+\sqrt{\mu_i^2-4(t_i^2-\Delta_i^2)}}{2(\Delta_i\pm t_i)}\bigg{)}^x\\
-\bigg{(}\frac{\mp\mu_i-\sqrt{\mu_i^2-4(t_i^2-\Delta_i^2)}}{2(\Delta_i\pm t_i)}\bigg{)}^x\bigg{]}
\begin{pmatrix}
a\\
b\\
c\\
d
\end{pmatrix}.
\end{aligned}
\end{equation}
where $(i=1,2)$ corresponds to the parent which is topological and depending on the parametric regime and assuming $\Delta_i>0$, the $\mp\mu_i$ signs correspond to the regions $\text{sgn}(t_i)=\pm\text{sgn}(\Delta_i)$ which is basically the two sets of possible Majorana edge modes. We keep aside the full expression for all the possible cases along with the eigenvectors until we discuss the MKC parallel Hamiltonian in the real space in Sec. D.

\subsubsection{Spectral dependence on chain length}

While the finite Kitaev chain realizes unpaired Majorana \textit{zero-modes} when these bound states do not overlap and hybridize, in general there is a finite split in energy between the topologically-protected bound states due to wavefunction overlap. The dependence of the finite Kitaev chain spectrum for open boundary conditions is therefore typically studied to demonstrate that this splitting decreases exponentially with increasing system size. We therefore study the spectral dependence of the MKC with open boundary conditions as a function of chain length for direct comparison.

\begin{figure}[h]
\includegraphics[width=8cm]{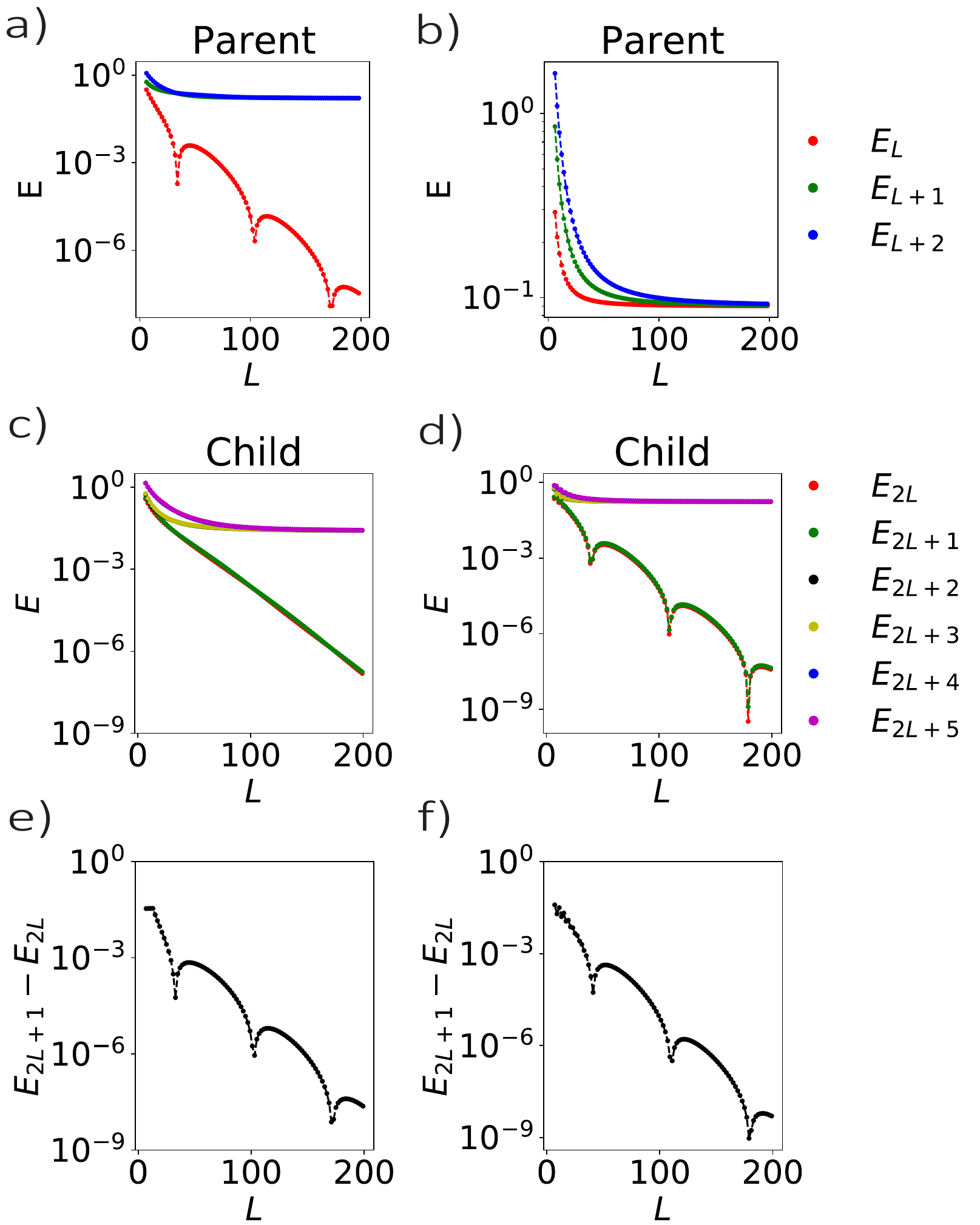}
\caption{Low-energy spectrum versus chain length $L$ for the Kitaev chain shown in (a) and (b) and for the parallel MKC Hamiltonian $H^c_{MKC,||}(k)$ shown in (c-f). Each plot shows the spectrum only for even values of $L$. (a) The three lowest-energy modes of the Kitaev chain Hamiltonian in the topological phase with open boundary conditions, corresponding to $t=1$, $\Delta=0.08$, $\mu=0.09$, as a function of chain length $L$. (b) The three lowest-energy modes of the Kitaev chain Hamiltonian in the trivial phase corresponding to $t=1$, $\Delta=0.08$, $\mu=2.09$. (c) The six lowest-energy modes of the MKC Hamiltonian with two parent Kitaev chains that each have a parameter set corresponding to (a). (d) The six lowest-energy modes of the MKC Hamiltonian with a parent Kitaev chain with parameter set corresponding to subfigure (a) and the second parent Kitaev chain with parameter set corresponding to subfigure (b). (e-f) Energy difference between the two lowest energies in subfigure (c-d), indicating the non-degeneracy of these.}
\label{fig:spectral-dependence-L}
\end{figure}

For the topologically-protected pair of low-energy modes localized on the boundary of the finite-length Kitaev chain described by $H_{BdG}$ given in Eq.~\ref{eq:Hbdg_kitaev} for OBC to be at $E=0$, the parameters $t$, $\mu$ and $\Delta$ need to be fine-tuned~\cite{Leumer_2020}. Otherwise, as shown in Fig.~\ref{fig:spectral-dependence-L}a), these boundary mode energies oscillate as a function of chain length $L$ with a period determined by $\mu/\Delta$~\cite{Leumer_2020} while also decreasing overall in exponential fashion.

The finite multiplicative chain also presents an oscillatory dependence on ground state energy with respect to chain length when at least one of the parents is in the topological phase. Fig.~\ref{fig:spectral-dependence-L} shows the spectral dependence of the finite multiplicative Kitaev chain $H^c_{MKC,||}(k)$ in Eq.~\ref{KCparallel} for two key cases:
\begin{enumerate}
    \item The parameter set of parent $1$, $\left\{t_1,\Delta_1, \mu_1 \right\}$, and that of parent $2$, $\left\{t_2,\Delta_2, \mu_2 \right\}$, are equal, meaning $t_1 = t_2$, $\Delta_1 = \Delta_2$, and $\mu_1 = \mu_2$, and each parent is topologically non-trivial. The low-energy spectrum of the parents for this case is shown in Fig.~\ref{fig:spectral-dependence-L} (a), and the corresponding low-energy spectrum of the child Hamiltonian is shown in Fig.~\ref{fig:spectral-dependence-L}(c) and (e).
    \item Parent $1$ is topologically non-trivial and Parent $2$ is topologically trivial. The low-energy spectra of parents $1$ and $2$ for this case are shown in Fig.~\ref{fig:spectral-dependence-L} (a) and (b), respectively. The corresponding low-energy spectrum of the child Hamiltonian is shown in Fig.~\ref{fig:spectral-dependence-L} (d) and (f).
\end{enumerate}

In each case, the child Hamiltonian exhibits oscillations in the two lowest-energy modes $E_{2L+1}-E_{2L}$, indicating splitting of the ground state degeneracy due to finite-size effects. A key difference is that the child exhibits negligible Friedel oscillations relative to zero energy in case $1$ as shown in Fig.~\ref{fig:spectral-dependence-L}(c), although there is evidence of Friedel oscillations in the splitting in energy between these two lowest energy states as shown in Fig.~\ref{fig:spectral-dependence-L}(e). Friedel oscillations are much more noticeable in the low-energy child spectrum for case $2$ as shown in Fig.~\ref{fig:spectral-dependence-L}(d), although splitting in energy between the two-lowest energy states is very similar to case $1$ as shown in Fig.~\ref{fig:spectral-dependence-L}(f).

\subsubsection{Spectral dependence on chemical potential of finite MKC}

\begin{figure}[h]
\includegraphics[width=8cm]{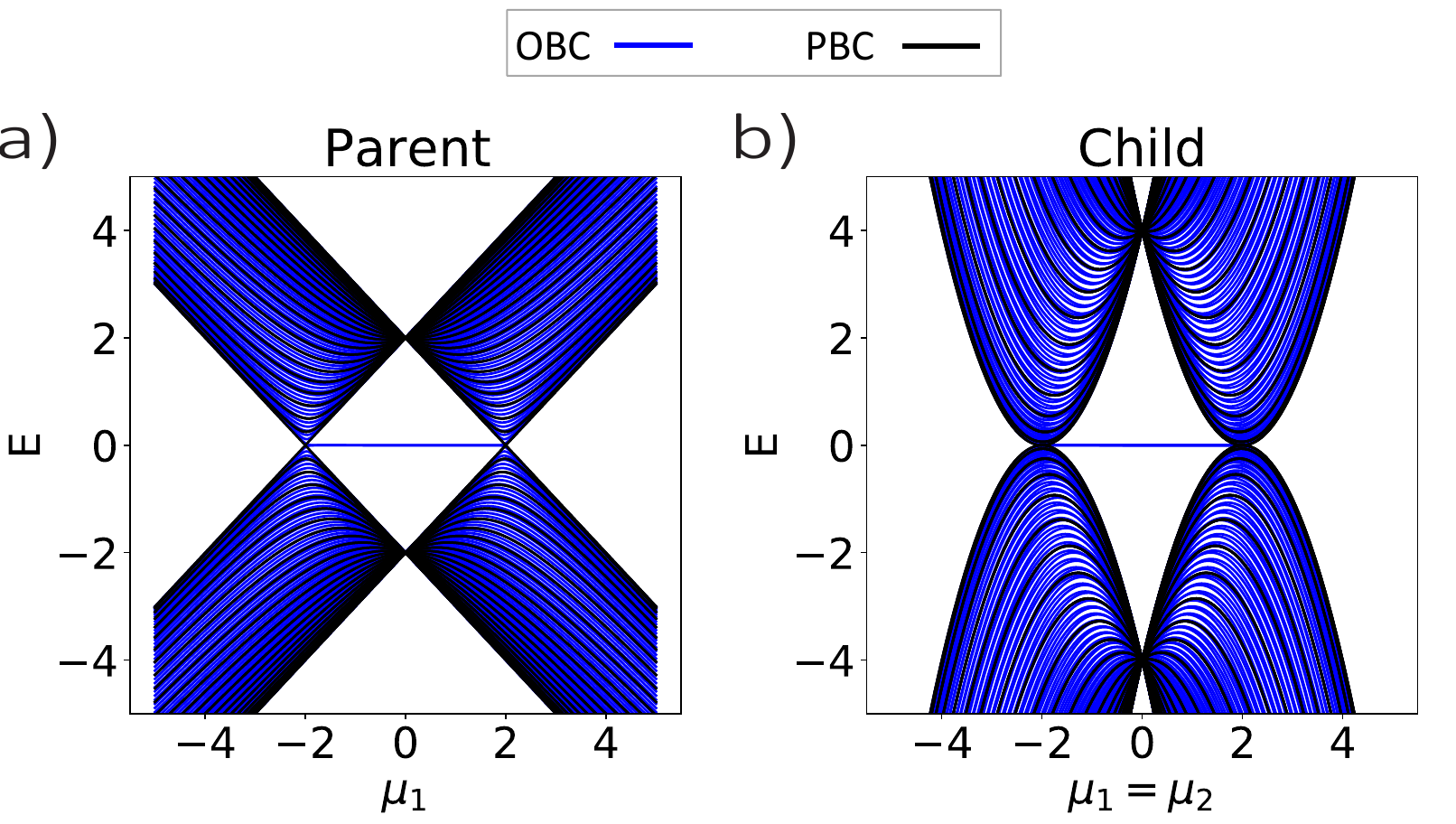}
\caption{Spectra of the parent and child Hamiltonians as a function of chemical potential for relatively long chain length $L=80$, shown in black for periodic boundary conditions and blue for open boundary conditions, respectively. The spectra for parent 1 and 2 are identical as their parameter sets are identical, thus the spectrum for parent 1 is shown in (a), for $t_1=t_2=1$, $\Delta_1=\Delta_2=1$, $\mu_1=\mu_2$. The corresponding child MKC spectrum is shown in (b) as a function of $\mu_1$, with $\mu_2=\mu_1$.}
\label{fig:parent-child-vs-mu-L80}
\end{figure}

Tuning chemical potential is a physically-relevant mechanism for exploring the phase diagram of parent Kitaev chains and therefore also important in understanding behaviour of the MKC. Much can be learned, in particular, by studying the spectra of the parent Kitaev chains and MKC as a function of chemical potential. These results are shown for the parent and child in Fig.~\ref{fig:parent-child-vs-mu-L80} (a) and (b), respectively, for a long chain length of $L=80$. Importantly, we observe a topological phase transition in the parent for $\mu_1 = \pm|2 t_1|$ due to closing of the bulk gap as expected, with states dispersing linearly when tuning $\mu_1$ away from these critial values. For $-2 t_1 < \mu_1 < 2 t_1$, we see low-energy modes inside the bulk gap, corresponding to the unpaired Majorana zero-modes localized at each end of the chain. Comparing this to the spectrum for the MKC, we see clear similarities for $\mu_2$ fixed in value to $\mu_1$ : the bulk gap also closes at $\mu_1 = \pm|2 t_1|$ as the system undergoes topological phase transitions, with $-2 t_1 < \mu_1 < 2 t_1$ again corresponding to a topologically non-trivial phase and the presence of topologically-protected boundary modes. The spectrum instead disperses quadratically as $\mu_1$ and $\mu_2$ are tuned away from the critical values, and the maximum bulk gap is larger, being the product of the maximum bulk gaps of the parents. This multiplicative structure also yields a four-fold degeneracy of the in-gap states, compared with a two-fold degeneracy of the in-gap states for the parents. More generally, the degeneracy of states for the child is twice that of each parent.

\begin{figure}[h]
\includegraphics[width=8cm]{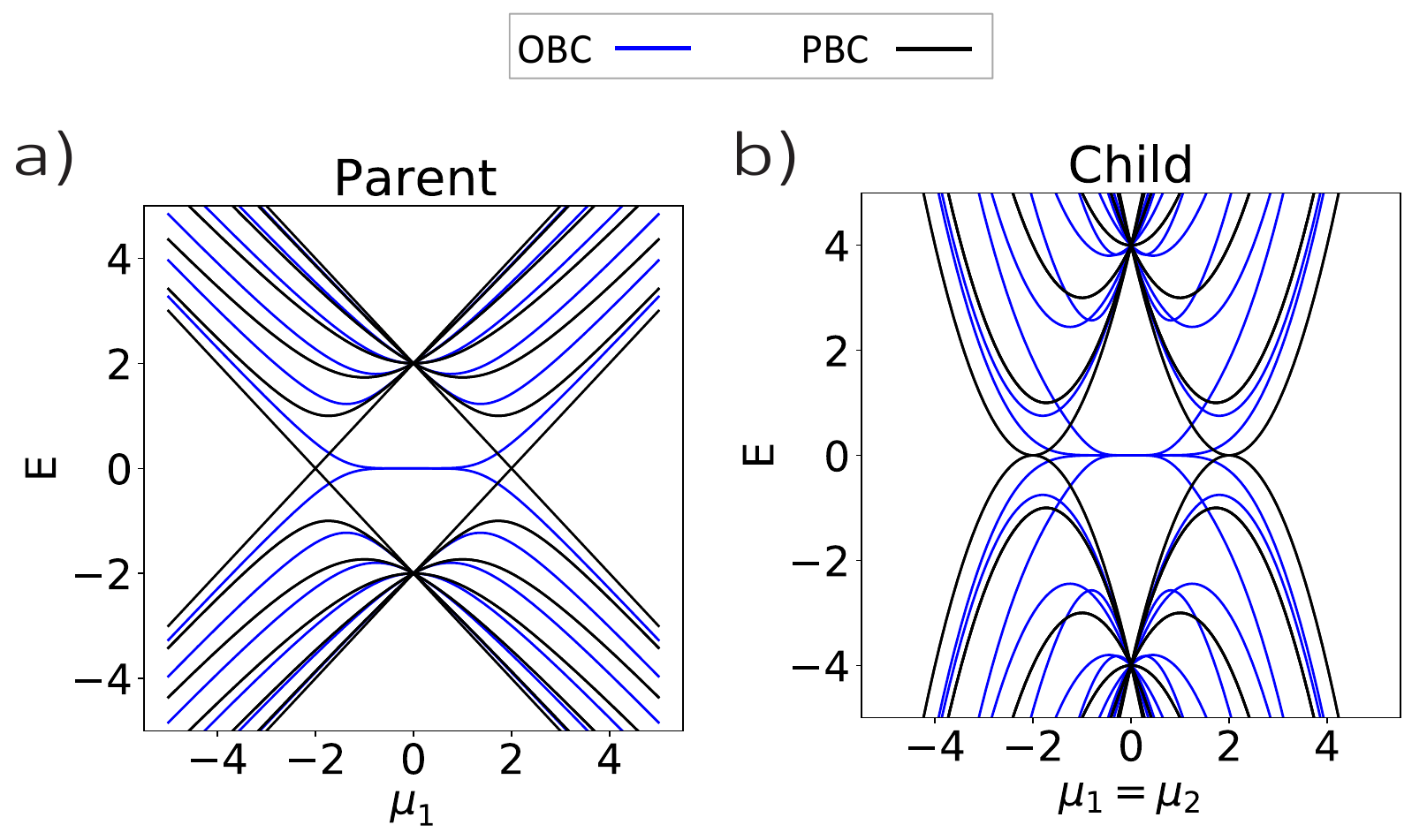}
\caption{Spectra of the parent and child Hamiltonians as a function of chemical potential for relatively short chain length $L=6$, with black lines depicting spectra for periodic boundary conditions and blue lines depicting spectra for open boundary conditions, respectively. The spectra for parent 1 and 2 are identical as their parameter sets are identical, thus the spectrum for parent 1 is shown in (a), for $t_1=t_2=1$, $\Delta_1=\Delta_2=1$, $\mu_1=\mu_2$. The corresponding child MKC spectrum is shown in (b) as a function of $\mu_1$, with $\mu_2=\mu_1$.}
\label{fig:parent-child-vs-mu-1}
\end{figure}

We also explore the dependence of the multiplicative spectrum on chemical potential for relatively short chain lengths, where finite-size topology~\cite{cook2022} is more prominent. These results are shown in Fig.~\ref{fig:parent-child-vs-mu-1}. While the spectra for periodic boundary conditions display bulk gap closings at the same values of $\mu_1$ and states disperse linearly as $\mu_1$ is tuned away from these critical values between the $L=80$ case and $L=6$ case, striking differences are observed for open boundary conditions. In particular, gap-closings occur at $\mu_1=0$ rather than $\mu_1 = \pm |2t_1|$ in the parents, as shown in Fig.~\ref{fig:parent-child-vs-mu-1}(a), due only to destructive interference between states resulting from bulk-boundary correspondence. In addition, the four-fold degeneracy of the in-gap states for the child, shown in Fig.~\ref{fig:parent-child-vs-mu-1}(b), is split away from $\mu_1=0$, with the energy gap between two states increasing more rapidly with increasing $|\mu_1|$ than for the other two states.

\begin{figure}[h]
\includegraphics[width=7cm]{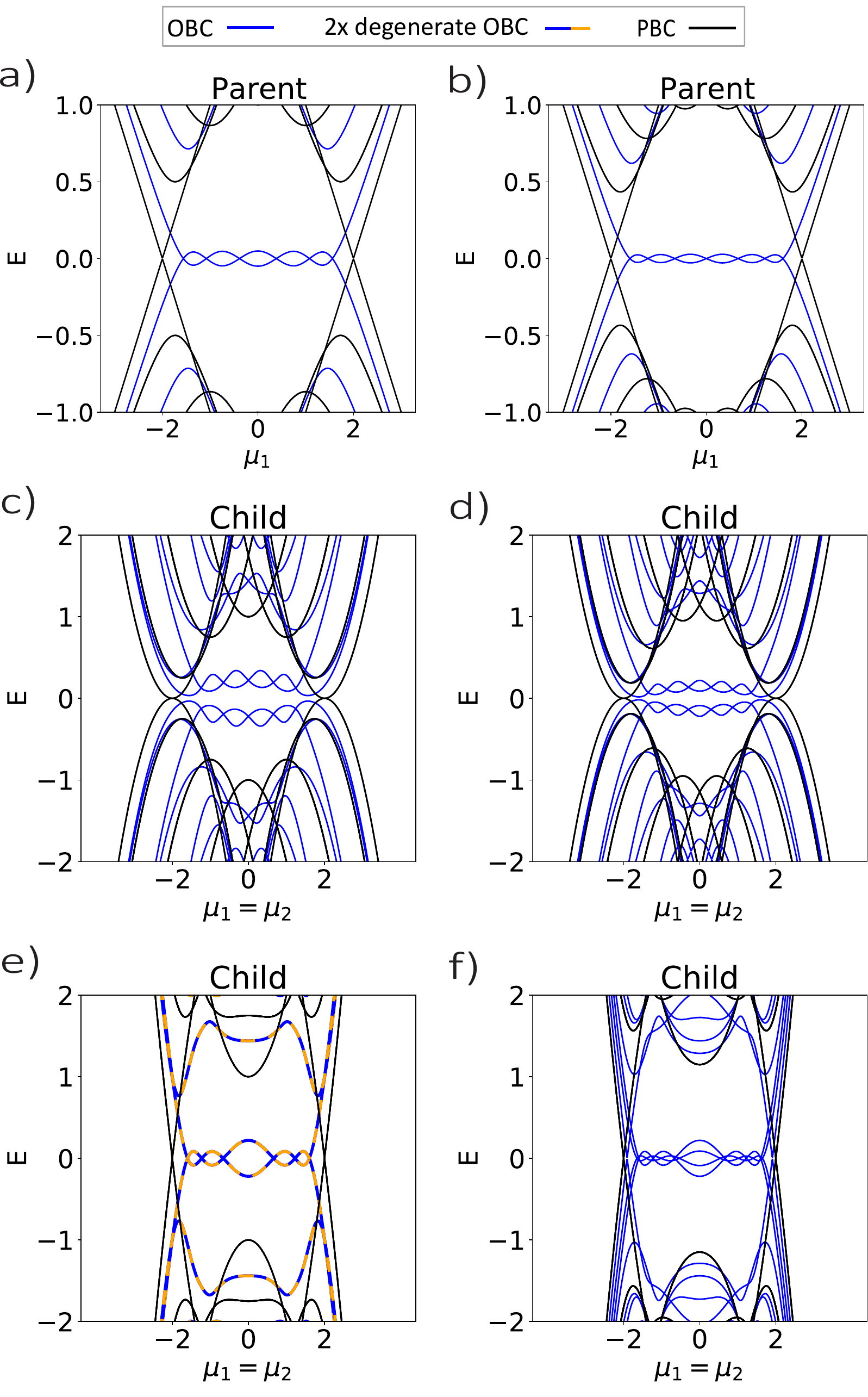}
\caption{Child and parent spectrum dependence on the chemical potential. Kitaev chain spectrum for parameters $\lvert t \rvert =1$, $\Delta=0.5$ for even and odd chains ($L=6,7$) is shown in (a) and (b), respectively.  Multiplicative chain spectrum for two identical parents as in (a) and (b) is shown in (c) and (d), respectively. Multiplicative chain spectrum for two different parents with $t_1=-t_2=1$, $\Delta_1=\Delta_2=0.5$ and even and odd chains ($L=6,7$) is shown in (e) and (f), respectively. The black curves correspond to solutions under periodic boundary conditions, which are all doubly-degenerate for the children. The blue curves  correspond to open boundary conditions, with double-degeneracy indicated by dashed blue-orange curves.}
\label{fig:parent-child-vs-mu-2}
\end{figure}

While the finite Kitaev chain is known to have exact zero energy modes for discrete values of the chemical potential~\cite{Leumer_2020} given by $\mu_n = 2\sqrt{t^2-\Delta^2} \cos \big( \frac{n\pi}{L+1} \big)$ with $n \in \{1,\dots, L\}$, which we refer to as \textit{Majorana points}, the multiplicative finite chain does not present exact zero energy modes for identical parent Hamiltonians with equal parameter sets such that $t_1 = t_2$, $\Delta_1 = \Delta_2$, and $\mu_1 = \mu_2$, unless $\frac{t_1}{\Delta_1}=\frac{t_2}{\Delta_2}=1$. The latter configuration is represented in Fig.~\ref{fig:parent-child-vs-mu-1}, where the exact zero energy dependence on the chemical potential of a multiplicative chain is qualitatively similar to the behavior of its two identical parents. Finite-size effects can lead to more significant differences between parent and child spectra, however. As shown in Fig.~\ref{fig:parent-child-vs-mu-2} for $\Big \lvert \frac{t_1}{\Delta_1} \Big \rvert =\Big \lvert \frac{t_2}{\Delta_2} \Big \rvert=2$, the presence of exact zero modes in both identical parents, shown in Fig.~\ref{fig:parent-child-vs-mu-2}(a,b) for different chain lengths, does not imply that a finite multiplicative chain also possesses exact zero modes. In fact, we observe in Fig.~\ref{fig:parent-child-vs-mu-2}(c-f) that if $t_i\neq\Delta_i$, the parents must be non-identical  for the child to have exact zero modes, considering the example for which $\text{sign} (\frac{t_1}{\mu_1})=-\text{sign}(\frac{t_2}{\mu_2})$. Identical parents are shown by Fig.~\ref{fig:parent-child-vs-mu-2}(c-d) for different chain lengths. In these cases exact zero energies are not obtained in finite chains.

The spectral dependence on the chemical potential reveals some interesting differences between the multiplicative chain and its parents. In the latter, the number of zero modes is given by the chain's length $L$ (see Fig.~\ref{fig:parent-child-vs-mu-2}a-b), while in the former the parity of the number of the zero modes is always even when the necessary conditions  $\frac{t_1}{\Delta_1}\neq 1$ and $\frac{t_2}{\Delta_2} \neq 1$ for exact zero modes with distinct parents is satisfied, regardless of the chain length's parity (see Fig.~\ref{fig:parent-child-vs-mu-2}(e-f)). Furthermore, dependence of the child spectra on free parameters shows greater variety than expected: the spectra shown in Fig.~\ref{fig:parent-child-vs-mu-2}(c) and (d), for instance, display a quadratically dispersing child spectrum, which results quite naturally from the child's tensor product combination of two linearly dispersing Kitaev Hamiltonians. This is a fairly general characteristic of multiplicative models. However, a linear dispersion can be obtained when $\text{sign}(\frac{t_1}{\mu_1})=-\text{sign}(\frac{t_2}{\mu_2})$ as shown in Fig.~\ref{fig:parent-child-vs-mu-2}(e) and Fig.~\ref{fig:parent-child-vs-mu-2}(f). Both the quadratic and linear dispersions are explained in Supplementary section~\ref{Evsmuexpl} in Eqn.~\eqref{Evsmuexplplus} and Eqn.~\eqref{Evsmuexplminus}. Such results demonstrate the rich interplay between finite-size topology and multiplicative topological phases.

\subsubsection{Localization of topologically-protected boundary modes of the MKC}

\begin{figure}[h]
\includegraphics[width=8cm]{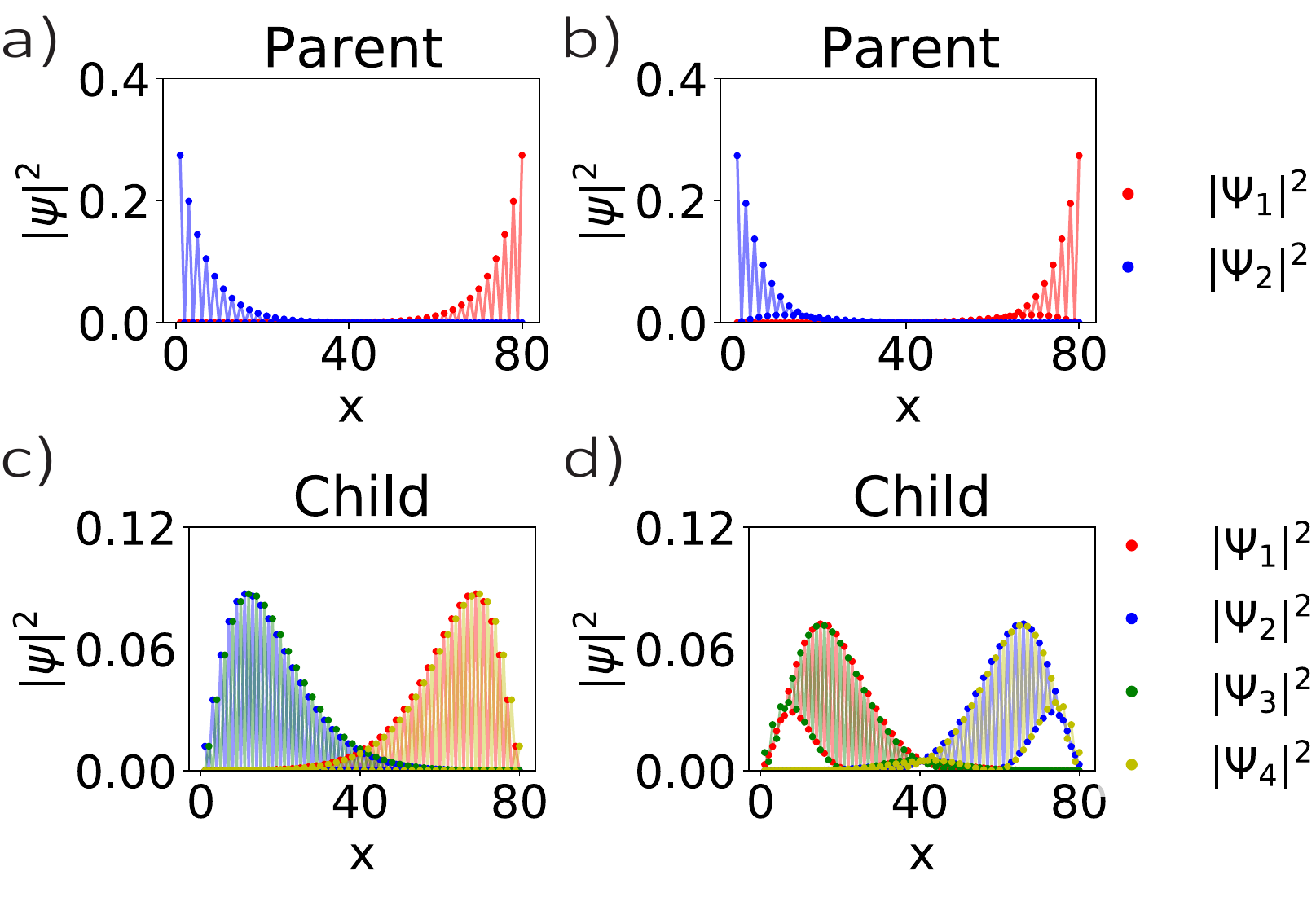}
\caption{a) Bound states for the Kitaev chain in the topological region with $t=1$, $\Delta=0.08$ and $\mu=0$. b) Ground states for the Kitaev chain for $t=1$, $\Delta=0.08$ and $\mu=0.09$. c) Bound states for a multiplicative chain with two identical parents in the topological phase, each shown in subfigure (a). Here $\Psi_{1,2} = \Psi'_1 \pm \Psi'_3$ and $\Psi_{3,4} = \Psi'_2 \pm \Psi'_4$ with $\Psi'$ an eigenvector of the finite Hamiltonian.
d) Bound states for a multiplicative chain with two identical parents in the topological phase, each shown in subfigure (b). Here $\Psi_{1,2} = \Psi'_1 \pm \Psi'_4$ and $\Psi_{3,4} = \Psi'_2 \pm \Psi'_3$ with $\Psi'$ an eigenvector of the finite Hamiltonian.
}
\label{fig:parent-child-vs-x-1}
\end{figure}

Spectral properties of the Kitaev chain are generally studied in combination with additional characteristics of the unpaired Majorana zero-mode states to more fully characterize the topologically non-trivial phase of the model. In particular, probability density of the in-gap state wavefunctions is an important measure of localization and  robustness of the Majorana zero-modes in the topologically non-trivial phase.
We therefore also compare and contrast the Kitaev chain and the MKC in terms of probability density distributions for topologically-protected in-gap states. These results are shown in Fig.~\ref{fig:parent-child-vs-x-1}. Similarly to the Kitaev chain, we observe that the MKC zero-modes can be spatially separated from each other, with their probability densities peaking near opposite ends of the chain and on sites of different parity. When there are four degenerate zero-modes in the MKC, two are localized at each end of the chain, instead of one zero-mode localized at each end of the Kitaev chain.

Interestingly, two parent Hamiltonians with mid-gap states that decay exponentially towards the bulk do not give rise to the same behavior in the multiplicative chain. As shown in Fig.~\ref{fig:parent-child-vs-x-1}, the boundary modes of the child Hamiltonian peak in probability density away from the ends of the chain, though still predominantly near one end or the other. The nature of the decay depends on the size of the bulk gap, which is naturally smaller for the multiplicative model than the parents for small gaps.

\subsubsection{Robustness of the MKC parallel MZMs:}
Before proceeding further, one must check for the robustness of the MZMs for the MKC parallel system to local disorder. We know that for the two-band Kitaev Chain, the MZMs persist when subject to local disorder proportional to $\sigma^z$ and $\sigma^y$ in the particle-hole basis. Only when the local disorder is proportional to $\sigma^x$ in the particle-hole basis are the MZMs shifted from zero energy.
\begin{figure}
\includegraphics[scale=0.5]{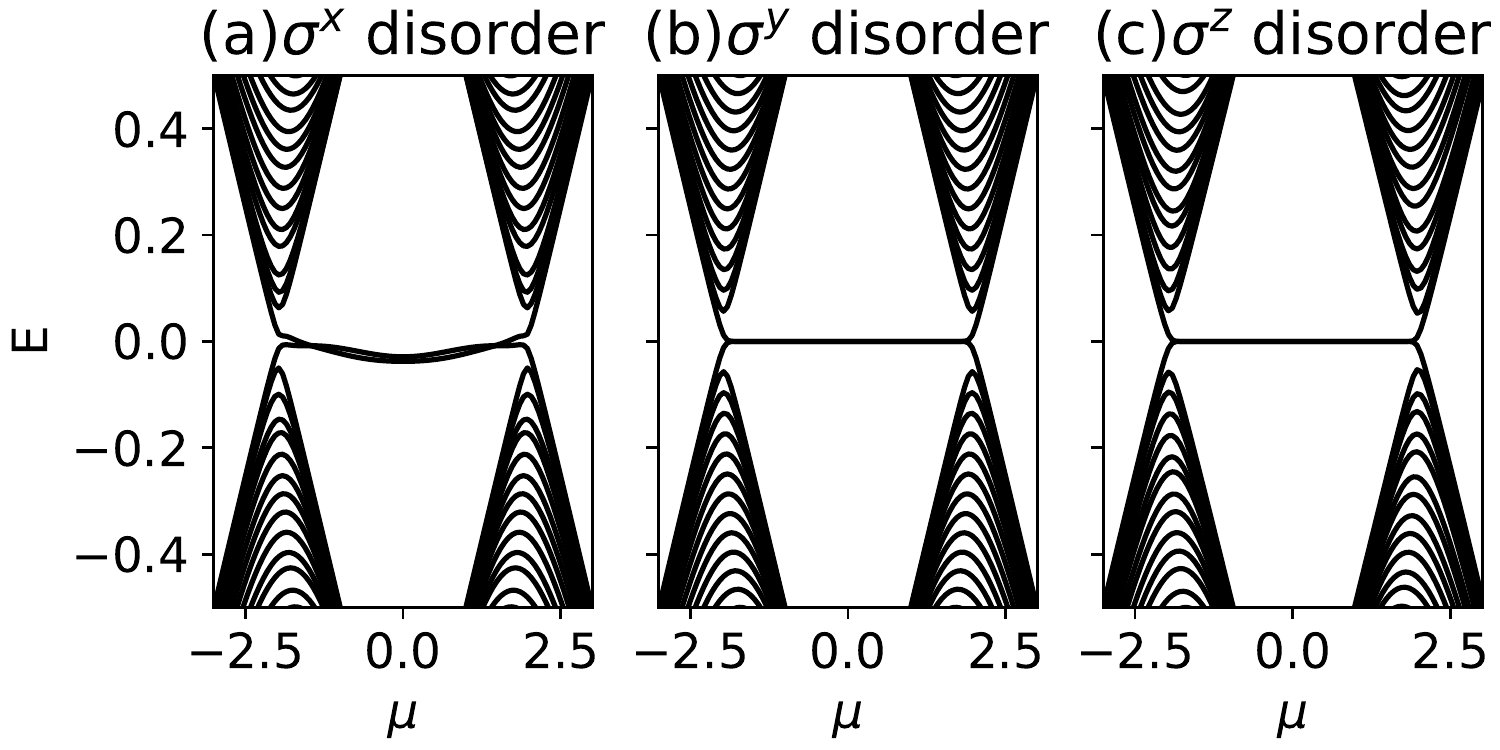}
\caption{Checking the robustness of Kitaev chain to disorder proportional to (a)$\sigma^x$, (b)$\sigma^y$ and (c)$\sigma^z$. MZMs are robust for $\sigma^y$ and $\sigma^z$ disorder while they break off from zero energy for $\sigma^x$ disorder.}
\label{KCrobust}
\end{figure}
We similarly investigate effects of myriad disorder terms for the MKC parallel system.
\begin{figure}[htb!]
    \includegraphics[scale=0.55]{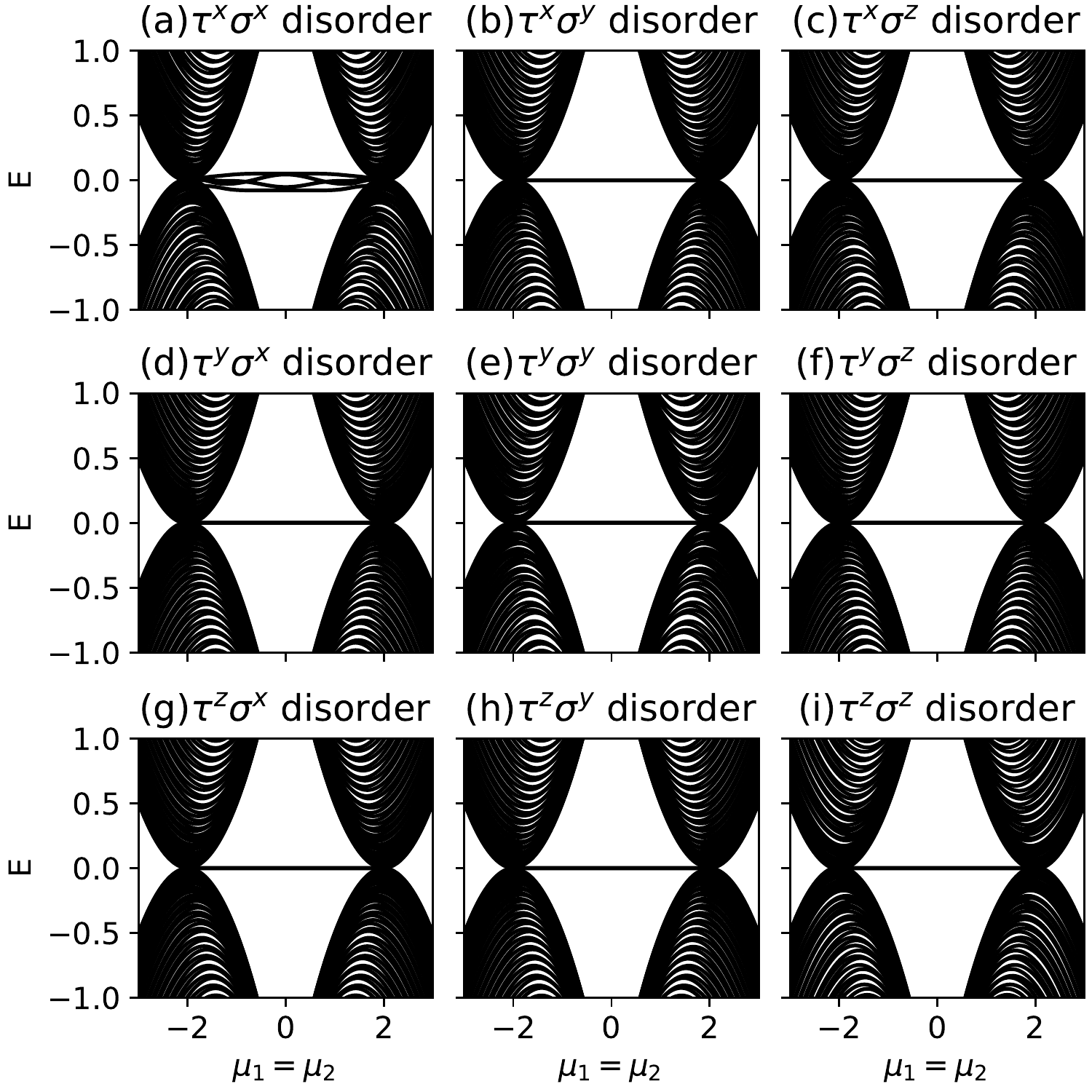}
    \caption{Checking for robustness of the MKC parallel MZMs in the presence of various onsite disorders with magnitude $0.2t$. In the case $\mu_1=\mu_2$, only the onsite disorder proportional to $\tau^x\sigma^x$ perturbs the MZMs from zero energy, signifying that the MZMs in the parallel MKC system are naturally more robust than their consitutent parents.}
    \label{MKCpllrobustmu1eqmu2}
\end{figure}
We similarly investigate effects of myriad disorder terms for the MKC parallel system. We have both the particle-hole and spin basis in this case, however, so we must check for all possible tensor-product combinations of local disorder. We observe that the MZMs persist at zero energy for local disorder proportional to any of the combinations $\tau^i\sigma^j$, where $i,j\in\{0,y,z\}$ if at least one of the parents is topological. Also it is robust to local disorder proportional to $\tau^z\sigma^x$, $\tau^x\sigma^z$, $\tau^y\sigma^x$, and $\tau^x\sigma^x$ if both the parents are topological. The flat bands corresponding to the MZMs only break down when the local disorder is proportional to $\tau^x\sigma^x$ even if one of the parents is topological. This suggests that the MKC parallel child MZMs are more robust than those of its parents.

\subsection{Parallel MKC Hamiltonian in real-space}\label{MKC_pll_realsp}

\begin{figure*}[htb]
    \hspace*{-1.2cm}
    \includegraphics[scale=0.47]{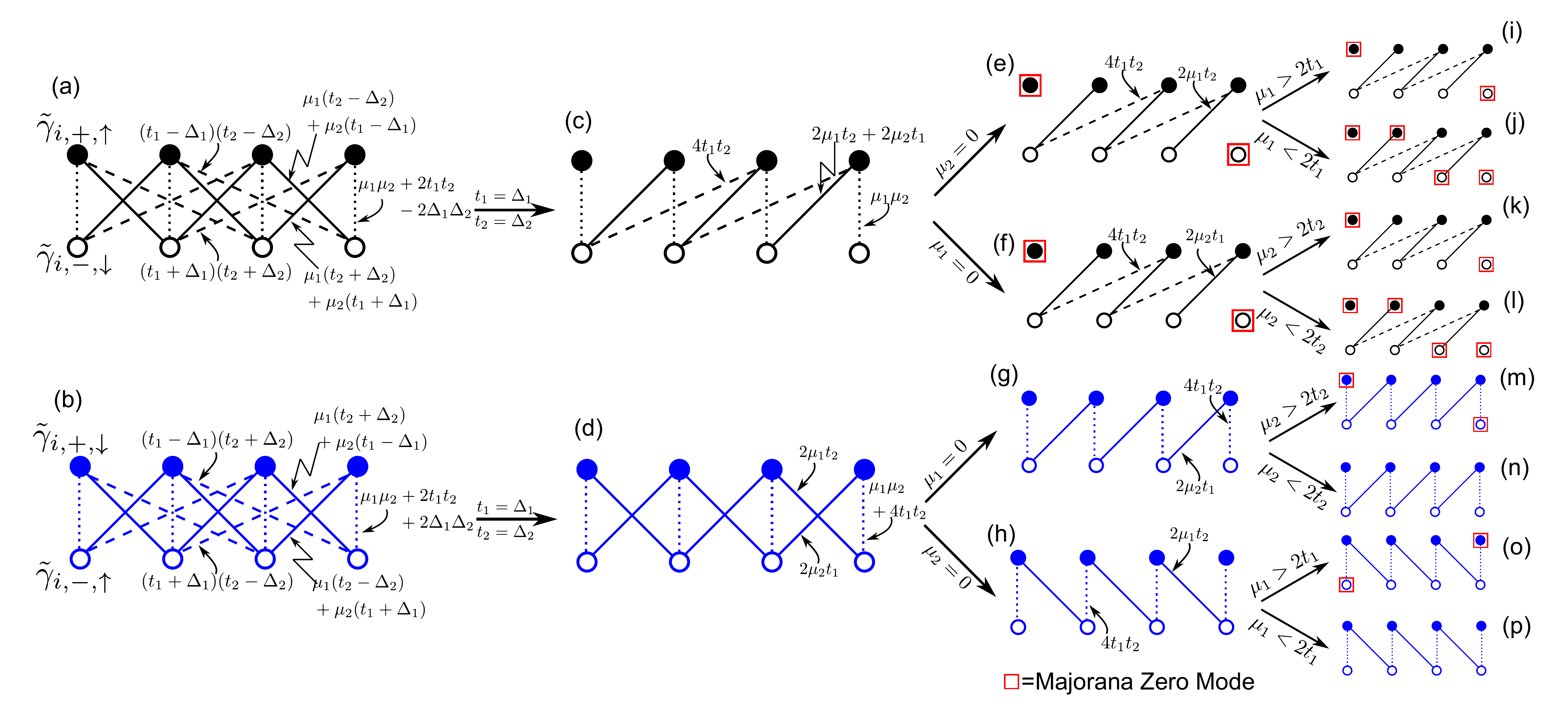}
    \caption{MKC parallel Hamiltonian in the Majorana basis. The exact Majorana zero modes for the cases where where one parent is topological and the other trivial and the case where both parents are topological are shown schematically. The two Hamiltonians $H_{1,||}$ and $H_{2,||}$ are represented in colors (black) and (blue) in (a) and (b) respectively. (c) and (d) refers to the modified system when one imposes the condition, $t_i=\Delta_i$, ($i=\{1,2\}$). (f) and (g) show the respective systems when the first parent is topological with $\mu_1=0$. (e) and (h) show the respective systems when the second parent is topological with $\mu_2=0$. The last column shows different outcomes for the positions of Majorana Zero modes when the either only one parent is topological or both of them are topological. The (red) square indicates the position of the Majorana Zero Mode.}
    \label{MKCpllphases}
\end{figure*}

The lattice Hamiltonian in the Majorana representation shows the different phases of the Kitaev Chain as well as the MKC rewritten in terms of different SSH models. We utilise a diagrammatic approach in Fig. \ref{MKCpllphases} to provide a clear description about the position of the Majorana zero modes and also an analytical explanation of the features we have shown numerically.

We have numerically observed that the MKC has unpaired Majorana bound states in even quantities. We use this fact to analytically characterize the MKC, by defining spinful Majoranas via the following expression: $c_{j,\sigma}=\frac{1}{2}(\gamma_{j,+,\sigma}+i\gamma_{j,-,\sigma})$. We may then, for a given lattice site, group two such Majoranas with opposite spins into the two-component vectors, $\gamma_{j,+}=(\gamma_{j,+,\uparrow},\gamma_{j,+,\downarrow})$ and $\gamma_{j,-}=(\gamma_{j,-,\uparrow},\gamma_{j,-,\downarrow})^T$, so that we can visualize any analysis of the possible phases. The MKC parallel Hamiltonian is then shown as follows,
\begin{equation}
\begin{split}
H^c_{\text{MKC},||} =&\frac{i}{2}\sum_{j}-\gamma_{j,+}[(\mu_1\mu_2+2t_1t_2)\sigma^z-2i\Delta_1\Delta_2\sigma^y]\gamma_{j,-}\\
&-\gamma_{j,+}[((t_2\mu_1+t_1\mu_2)-\mu_2\Delta_1)\sigma^z-i\mu_1\Delta_2\sigma^y]\gamma_{j+1,-}\\
&-\gamma_{j+1,+}[((t_2\mu_1+t_1\mu_2)+\mu_2\Delta_1)\sigma^z+i\mu_1\Delta_2\sigma^y]\gamma_{j,-}\\
&-(t_1-\Delta_1)\gamma_{j,+}(t_2\sigma^z-i\Delta_2\sigma^y)\gamma_{j+2,-}\\
&-(t_1+\Delta_1)\gamma_{j+2,+}(t_2\sigma^z+i\Delta_2\sigma^y)\gamma_{j,-}.
\end{split}
\end{equation}
In this form, three kinds of interaction terms are distinguishable, which are the onsite-interaction, the nearest-neighbour interaction and the next-nearest-neighbour interaction. The matrix structure of the coefficients imply the presence of inter-spin interactions.

To visualize the Majorana bound states, we perform a similarity transformation, $h\rightarrow UhU^\dagger$, $\gamma_{j,+}\rightarrow \tilde{\gamma}_{j,+}=\gamma_{j,+}U^\dagger$ and $\gamma_{j,-}\rightarrow \tilde{\gamma}_{j,-}=U\gamma_{j,-}$ where, $U=\frac{1}{\sqrt{2}}(\sigma^0-i\sigma^y)$, after which the two components of $\tilde{\gamma}_{j,\pm}$ still satisfy the Majorana anti-commutation relations, $\{\tilde{\gamma}_{j},\tilde{\gamma}_{k}\}=2\delta_{jk}$. The transformation $U$ changes $\sigma^z$ to $\sigma^x$, so that the resulting Hamiltonian is off-diagonal and it can be separated into two separate inter-spin coupling parts,
\begin{equation}\label{MKCpll_Hcomponents}
\begin{split}
    H^c_{\text{MKC},||} =& \frac{i}{2}\sum_j[-(\mu_1\mu_2+2t_1t_2-2\Delta_1\Delta_2)\tilde{\gamma}_{j,\uparrow,+}\tilde{\gamma}_{j,\downarrow,-}\\
    &-(\mu_1(t_2-\Delta_2)+\mu_2(t_1-\Delta_1))\tilde{\gamma}_{j,\uparrow,+}\tilde{\gamma}_{j+1,\downarrow,-}\\
    &-(\mu_1(t_2+\Delta_2)+\mu_2(t_1+\Delta_1))\tilde{\gamma}_{j+1,\uparrow,+}\tilde{\gamma}_{j,\downarrow,-}\\
    &-(t_1-\Delta_1)(t_2-\Delta_2)\tilde{\gamma}_{j,\uparrow,+}\tilde{\gamma}_{j+2,\downarrow,-}\\
    &-(t_1+\Delta_1)(t_2+\Delta_2)\tilde{\gamma}_{j+2,\uparrow,+}\tilde{\gamma}_{j,\downarrow,-}]\\
    &+\frac{i}{2}\sum_j[-(\mu_1\mu_2+2t_1t_2+2\Delta_1\Delta_2)\tilde{\gamma}_{j,\downarrow,+}\tilde{\gamma}_{j,\uparrow,-}\\
    &-(\mu_1(t_2+\Delta_2)+\mu_2(t_1-\Delta_1))\tilde{\gamma}_{j,\downarrow,+}\tilde{\gamma}_{j+1,\uparrow,-}\\
    &-(\mu_1(t_2-\Delta_2)+\mu_2(t_1+\Delta_1))\tilde{\gamma}_{j+1,\downarrow,+}\tilde{\gamma}_{j,\uparrow,-}\\
    &-(t_1-\Delta_1)(t_2+\Delta_2)\tilde{\gamma}_{j,\downarrow,+}\tilde{\gamma}_{j+2,\uparrow,-}\\
    &-(t_1+\Delta_1)(t_2-\Delta_2)\tilde{\gamma}_{j+2,\downarrow,+}\tilde{\gamma}_{j,\uparrow,-}],\\
    =& H_{||,1}+H_{||,2}.
\end{split}
\end{equation}
It is thus possible to view the problem as two separate systems as shown in Fig.~\ref{MKCpllphases}(a) and (b) and then consider a case-by-case approach. We denote these two commuting parts, the \textit{component Hamiltonians}, by $H_{||,1}$ and $H_{||,2}$. We assume that $t_i=\Delta_i$, $i\in \{1,2\}$ and explore the different phases derived thereof from Fig.~\ref{MKCpllphases}(c) and (d) corresponding to the phases of the parent Hamiltonians.
\\
\textit{Case 1}: The first parent is topological with $\mu_1=0$ and ths second one is trivial with $\mu_2>2t_2$. This is illustrated in Fig.~\ref{MKCpllphases}(f), (k) and (g), (m) for components $H_{||,1}$ and $H_{||,2}$ respectively. The condition, $\mu_2>2t_2$ implies that the KC in (g) is topological with two Majorana zero modes and $\mu=0$ already provides two Majorana zero modes in (f). Therefore we have four Majorana edge modes, all situated at the first and last sites of the MKC parallel system.\\
\linebreak
\textit{Case 2}: The second parent is topological with $\mu_2=0$ and the first one is trivial with $\mu_1>2t_1$. This leads to a similar situation as in Case 1 with respect to the position of the Majorana zero modes and is illustrated by Fig.~\ref{MKCpllphases}(e), (i) and (h), (o) for components $H_{||,1}$ and $H_{||,2}$ respectively. We again have four Majorana zero modes, two from each component at the first and last sites of the MKC parallel system. One must however notice that the spin configuration at the first site and the last sites are parallel unlike Case 1 where the spins are anti-parallel.\\
\linebreak
\textit{Case 3}: Both parents are topological, i.e. $\mu_1=0$, $\mu_2<2t_2$ and $\mu_1<2t_1$, $\mu_2=0$. This case is illustrated by Fig. \ref{MKCpllphases}(k), (m) and (n), (p) for the components $H_{||,1}$ and $H_{||,2}$ respectively. Observe that no Majorana zero modes are present in $H_{||,2}$ while, for $H_{||,1}$, we have Majorana zero modes at positions 1, 2 and $L-1$, $L$ for $L$ sites.

\subsubsection{Topology of the MKC parallel from the component Hamiltonians:}
It is possible to derive Bloch Hamiltonians from the component Hamiltonians which should look like our usual two-band Kitaev chains but with next-nearest neighbour coupling. We define $\tilde{c}_{j,\sigma}=\frac{1}{2}(\tilde{\gamma}_{j,\sigma,+}+i\tilde{\gamma}_{j,\sigma,-})$, and then from Eqn.~\eqref{MKCpll_Hcomponents}, we write,
\begin{equation}
H_{\text{MKC},||}^c = \frac{1}{2}\sum_{k}\tilde{\mathbf{c}}^\dagger_{k,1}\mc{H}_{||,1}(k)\tilde{\mathbf{c}}_{k,1}+\frac{1}{2}\sum_{k}\tilde{\mathbf{c}}^\dagger_{k,2}\mc{H}_{||,2}(k)\tilde{\mathbf{c}}_{k,2},
\end{equation}
\begin{subequations}
\begin{equation}
\begin{split}
\mc{H}_{||,1}(k) =& -[2(\mu_1t_2+\mu_2t_1)\cos k+2(t_1t_2+\Delta_1\Delta_2)\cos 2k\\
&+\mu_1\mu_2+2t_1t_2-2\Delta_1\Delta_2]\sigma^z\\
&+[2(\mu_2\Delta_1+\mu_1\Delta_2)\sin k\\
&+2(t_2\Delta_1+t_1\Delta_2)\sin 2k]\sigma^y=\mathbf{d}_1(k)\cdot \boldsymbol{\sigma},
\end{split}
\label{Hpll1Bloch}
\end{equation}
\begin{equation}
\begin{split}
\mc{H}_{||,2}(k) =& -[2(\mu_1t_2+\mu_2t_1)\cos k+2(t_1t_2-\Delta_1\Delta_2)\cos 2k\\
&+\mu_1\mu_2+2t_1t_2+2\Delta_1\Delta_2]\sigma^z\\
&+[2(\mu_2\Delta_1-\mu_1\Delta_2)\sin k\\
&+2(t_2\Delta_1-t_1\Delta_2)\sin 2k]\sigma^y=\mathbf{d}_2(k)\cdot \boldsymbol{\sigma},
\end{split}
\label{Hpll2Bloch}
\end{equation}
\end{subequations}
where $\tilde{c}_{k,1}=(\tilde{c}_{k,\uparrow}, \tilde{c}^\dagger_{-k,\downarrow})^T$ and $\tilde{c}_{k,2}=(\tilde{c}_{k,\downarrow},\tilde{c}^\dagger_{-k,\uparrow})^T$. Here, each of the component Hamiltonians Eq.~\ref{Hpll1Bloch} and Eq.~\ref{Hpll2Bloch} result in the non-degenerate energy dispersion $E(k)$ from Eqn.~\eqref{MKCplldispersion} which are equivalent to the MKC parallel dispersion. Next, we study the winding number for the two component Bloch Hamiltonians by constructing the parametric curves $\mathbf{d}_1(k)$ and $\mathbf{d}_2(k)$ from Eqn.~\eqref{Hpll1Bloch} and Eqn.~\eqref{Hpll2Bloch} when $k$ is varied in the interval $[0,2\pi)$. For each of the parent Kitaev chains, the system is said to be in the topological phase with winding number $\mc{W}=\pm 1$ if the parametric curve winds around the origin once. At the critical point, the parametric curve intersects the origin while, in the trivial phase, it does not wind around the origin at all. Based on similar views, we try to infer the parametric curves due to our component Hamiltonians.

\begin{center}
\begin{figure}
    \hspace*{-1cm}
    \includegraphics[scale=0.6]{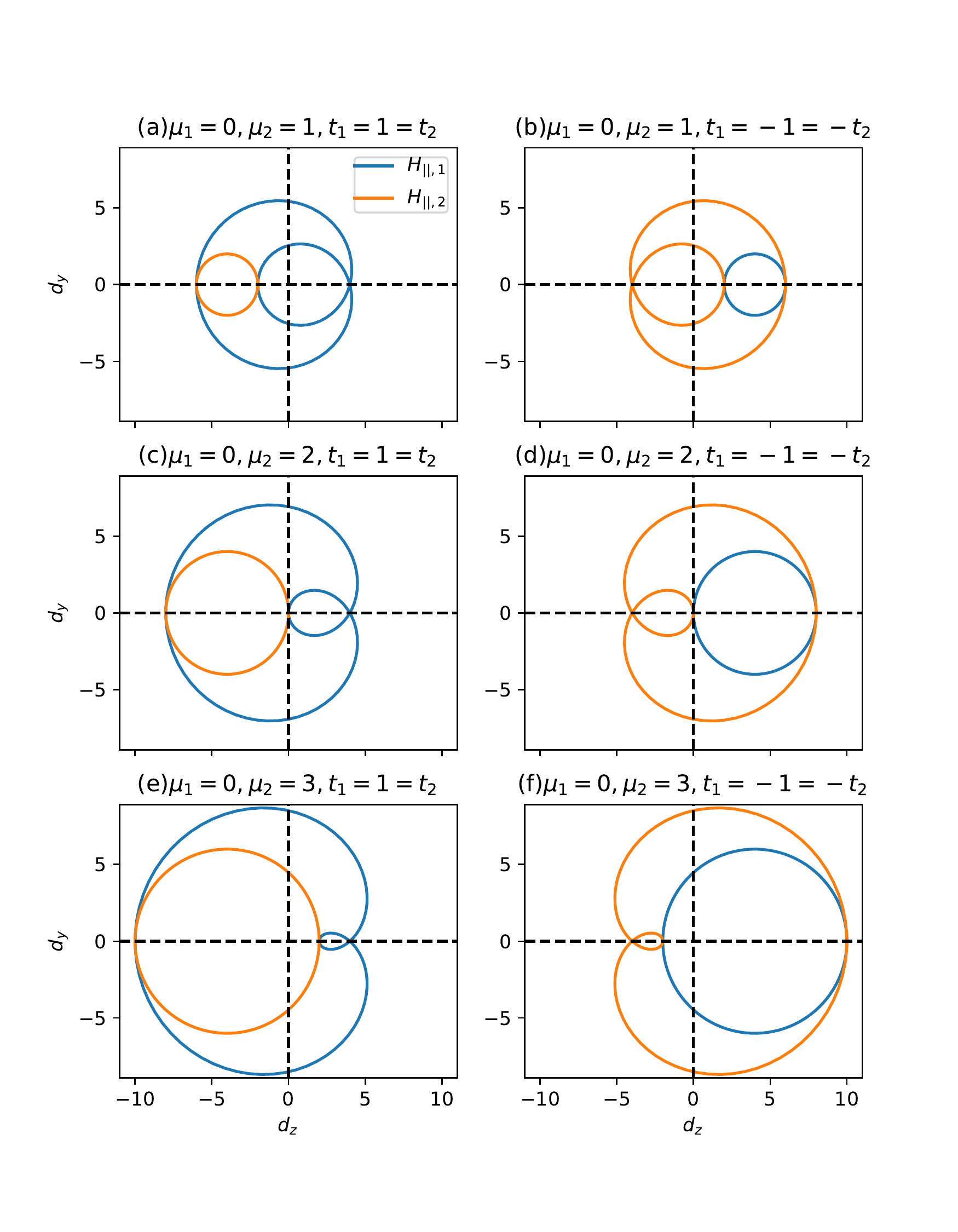}
    \caption{Parametric curves $\mathbf{d}_1(k)$ and $\mathbf{d}_2(k)$ for the Hamiltonian components, $H_{||,1}$(blue) and $H_{||,2}$(orange) respectively as $k$ is varied in the interval $[0,2\pi)$. The winding of the curves around the origin show the different topological characteristics for different values of $\mu_2$ with $\mu_1=0$ for the cases, $t_1=1=t_2$(first column, (a),(c),(e)) and $t_1=-1=t_2$(second column, (b),(d),(f)) at $\Delta_1=\Delta_2=1$ for all cases.}
    \label{MKCpll_winding}
\end{figure}
\end{center}
From Fig.~\ref{MKCpll_winding}(a) and (b), we observe that for $t_1=t_2=1=\Delta_1=\Delta_2$, when both the parent KCs are topological, i.e., $\mu_i<2t_i$, $i\in\{1,2\}$, the curve due to $\mathbf{d}_1(k)$ winds around the origin twice while the curve from $\mathbf{d}_2(k)$ does not wind around the origin at all, giving rise to an overall winding number, $\mc{W}=2$ and this is exactly as we expected from our earlier analysis from Fig.~\ref{MKCpllphases} which shows that $H_{||,1}$ contains two pairs of MZMs while $H_{||,2}$ contains no MZMs. We also check all the three critical points - when either one of the parents are critical or both of them are, in which case both the parametric curves intersect the origin, albeit in different configurations. For example, Fig.~\ref{MKCpll_winding}(c) and (d) show the case when one parent is topological while the other is critical. Finally, we consider the case in which one parent is topological while the other is trivial ($\mu_1<2t_1$ and $\mu_2>2t_2$ or vice-versa). The winding for each component Hamiltonian in this case is shown in Fig.~\ref{MKCpll_winding}(e) and (f). We see that both curves derived from $\mathbf{d}_1(k)$ and $\mathbf{d}_2(k)$ each wind around the origin once, giving rise to the winding number $\mc{W}=1\oplus 1$. This is again consistent with our discussion due to Fig.~\ref{MKCpllphases} where each of the component Hamiltonians carry one pair of MZMs each.\\

Here, one might think that the MZMs on the same site of the MKC in the topological-trivial case should hybridize. But, in  Sec. \ref{sec:parent_model_hamiltonians}, we had already discovered that we have an emergent unitary symmetry. One may block diagonalize in the Bell-state basis of this symmetry, $\mc{U}=\tau^x\sigma^x$ to recover the exact component Hamiltonians we have described in this section. Hence, it the presence of this unitary symmetry which protects the two MZMs on the same site of the MKC from hybridizing, leading to separate winding number descriptions.

\subsubsection{Explanation for Majorana points in the $t_i\neq\Delta_i$ case:}\label{schematictneqdel}
\begin{center}
\begin{figure}[htb]
    \includegraphics[scale=0.45]{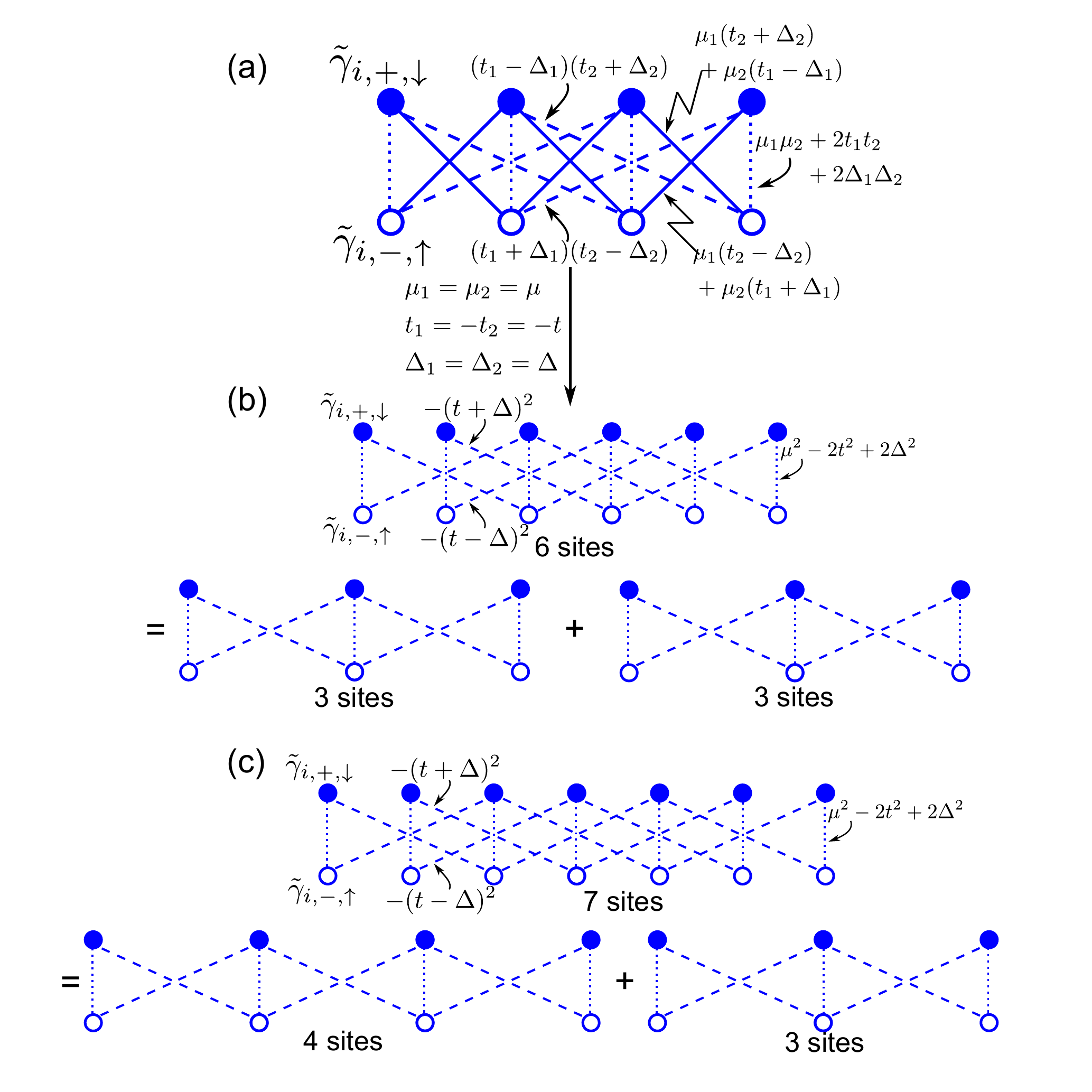}
    \caption{Schematic diagram of $H_{||,2}$(a) for the particular case, $\mu_1=\mu_2=\mu$, $t_1=-t_2=-t$ and $\Delta_1=\Delta_2=\Delta$ and $t\neq \Delta$. Illustrating for the cases (b)$L=6$ and (c)$L=7$, the system can now be broken down to two 3-site KCs or one 4-site and another 3-site KC. The zero energy Majorana points can be explained from here.}
    \label{MKCpllt_neq_delta}
\end{figure}
\end{center}

Using this diagrammatic approach, we can gain greater understanding of the exact zero-modes prominent for finite size MKC.
In Fig. \ref{fig:parent-child-vs-mu-2}(e) and (f), we observe that for the parameters, $|t_i|=2|\Delta_i|$, $i\in \{1,2\}$ and $t_1=-t_2$, one gets bubbles for the two energy levels near zero vs. $\mu_1=\mu_2$. Notably, there is a difference in the positions of the zero energy or Majorana points between systems with an even vs. odd number of lattice sites in a finite size MKC parallel system. Systems with an even number of sites, as shown in Fig. 6 (e), exhibit a two-fold degeneracy in the spectrum, here high-lighted by dashed blue and orange lines, while systems with an odd number of sites exhibit more complex structure for the low-energy states occurring for open-boundary conditions as shown in Fig.~6 (f). The rich structure in this case results because the full chain consists of effectively two decoupled subsystem chains derived in the schematic diagram Fig.~\ref{MKCpllt_neq_delta} from Fig.~\ref{MKCpllphases}(b) corrsponding to $H_{||,2}$. As shown in Fig.~\ref{fig:parent-child-vs-mu-2}(a) and (b), the number of Majorana points changes with chain length, so the spectra of the two subsystem chains will not coincide in this case.

As we can see, when $\mu_1=\mu_2=\mu$, $t_1=-t_2=-t$, $\Delta_1=\Delta_2=\Delta$, the component Hamiltonian, $H_{||,2}$ in Eq. ~\eqref{Hpll2Bloch} possesses only next-nearest neighbour interactions and thus can be split into two KCs so that the number of sites add up to the number of sites for the original system. Then for a system with $2L$ lattice sites, we get two KCs of length $L$, while for a system with $2L+1$ sites, we get two KCs of length $L$ and $L+1$ respectively. We know for an $L$-site KC with parameters, $\mu^\prime$, $t^\prime$ and $\Delta^\prime$, the zero energy Majorana points are found at $\mu^\prime = 2\sqrt{{t^{\prime}}^2-{\Delta^\prime}^2}\cos \frac{n\pi }{L+1}$, where $n\in \{1,...,L\}$ i.e. there are $L$ Majorana points. We apply a similar calculation to our KCs with the mapping $\mu^\prime=\mu^2-2t^2+2\Delta^2$, $t^\prime+\Delta^\prime = -(t+\Delta)^2$, $t^\prime-\Delta^\prime = -(t-\Delta)^2$ derived from Fig.~\ref{MKCpllt_neq_delta}. Then we get the following identity,
\begin{equation}{\label{MKCMajPoints}}
\begin{split}
&\mu = \pm\sqrt{2(t^2-\Delta^2)\big{(}1+\cos \frac{n\pi}{L+1}\big{)}},\quad n\in \{1,...,L\},\\
\implies &\mu = 2\sqrt{t^2-\Delta^2}\cos\frac{n\pi}{2L+2},\quad n\in\{1,...,2L+2\}.
\end{split}
\end{equation}

for exact zero-modes in each of the split KC systems of length $L$. The square root explains why only even number of Majorana points are observed and why a KC of size $L$ produces $2L$ Majorana points. From this calculation, we infer that $H_{||,2}$ with $2L$ sites corresponds to $2L$ Majorana points, which are two-fold degenerate. Similarly, $H_{||,2}$ with $2L+1$ sites produces $2L\oplus 2(L+1)$ Majorana points due to contributions from each of the two subsystem KCs. The set of $\mu$ values corresponding to Majorana points  derived from Eq.~\eqref{MKCMajPoints},  $\{\mu_i \}$, agrees with the Majorana points shown from the numerical simulation in Fig. \ref{fig:parent-child-vs-mu-2}(e) and (f).

\subsubsection{Edge states of the MKC parallel system from the component Hamiltonians and entanglement:}
 As the topologically-protected bound states obtained from the MKC parallel system are distinct from the topologically-protected bound states of the constituent parents, both in their existence in parameter space and their entanglement structure, we define them separately as \textit{Multiplicative Majorana Zero Modes} or \textit{MMZMs} in short. We are finally in the position to discuss the full analytical expressions for the MMZMs. The component Hamiltonians $H_{||,1}$ and $H_{||,2}$ derived from the MKC parallel system each satisfy conditions for the null eigenvalue such that the four conditions outlined in Sec.~\ref{Edge_MKCpll} are subdivided into two conditions at a time for each of the component systems. As derived in~\ref{Sup_subsec_A1}, after localization $k\rightarrow iq$, $H_{||,1}$ and $H_{||,2}$ are given as,
\begin{subequations}
\begin{equation}
\begin{split}
\mc{H}_{||,1}(iq) =& -[(\mu_1+2t_1\cosh q)(\mu_2+2t_2\cosh q)\\
&+4\Delta_1\Delta_2\sinh^2q]\sigma^z\\
&+i[2\Delta_1\sinh q(\mu_2+2t_2\cosh q)\\
&+2\Delta_2\sinh q(\mu_1+2t_1\cosh q)]\sigma^y
\end{split}
\label{Hpll1loc}
\end{equation}
\begin{equation}
\begin{split}
\mc{H}_{||,2}(iq) =& -[(\mu_1+2t_1\cosh q)(\mu_2+2t_2\cosh q)\\
&-4\Delta_1\Delta_2\sinh^2q]\sigma^z\\
&+i[2\Delta_1\sinh q(\mu_2+2t_2\cosh q)\\
&-2\Delta_2\sinh q(\mu_1+2t_1\cosh q)]\sigma^y
\end{split}
\label{Hpll2loc}
\end{equation}
\end{subequations}
The condition to get null eigenvalues from the above expressions is,
\begin{equation}
\begin{split}
&[(2t_1\cosh q+\mu_1)\mp 2\Delta_1\sinh q]\\
&\times[(2t_2\cosh q+\mu_2)\mp 2\Delta_2\sinh q]=0.
\end{split}
\label{Hpll1cond}
\end{equation}
for the component $H_{||,1}$ and,
\begin{equation}
\begin{split}
&[(2t_1\cosh q+\mu_1)\mp 2\Delta_1\sinh q]\\
&\times[(2t_2\cosh q+\mu_2)\pm 2\Delta_2\sinh q]=0.
\end{split}
\label{Hpll2cond}
\end{equation}
for the component $H_{||,2}$. From the schematic diagram Fig.~\ref{MKCpllphases}, it may be observed that based on the topological nature of the two parents, the MMZMs are localized in different ways. Let us consider the condition Eqn.~\ref{Hpll1cond} for $\text{sgn}(t_i)=\text{sgn}(\Delta_i)$, $i\in\{1,2\}$,
\begin{equation}\label{MZMcase1}
\begin{split}
&[(2t_1\cosh q+\mu_1)-2\Delta_1\sinh q]\\
&\times[(2t_2\cosh q+\mu_2)-2\Delta_2\sinh q]=0.
\end{split}
\end{equation}
\textit{If both the parents are topological} we have $2t_1\cosh q+\mu_1=2\Delta_1\sinh q$ and $2t_2\cosh q+\mu_2=2\Delta_2\sinh q$, which if substituted into Eqns.~\ref{Hpll1loc} and \ref{Hpll2loc} shows that $\mc{H}_{||,2}$ vanishes. The full basis of the MKC parallel system is given by four degrees of freedom,  $(\tilde{c}_{k,\uparrow},\tilde{c}_{k,\downarrow},\tilde{c}^\dagger_{-k,\uparrow},\tilde{c}^\dagger_{-k,\downarrow})^T$, by combining the degrees of freedom of the two components. In this basis, the null eigenvectors derived from $\mc{H}_{||,1}(iq)$ are given as,
\begin{equation}
\ket{\Psi}_{MMZM}=\{\frac{1}{\sqrt{2}}(\ket{00}-\ket{11}),\ket{01},\ket{10}\},
\end{equation}
where $\ket{0}=(1,0)^T$, $\ket{1}=(0,1)^T$.\\
Again, \textit{say only parent 1 is topological and parent 2 is trivial}, i.e. we only have the condition $2t_1\cosh q+\mu_1=2\Delta_1\sinh q$ to fulfil. Substituting into Eqns.~\ref{Hpll1loc} and \ref{Hpll2loc}, the null eigenvectors in the full basis with four degrees of freedom are shown to be,
\begin{equation}
\ket{\Psi}_{MMZM}=\{\frac{1}{\sqrt{2}}(\ket{00}-\ket{11}),\frac{1}{\sqrt{2}}(\ket{01}-\ket{10})\}.
\end{equation}
We would get the same null eigenvectors if parent 2 had been the only one topological. Detailed calculations can be found in Supplementary section~\ref{Sup_subsec_A1}. The interesting point to note here is that by changing the topological character of one of the parents it is possible to transition from a product state to a maximally entangled Bell state. We list all the possible eigenvectors for different combinations of topology of the parents and signs of $t_i$ compared to $\Delta_i$ in Table~I. Here it is important to remember that in each case, one has \textit{four} MMZMs. The table lists only the MMs at edge $x=0$. The eigenvectors at the other edge can be found by changing, $\frac{\text{sqn}(t_i)}{\text{sgn}(\Delta_i)}$ from $+$ to $-$ and vice-versa for both the parents. We will recover a total of four eigenvectors with two common eigenvectors for both signs when both parents are topological.\\\\


\begin{table*}[ht!]
\begin{tabular}{|c|c|c|c|c|}
\hline
\multicolumn{2}{|c|}{Parent 1} & \multicolumn{2}{c|}{Parent 2} & \multirow{2}{*}{MZM Eigenvectors} \\
\cline{1-4}
Phase & $\frac{\text{sgn}(t_1)}{\text{sgn}(\Delta_1)}$ & Phase & $\frac{\text{sgn}(t_2)}{\text{sgn}(\Delta_2)}$ & \\
\hline
\multirow{4}{*}{topo} & + & \multirow{4}{*}{topo} & + & $\{\frac{1}{\sqrt{2}}(\ket{00}-\ket{11}),\ket{01},\ket{10}\}$ or $\{\frac{1}{\sqrt{2}}(\ket{00}-\ket{11}),\frac{1}{\sqrt{2}}(\ket{01}-\ket{10})\}$\\
 & + &  & - & $\{\frac{1}{\sqrt{2}}(\ket{01}-\ket{10}),\ket{00},\ket{11}\}$ or $\{\frac{1}{\sqrt{2}}(\ket{01}-\ket{10}),\frac{1}{\sqrt{2}}(\ket{00}+\ket{11})\}$\\
 & - &  & + & $\{\frac{1}{\sqrt{2}}(\ket{01}+\ket{10}),\ket{00},\ket{11}\}$ or $\{\frac{1}{\sqrt{2}}(\ket{01}+\ket{10}),\frac{1}{\sqrt{2}}(\ket{00}-\ket{11})\}$\\
 & - &  & - & $\{\frac{1}{\sqrt{2}}(\ket{00}+\ket{11}),\ket{01},\ket{10}\}$ or $\{\frac{1}{\sqrt{2}}(\ket{00}+\ket{11}),\frac{1}{\sqrt{2}}(\ket{01}+\ket{10})\}$\\
\hline
\multirow{2}{*}{topo} & + & \multirow{2}{*}{triv} & & $\{\frac{1}{\sqrt{2}}(\ket{00}-\ket{11}),\frac{1}{\sqrt{2}}(\ket{01}-\ket{10})\}$\\
 & - &  &  & $\{\frac{1}{\sqrt{2}}(\ket{00}+\ket{11}),\frac{1}{\sqrt{2}}(\ket{01}+\ket{10})\}$\\
\hline
\multirow{2}{*}{triv} &  & \multirow{2}{*}{topo} & + & $\{\frac{1}{\sqrt{2}}(\ket{00}-\ket{11}),\frac{1}{\sqrt{2}}(\ket{01}+\ket{10})\}$\\
 &  &  & - & $\{\frac{1}{\sqrt{2}}(\ket{00}+\ket{11}),\frac{1}{\sqrt{2}}(\ket{01}-\ket{10})\}$\\
\hline
\end{tabular}
\caption{Null eigen-vectors of the MKC parallel system for different topological characterizations of the two parent systems, ratio of signs of $t_i$ and $\Delta_i$, $i\in\{1,2\}$, and boundary conditions.}
\label{MMZMpll}
\end{table*}

\subsubsection{Spatial distribution of MMZM wavefunctions for $N$-site MKC}

\begin{figure}[h!]
    \includegraphics[scale=0.6]{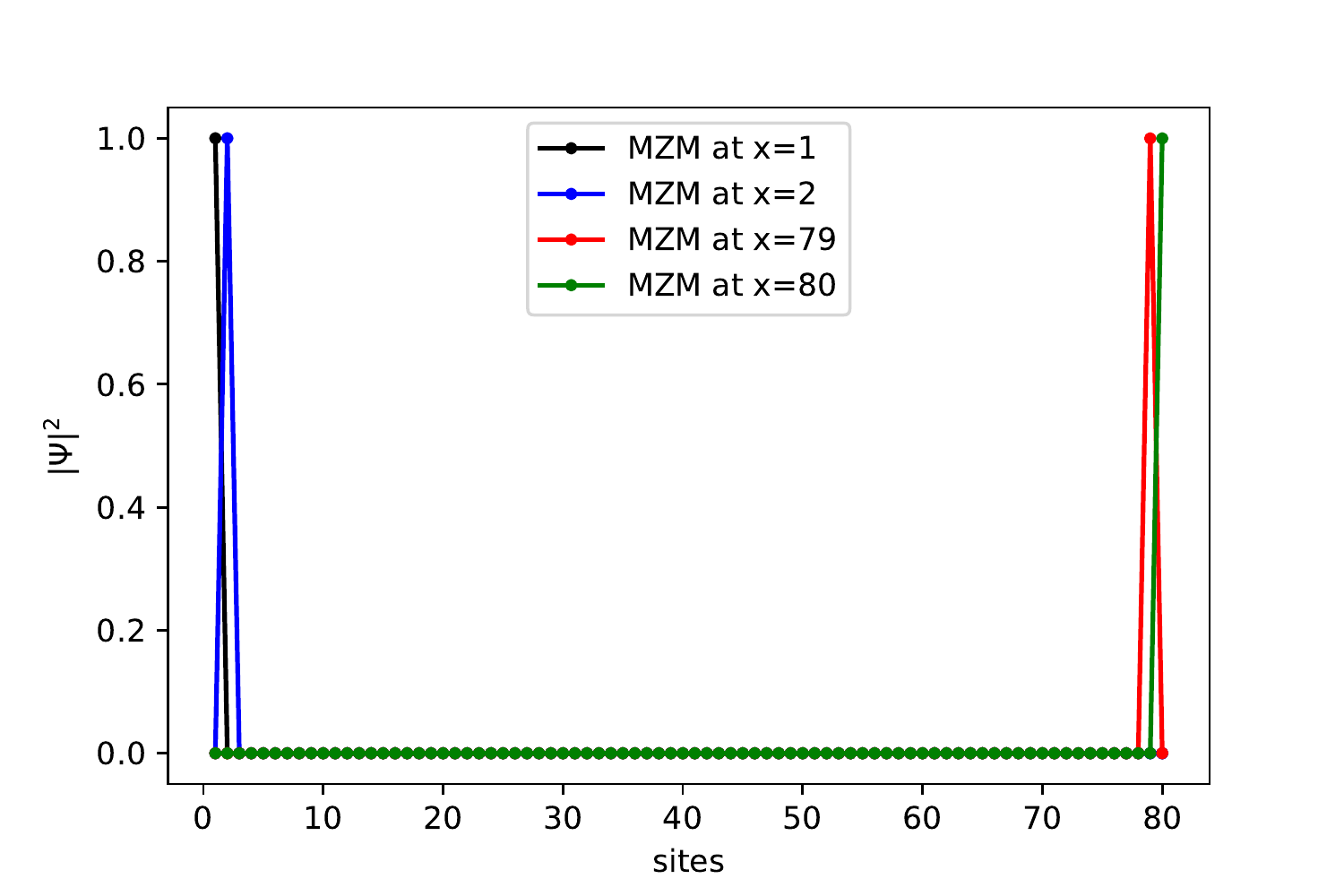}
    \caption{MMZMs for the MKC parallel system with $N=80$ sites for the parameter values $\mu_1=\mu_2=0$, $t_1=\Delta_1=1$ and $t_2=\Delta_2=1$ obtained numerically. We observe MMZMs at site indices 1, 2, 79 and 80 as inferred previously from the schematic diagram.}
    \label{MZMsp_wf}
\end{figure}

We now further characterize MMZM wavefunctions in the MKC parallel lattice with N sites by computing the associated spatially-resolved probability density for these states. For the specific Majorana point, $\mu_1=\mu_2=0$ for $t_1=\Delta_1$ and $t_2=\Delta_2$, the wavefunction must be a delta function at the two edges (site indices $j=1$ and $j=N$). As seen from the schematic diagram Fig.~\ref{MKCpllphases}, two more delta functions at site indices $j=2$ and $j=N-1$. We illustrate this with a numerical simulation for this specific case in Fig.~\ref{MZMsp_wf}.
For cases where $t_1\neq \Delta_1$ and/or $t_2\neq\Delta_2$, in finite size lattices, we have already seen numerically in Fig.~\ref{fig:parent-child-vs-mu-2}(c) and (d), that there are no Majorana zero points for the case $t_1=t_2$ and $\Delta_1=\Delta_2$. We therefore construct the wavefunction for the case where we have Majorana points available, namely Fig.~\ref{fig:parent-child-vs-mu-2}(e) and (f) where $t_1=-t_2$ and $\Delta_1=\Delta_2$. We illustrate just for the case highlighted in  Eqn.~\ref{MZMcase1}. Here, we obtain four values for $e^{-q}$, namely $\frac{-\mu_1\pm\sqrt{\mu_1^2-4(t_1^2-\Delta_1^2)}}{2(t_1+\Delta_1)}$ and $\frac{\mu_2\pm\sqrt{\mu_2^2-4(t_2^2-\Delta_2^2)}}{2(t_2+\Delta_2)}$. We require standing wave solutions for the finite size lattice, which require that we write down our four $e^{-q}$ values as $R_1e^{\pm i\theta_1}$ and $R_2e^{\pm i\theta_2}$ respectively. We hence propose a general form for the wavefunction,
\begin{equation}\label{genwf}
\begin{split}
\Psi(l) =& A_1R_1^le^{il\theta_1}+A_2R_1^le^{-il\theta_1}\\
&+B_1R_2^le^{il\theta_2}+B_2R_2^le^{-il\theta_2},
\end{split}
\end{equation}
where $A_1$, $A_2$, $B_1$ and $B_2$ are constants, and $l$ is the site index. From recurrence relations derived from Eqn.~\ref{MZMcase1}(via the alternative equivalent chiral decomposition) \cite{leumer2021spectral}, one can have open boundary conditions at the artificial sites outside the lattice, i.e., $\Psi(l=0)=\Psi(l=-1)=0=\Psi(l=N+1)=\Psi(l=N+2)$. From these four boundary conditions it is possible to derive a quantization condition for the existence of any MMZM standing wave eigen-function on a finite lattice of size $N$,
\begin{equation}\label{quantcond}
\begin{split}
\frac{R_1^{2(N+2)}+R_2^{2(N+2)}-2R_1^{N+2}R_2^{N+2}\cos(2(N+2)\theta_+)}{R_1^2+R_2^2-2R_1R_2\cos 2\theta_+}\\
=\frac{R_1^{2(N+2)}+R_2^{2(N+2)}-2R_1^{N+2}R_2^{N+2}\cos(2(N+2)\theta_-)}{R_1^2+R_2^2-2R_1R_2\cos 2\theta_-},
\end{split}
\end{equation}
where $\theta_\pm=\frac{1}{2}(\theta_1\pm\theta_2)$. We have explained in the previous subsection, why we get Majorana points at all for the parameter values $t_1=-t_2=-t$, $\Delta_1=\Delta_2=\Delta$ and $\mu_1=\mu_2=\mu$. For this specific case, $R_2e^{i\theta_2}=R_1e^{i\pi}e^{i\theta_1}$ derived from the conditions for component Hamiltonian 2. Substituting this into the quantization condition Eqn.~\ref{quantcond} above, we can obtain the same values of $\mu$ one obtained in sub-section ~\ref{schematictneqdel} with $\mu=2\sqrt{t^2-\Delta^2}\cos\frac{n\pi}{N+2}$, $n\in\{1,...,N+1\}$ for $N=even$ and $\mu=2\sqrt{t^2-\Delta^2}\cos\frac{n\pi}{N+1}$ $n\in\{1,...,N\}$ and $\mu=2\sqrt{t^2-\Delta^2}\cos\frac{n\pi}{N+3}$ $n\in\{1,...,N+2\}$ for $N=odd$. One may look into the Supplementary materials \ref{Sup_subsec_A1} for more detailed calculations. Hinging on the same schematic foundation, and adjusting with the form Eqn.~\ref{genwf}, one can show that we get two eigen-functions for the MMZMs at the Majorana points are of the form,
\begin{subequations}
\begin{equation}
\Psi_1(l) \sim R_1^l(1+(-1)^l)e^{il\theta_1^1}-R_1^l(1+(-1)^l)e^{-il\theta_1^1},
\end{equation}
\begin{equation}
\Psi_2(l) \sim R_1^{l+1}(1+(-1)^{l+1})e^{i(l+1)\theta_1^2}-R_1^{l+1}(1+(-1)^{l+1})e^{-i(l+1)\theta_1^2},
\end{equation}
\end{subequations}
where for $N=even$, we have $\theta_1^1=\theta_1^2=\frac{n\pi}{N+2}$ and for $N=odd$, we have $\theta_1^1=\frac{n\pi}{N+1}$ and $\theta_1^2=\frac{n\pi}{N+3}$.
\begin{figure}[htb!]
    \includegraphics[scale=0.6]{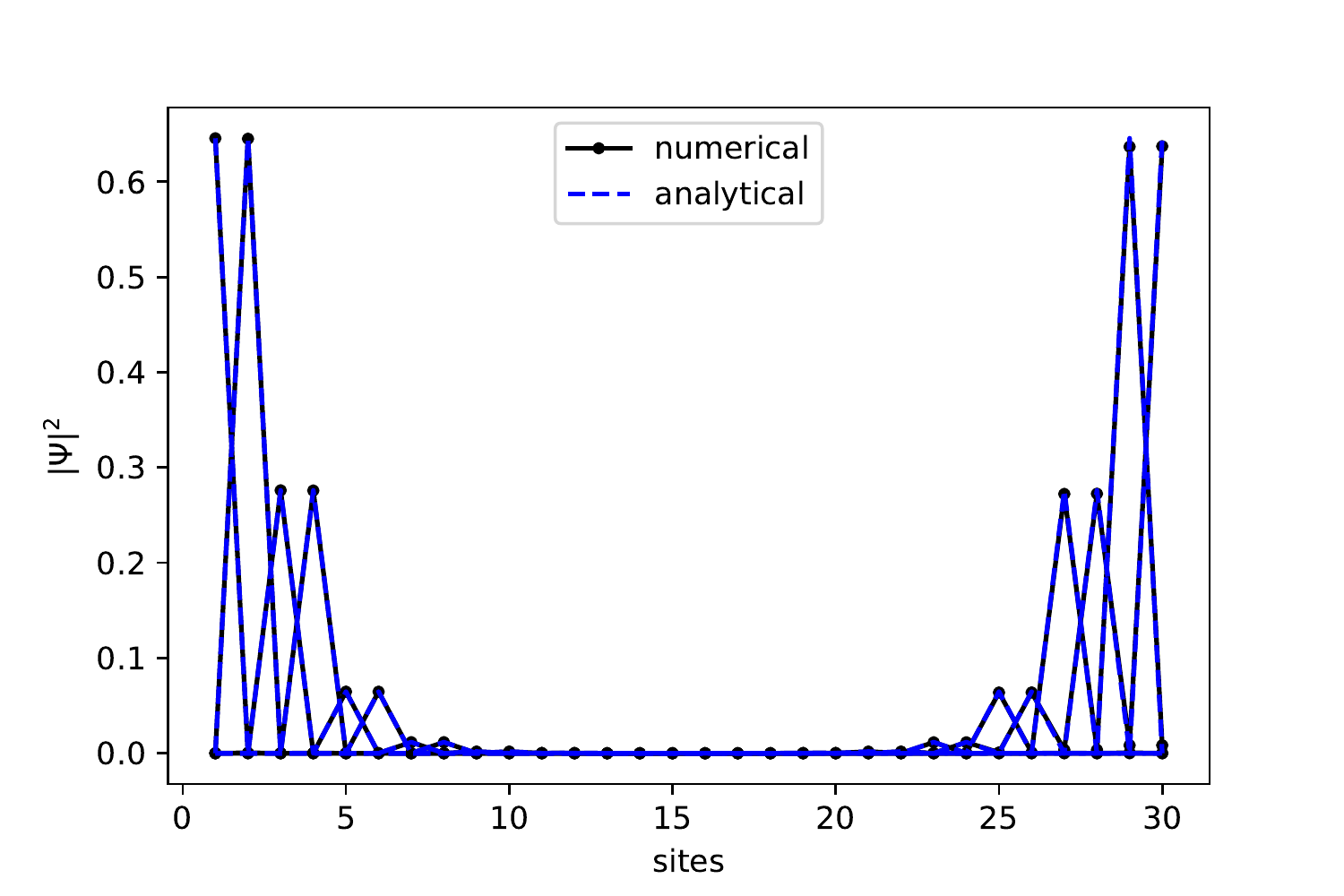}
    \caption{We compare the numerically and analytically derived wavefunction probability density for the MMZMs in the parameteric range $t_1=-t_2=-1$, $\Delta_1=\Delta_2=0.5$, and $\mu_1=\mu_2=2\sqrt{t_1^2-\Delta_1^2}\cos(\pi/(N+2))$ for a lattice size, $N=30$. We see that the numerical and analytical expressions match.}
    \label{MZMpllnumvsanalytic}
\end{figure}
The above expressions include only eigenfunctions localised at or near the left edge of the system. The eigenfunctions for the multiplicative Majoranas localised at or near the right edge can be derived analogously by the transformation, $l\rightarrow N+1-l$. Here it remains to be said that the quantization condition is a much more general statement than the specific case we just dealt with and one can derive conditions for $\mu$s at different values of $\theta_2-\theta_1=\delta$ for $R_1=R_2$ which is found for $|t_1/\Delta_1|=|t_2/\Delta_2|$. We illustrate the case for $\delta=\frac{2\pi}{3}$, in the Supplementary materials \ref{Sup_subsec_A1}.

\subsubsection{Quantum gate operations without braiding}
 According to Table~I, myriad separable and maximally-entangled two-qubit states are realized by the MKC. For instance, if each parent KC is in the topological phase and the sign of $\frac{t_i}{\Delta_i}$ is $+$ for each $i$, with $i\in\{1,2\}$, one realizes the Bell state, $\frac{1}{\sqrt{2}}(\ket{00}-\ket{11})$ and the separable states $\{\ket{01},\ket{10}\}$. This situation can be easily reversed by changing the sign of $\frac{t_2}{\Delta_2}$ to $-$, so that  Bell state instead takes the form, $\frac{1}{\sqrt{2}}(\ket{01}-\ket{10})$, while the  separable states are instead $\{\ket{00},\ket{11}\}$. Other combinations of separable state sets or maximally-entangled states are possible, although only one parity possesses entanglement at a given point in phase space when both the parent systems are topological. Moreover, if one wants to retain the entanglement of the complement parity while converting the separable set of states to a Bell state, one tunes one of the parents through phase space until it undergoes a topological phase transition to its trivial phase. As transport of the MKC through phase space corresponds to preparation of particular two-qubit states, including qubit entanglement, multiplicative topological phases have some potential as platforms for topologically-protected quantum computation schemes. First, there is the interesting possibility of using the degenerate manifold of states for the case of each parent topological, in braiding-based topological quantum computation schemes, despite the resultant MKC corresponding to an even number of particles in the ground state. Second, there also appears to be the potential for topological quantum computation schemes based on tuning the system through topological phase transitions of the parents in combination with changes in parity of certain parameter ratios. This possibility of ``phase space'' topological quantum computation schemes will be explored in future work.


\section{Child Hamiltonian for perpendicular parent chains}
\label{sec:child_model_hamiltonian_perpendicular}

To further explore the potential for multiplicative phases to realize exotic phenomena, we now characterize an MKC Hamiltonian with the two parent Kitaev chains which are \textit{perpendicular} to one another, constructing a two-dimensional rather than one-dimensional MKC. That is, we take one parent Kitaev chain to lie along the $\hat{x}$-axis, and the second parent Kitaev chain to lie along the $\hat{y}$-axis, respectively. The parent Hamiltonians and child Hamiltonian then take the following forms:

\begin{subequations}
\begin{equation}
H_{p,1}(k_x) = -(2t_1\cos k_x+\mu_1)\tau^z+2\Delta_1\sin k_x\tau^y,
\end{equation}
\begin{equation}
H_{p,2}(k_y) = -(2t_2\cos k_y+\mu_2)\sigma^z+2\Delta_2\sin k_y\sigma^y,
\end{equation}
\begin{equation}
\begin{split}
H^c_\perp(k_x,k_y) =& [-(2t_1\cos k_x+\mu_1)\tau^z+2\Delta_1\sin k_x\tau^y]\\
&\otimes [(2t_2\cos k_y+\mu_2)\sigma^z+2\Delta_2\sin k_y\sigma^y]
\label{MKCperpHamiltonian}
\end{split}
\end{equation}
\end{subequations}

This system is significantly different from the parallel MKC not only because the perpendicular orientation of the two parent chains yields next-nearest-neighbor (NNN) hopping along $(\hat{x}\pm\hat{y})$ and $-(\hat{x}\pm\hat{y})$ directions, but also due to the absence of correlation between the two parent Hamiltonians in the expression for the edge modes as we shall show. We characterize the perpendicular MKC specifically by starting with the bulk spectrum and then trying to infer about its topology via the Wilson loop method. The quasiparticle velocities near the critical points are mentioned next after which we delve into the perpendicular MKC under open boundary conditions. We start to analyse the Majorana zero modes which might be obtained as edge modes in this situation in specific parametric windows but we must instead look into the real space description to actually understand how the edge modes are localized which are further explained both schematically and numerically. The analytical expressions for the edge states and the corresponding quantization conditions then naturally arise from the real space decomposition. We will observe that although MZMs in this case are more attuned to the parametric regimes of the constituent parents, it is similar to the MMZMs we encountered in the MKC paralle case, so that we may also refer to the MZMs obtained for the perpendicular case as MMZMs. We of course defer it to a later part after it similarity with the MMZM in the paralle case has been proven.\\
\begin{figure}
    \includegraphics[scale=0.55]{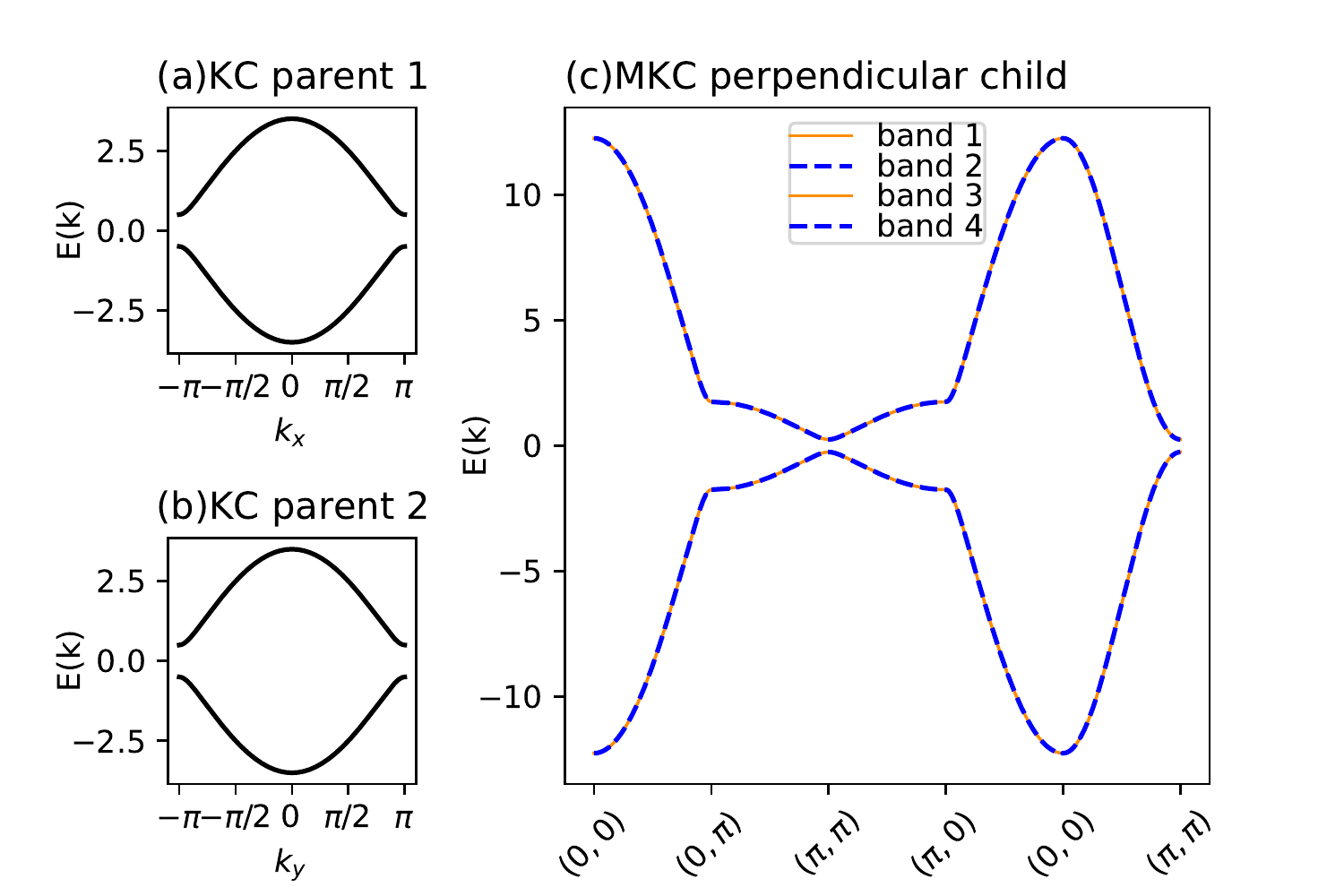}
    \caption{The dispersion for (a) parent KC 1 ($t_1=1.0$, $\mu_1=1.5$, $\Delta_1=1.0$) along the $k_x$ axis, (b) parent KC 2 ($t_2=1.0$, $\mu_2=1.5$, $\Delta_2=1.0$) along the $k_y$ axis, and (c) the MKC perpendicular child Hamiltonian from the two parents.}
    \label{MKCperpt11del11mu11}
\end{figure}

\subsection{Bulk spectrum of perpendicular multiplicative Kitaev chain}

Similarly to the case of the parallel MKC, we first characterize spectral properties of the perpendicular MKC bulk.  We consider the simplest case here of two parent Hamiltonians with identical parameter sets but differing in that one is a function of momentum in the $\hat{x}$-direction, $k_x$, and the other is a function of momentum in the $\hat{y}$-direction, $k_y$. Each parent KC is in the topologically non-trivial phase, with a minimum direct gap of $2(2t_i-\mu_i)$, (i=1,2)  at the edge of the Brillouin zone which is 1 in this case, as shown in Fig.~\ref{MKCperpt11del11mu11} (a) and (b). Bands disperse quadratically near high-symmetry points $0$ and $\pi$, respectively. The minimum direct band gap of the perpendicular MKC is analogously at $(k_x,k_y) = (\pi,\pi)$, and $2(2t_1-\mu_1)(2t_2-\mu_2)$, which in this case is 0.5. This already shows greater variety in spectra of the perpendicular MKC when compared with the parallel case, where the eigenvalues of the MKC in the bulk are products of eigenvalues of the parent Kitaev chains in the bulk. The direct gap widens at $(k_x,k_y) = (\pi,0)$ and $(0,\pi)$, approximately matching the value of each of the parent direct gaps, at $k_x=\pi$ or $k_y = \pi$, respectively. However, the direct gap widens significantly beyond the maximum direct gap of the parents at $(k_x, k_y) = (0,0)$. This value reflects the multiplicative nature of the spectrum, being the square of the maximum direct gap of each parent.

\subsection{Wilson loops and Wannier spectrum for the perpendicular MKC}

\begin{figure}[htb!]
    \includegraphics[scale=0.6]{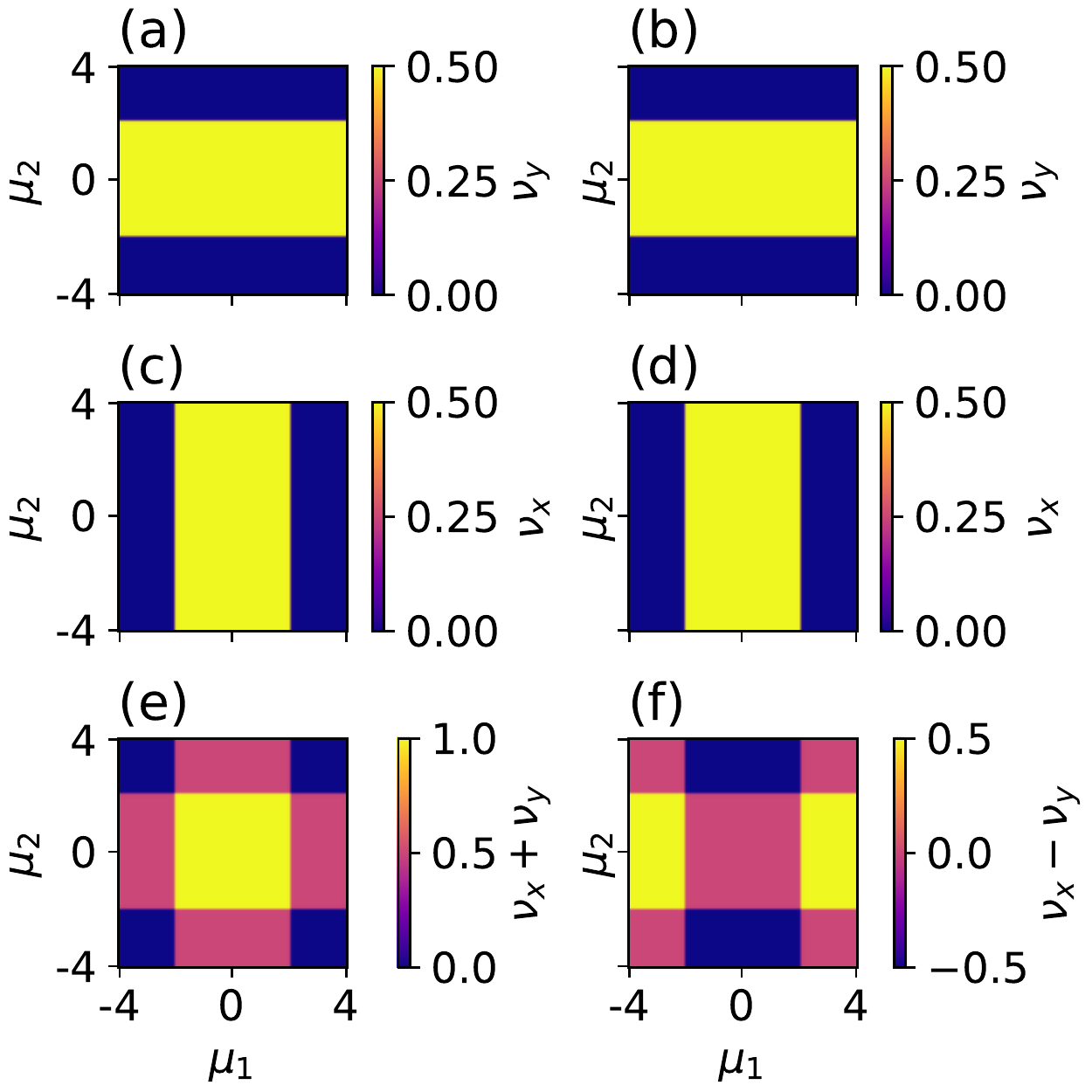}
    \caption{Wannier spectra $\nu_{x/y}$(colorbar) for MKC perpendicular system derived from Wilson loop operators, $\mathcal{W}_x$((a) and (b)) and $\mathcal{W}_y$((c) and (d)) respectively. We also plot $\nu_x\pm\nu_y$ in row 3((e) and (f)) in the left and right respectively.}
    \label{MKCperpWannier}
\end{figure}

As in the case of the parallel MKC, we now characterize topology of the child Hamiltonian without assuming knowledge of how the child Hamiltonian is constructed from parent Hamiltonians, nor how its topology is determined by topological invariants of the parents. For this reason, we calculate the Wannier spectra for the different topological phases of the perpendicular MKC. For one-dimensional systems, a Wilson loop is expressed as in Eqn.~\eqref{WLdefA}, but can be generalized for the two-dimensional Brillouin zone of the perpendicular MKC as the Wilson loop across the $k_x$ BZ for a given $k_y$ and across the $k_y$ BZ for a given $k_x$. We use the alternative definition of Wilson loop matrix in terms of the occupied state projectors,
\begin{equation}
\mathcal{W}_{mn} = \bra{u_{m}(\bm k_0)} \lim_{R \to \infty} \prod_{i=R}^1  P(\bm k_i) \ket{u_n(\bm k_0)},
\end{equation}
and calculate the matrix components for the case with loop along the $k_x$ BZ for a given $k_y$ as shown explicitly in Supplementary Section \label{app:wilson_loop} in Eqn. \eqref{WLMKCperp},
\begin{equation}
\begin{split}
\mathcal{W}_{11} =& \bra{v_{1+}(k_{x0})}\lim_{R\rightarrow\infty}\bigg{[}\prod_{i=R}^1P_{1+}(k_{xi})\bigg{]}\ket{v_{1+}(k_{x0})},\\
\mathcal{W}_{22} =& \bra{v_{1-}(k_{x0})}\lim_{R\rightarrow\infty}\bigg{[}\prod_{i=R}^1P_{1-}(k_{xi})\bigg{]}\ket{v_{1-}(k_{x0})},\\
\mathcal{W}_{12} =& \mathcal{W}_{21}=0.
\end{split}
\end{equation}
One can similarly work out the alternative case where the loop is along the $k_y$ BZ for a given $k_x$ and the final Wannier spectra is given as
\begin{equation}
\begin{split}
\nu_{i}=\nu_x=&\nu^{(1)}\text{mod 1}\quad \text{for BZ along $k_x$ and given $k_y$},\\
\nu_{i}=\nu_y=&\nu^{(2)}\text{mod 1}\quad \text{for BZ along $k_y$ and given $k_x$},
\end{split}
\end{equation}
where $\nu^{(j)}$, $j\in\{1,2\}$ is the Wannier spectra due to the $i$-th parent Hamiltonian.\\
Topology of two-dimensional phases is then characterized in terms of the \textit{winding} of these two Wilson loops as a function of $k_x$ and $k_y$, respectively. We find, however, that these two quantities are each constant as a function of $k_x$ or $k_y$, and we therefore may characterize the topology entirely with $\mc{W}(k_x)$ ($\mc{W}(k_y)$), with $k_x$ ($k_y$) fixed and integration over $k_y$ ($k_x$). We therefore compute Wannier center charge spectra for Wilson loops computed by integrating over $k_x$ ($k_y$) for each $k_y$ ($k_x$) and shown in Fig.~\ref{MKCperpWannier}. We find the spectra exhibit topologically non-trivial Wannier charge center values when one of the parent Hamiltonians is in a topologically non-trivial state. The spectra are topologically trivial when both parents are topologically trivial, but \textit{also when both parents are topological, and the child is also actually topologically non-trivial}. In the regime, when both the parents are topological, we have $(\nu_x,\nu_y)\equiv (0.5,0.5)$, where $\nu_{x/y}$ refer to the Wannier spectra derived from Wilson loop operators $\mathcal{W}_x$ and $\mathcal{W}_y$ respectively.

\subsection{Quasiparticle velocity near critical points:}
The two band KC Dirac Hamiltonian near a critical point, say $\mu\sim -2t$, with $k\rightarrow 0$ has quasi-particles which propagate with fixed velocity along the length of the system. For the MKC with perpendicular axes, on the other hand, one has the following Dirac Hamiltonian, say for $\mu_1\sim -2t_1$ with $k_x\rightarrow 0$,

\begin{equation}
\begin{split}
H_{D,x}(k_x,k_y) =& -m_1(2t_2\cos k_y+\mu_2)\Gamma^{zz}\\
&+2\Delta_1(2t_2\cos k_y+\mu_2)k_x\Gamma^{yz}\\
&-2m_1\Delta_2\sin k_y\Gamma^{zy}+4\Delta_1\Delta_2k_x\sin k_y \Gamma^{yy},
\end{split}
\end{equation}
where $m_1=2t_1+\mu_1$ and $\Gamma^{ij}=\tau^i\sigma^j$. The quasi-particles at this critical point corresponding to the parent 1 system. The doubly degenerate energy is given as,
\begin{equation}
\begin{split}
E(k_x,k_y) = \pm &\sqrt{4\Delta_1^2k_x^2+m_1^2}\\ &\times \sqrt{4\Delta_2^2\sin^2 k_y+(2t_2\cos k_y+\mu_2)^2}.
\end{split}
\end{equation}
Again expanding in the vicinity of the critical point derived from parent 2 system, say $\mu_2\sim -2t_2$ with $k_y\rightarrow 0$, the Dirac Hamiltonian is shown to be,
\begin{equation}
\begin{split}
H_{D,y}(k_x,k_y) =& -m_2(2t_1\cos k_x+\mu_1)\Gamma^{zz} +2m_2\Delta_1\sin k_x\Gamma^{yz}\\
& -2\Delta_2(2t_1\cos k_x+\mu_1)k_y\Gamma^{zy}\\
&+4\Delta_1\Delta_1k_y\sin k_x\Gamma^{yy},
\end{split}
\end{equation}
where $m_2=2t_2+\mu_2$. The doubly degenerate energy in this case is,
\begin{equation}
\begin{split}
E(k_x,k_y)=\pm &\sqrt{4\Delta_1^2\sin^2k_x+(2t_1\cos k_x+\mu_1)^2}\\
&\times\sqrt{4\Delta_2^2k_y^2+m_2^2}.
\end{split}
\end{equation}
Finally we expand the MKC perpendicular Hamiltonian at the vicinity of the critical point, $\mu_1\sim -2t_1$ and $\mu_2\sim -2t_2$ with both $k_x,k_y\rightarrow 0$, so that the Dirac Hamiltonian is found to be,
\begin{equation}
\begin{split}
H_{D,x,y}(k_x,k_y) = &-m_1m_2\Gamma^{zz}-2m_1\Delta_2k_y\Gamma^{zy}+2\Delta_1m_2k_x\Gamma^{yz}\\
&+4\Delta_1\Delta_2k_xk_y\Gamma^{yy}.
\end{split}
\end{equation}
Again, from the last expression, the doubly degenerate energy is shown below,
\begin{equation}
E(k_x,k_y) = \pm\sqrt{4\Delta_1^2k_x^2+m_1^2}\cdot\sqrt{4\Delta_2^2k_y^2+m_2^2}.
\end{equation}
As evident from all the Dirac Hamiltonian energies, the group velocity of the quasi-particles have both x and y components. We illustrate for the last case when both the parents are near criticality, when the group velocity turns out to be,
\begin{equation}
\mathbf{v}(k_x,k_y)=\pm 4\Delta_1\Delta_2(k_y\mathbf{e_x}+k_x\mathbf{e_y}).
\end{equation}
The velocity field in the $\boldsymbol{k}$-space for this case looks like an anti-vortex structure and may be helpful in creating further exotic phases by stacking a similar Bloch Hamiltonian structure as the MKC perpendicular system with coupling in the z-direction, as done in the case of the KC Bloch Hamiltonian while constructing a Chern insulator.
\\

\subsection{Perpendicular multiplicative Kitaev chain with open boundary conditions}

To begin characterizing the perpendicular MKC with open boundary conditions, we consider a slab geometry, with open boundary conditions in the $\hat{x}$-direction, and system width of $L_x$ finite, while keeping boundary conditions in the $\hat{y}$-direction periodic and $L_y$ infinite. We first characterize spectral properties of the system with these boundary conditions, finding evidence of additional topologically-protected boundary modes under these conditions. We then characterize these topologically-protected boundary states in greater detail focusing on localization of the states. We support numerical findings with additional analytical characterization of the boundary modes in a variety of limiting cases.

\subsubsection{Spectrum for open boundary conditions}

\begin{figure}[htb]
    \includegraphics[scale=0.4]{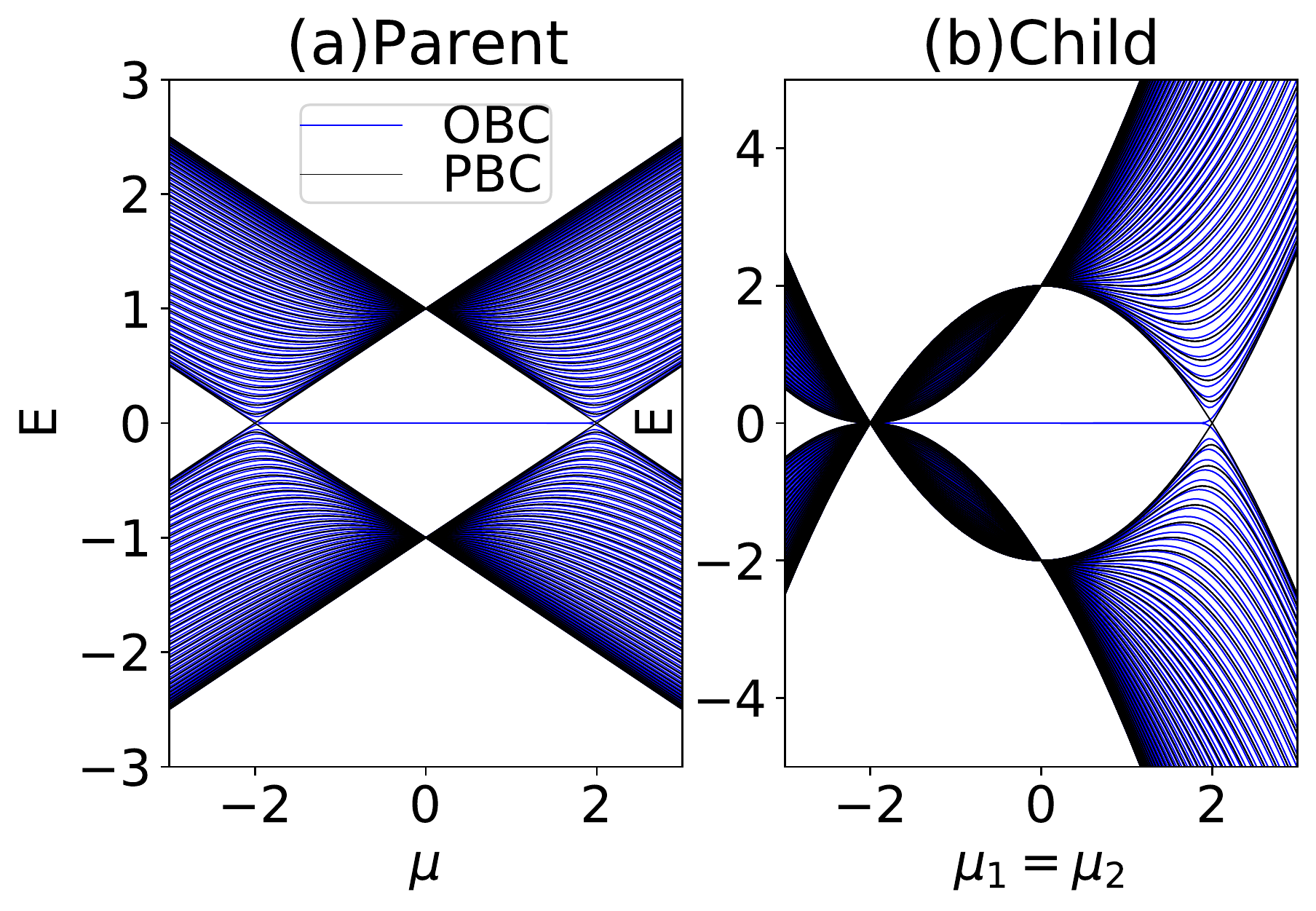}
    \caption{Slab spectra with $L_x=80$ for OBC along x direction(black) and PBC along x-direction(blue) for (a) Parent KC with $t=1=\Delta$ vs. $\mu$, and (b) Child MKC perpendicular with $t_1=t_2=1=\Delta_1=\Delta_2$ and $k_y=0$ vs. $\mu_1=\mu_2$.}
    \label{MKCperpXt11del11vsmueqL80}
\end{figure}

For comparison with the bulk properties, we also study spectra of the perpendicular MKC for open boundary conditions, first considering wide slab geometries with open boundary conditions in the $\hat{x}$-direction, corresponding to $L_x = 80$. These results are shown in Fig. ~\ref{MKCperpXt11del11vsmueqL80}.


\subsubsection{Edge modes for the perpendicular parent chains}
\label{MKCperp_edges}
We will first consider edge modes for the cases where the perpendicular MKC is finite in a single direction. The details of this process have been worked out in the Supplementary materials~\ref{Sup_subsec_A2}.
\begin{itemize}
\item For the Hamiltonian above, let us first have OBC in the $\hat{x}$-direction. To find the edge state expressions, we assume a bound-state ansatz wavefunction for the MKC perpendicular Hamiltonian, by taking $k_x\rightarrow iq_x$, and then looking for the null vectors. We arrive at the following condition for the existence of zero energy states,
\begin{equation}
2t_1\cosh q_x+\mu_1 = \pm 2\Delta_1\sinh q_x.
\label{MKCperp_reln_OBCx}
\end{equation}
The expression for the zero energy edge states taking into account the boundary conditions at $x=0$ and $x\rightarrow\infty$ are derived in Supplementary materials Sec.~\ref{Sup_subsec_A2}, and are provided below,
\begin{equation}
\begin{split}
\Psi(j,k_y)_\pm \sim &
\bigg{[}\bigg{(}\frac{-\mu_1+\sqrt{\mu_1^2-4(t_1^2-\Delta_1^2)}}{2(\Delta_1\pm t_1)}\bigg{)}^j\\
&-\bigg{(}\frac{-\mu_1-\sqrt{\mu_1^2-4(t_1^2-\Delta_1^2)}}{2(\Delta_1\pm t_1)}\bigg{)}^j\bigg{]}e^{ik_yy}
\begin{pmatrix}
a_1\\
a_2\\
a_3\\
a_4
\end{pmatrix}.
\end{split}
\end{equation}
\end{itemize}

It is interesting to notice here that the translational invariance along the $y$-direction indicates that Majorana modes are localized along the two edges parallel to the y-axis. Implementing PBCs in the $y$-direction simply quantizes the momenta $k_y$ and does not affect the analytical form of the edge states.

One can similarly calculate the edge state expressions for OBCs in the $y$-direction by the localization, $k_y\rightarrow iq_y$, as done in Supplementary materials Sec.~\ref{Sup_subsec_A2}. In this case, one arrives at the following relation for zero energy,
\begin{equation}
2t_2\cosh q_y+\mu_2=\pm 2\Delta_2\sinh q_y.
\label{MKCperp_reln_OBCy}
\end{equation}
In this case, we define $M_1=-(2t_1\cos k_x+\mu_1)$ and $R_1=2\Delta_1\sin k_x$ for ease of notation, and find edge states of the form, for the two signs in Eqn.~\eqref{MKCperp_reln_OBCy},
\begin{equation}
\begin{split}
\Psi(k_x,l)_\pm\sim &
e^{ik_xx}\bigg{[}\bigg{(}\frac{-\mu_2+\sqrt{\mu_2^2-4(t_2^2-\Delta_2^2)}}{2(\Delta_2\pm t_2)}\bigg{)}^l\\
&-\bigg{(}\frac{-\mu_2-\sqrt{\mu_2^2-4(t_2^2-\Delta_2^2)}}{2(\Delta_2\pm t_2)}\bigg{)}^l\bigg{]}
\begin{pmatrix}
b_1\\
b_2\\
b_3\\
b_4
\end{pmatrix}.
\end{split}
\end{equation}
Here, we notice that translational invariance in the $x$-direction (instead of the $y$-direction as in the previous case) corresponds to Majorana modes along the whole edge from left to right.  Implementing PBC along $x$ again does not change the analytical form of the edge mode expressions but simply quantizes $k_x$.\\

Finally, we consider OBC along both the $x$ and $y$ directions by localizing $k_x\rightarrow iq_x$ and $k_y\rightarrow iq_y$. As derived in Supplementary materials Sec.~\ref{Sup_subsec_A2}, we get the relation,
\begin{equation}
\begin{split}
&[(2t_1\cosh q_x+\mu_1)^2-4\Delta_1^2\sinh^2q_x]\\
&\times[(2t_2\cosh q_y+\mu_2)^2-4\Delta_2^2\sinh^2q_y]=0.
\end{split}
\end{equation}
The above condition yields four sign combinations so that the null vectors for the localized Hamiltonian are given as follows,
\begin{equation}
\Phi = \frac{1}{2}
\begin{pmatrix}
1\\
\pm 1
\end{pmatrix}
\otimes
\begin{pmatrix}
1\\
\mp 1
\end{pmatrix}.
\end{equation}
But there is also another set of eigen-vectors comprising the maximally entangled Bell states due to our tensor product structure. This ambiguity will be clarified once we derive the explicit form of the MZM eigen-vectors after working out the real space Hamiltonian for the MKC perpendicular system under different boundary conditions in Sec. E.

\subsection{Perpendicular MKC  Hamiltonian in real-space}
\begin{figure*}[t]
    \includegraphics[scale=0.5]{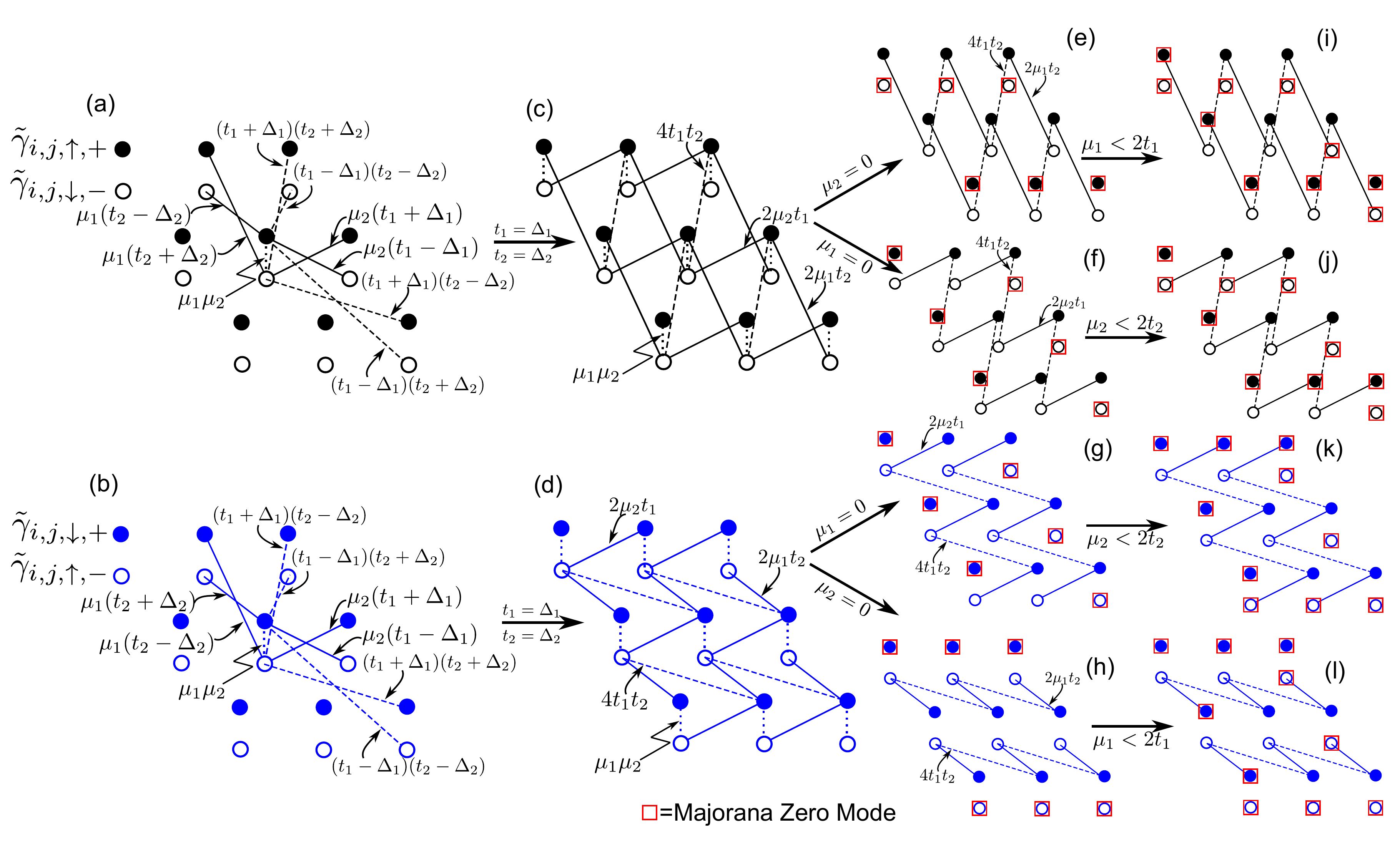}
    \caption{Schematic representation of the MKC perpendicular system in terms of Majorana fermionic interactions using the same color scheme and symbols as in Fig.~\ref{MKCpllphases}, (a) for $t_1=\Delta_1$ and (b) for $t_1=\Delta_1$ and $\mu_1=\mu_2 = 0$. One can see that for translational symmetry in the y-direction, one gets Majoarana edge modes along the whole y-axis for finite slab in the x-direction, as discussed beforehand.}
    \label{MKCperpschematic}
\end{figure*}

For convenience again, we redefine our notation from section \ref{MKC_pll_realsp}, $\gamma_{j,+}=(\gamma_{j,+,\uparrow},\gamma_{j,+,\downarrow})$ and $\gamma_{j,-}=(\gamma_{j,-,\uparrow},\gamma_{j,-,\downarrow})^T$. Then one can express the MKC  Hamiltonian in the case of perpendicular parents as follows,
\begin{equation}
\begin{split}
H^c_{MKC,\perp} =&\frac{1}{2}\sum_{i,j}-(t_1-\Delta_1)i\gamma_{i,j,+}(t_2\sigma^z-i\Delta_2\sigma^y)\gamma_{i+1,j+1,-}\\
&-(t_1+\Delta_1)i\gamma_{i+1,j+1,+}(t_2\sigma^z+i\Delta_2\sigma^y)\gamma_{i,j,-}\\
&-(t_1-\Delta_1)i\gamma_{i,j,+}(t_2\sigma^z+i\Delta_2\sigma^y)\gamma_{i+1,j-1,-}\\
&-(t_1+\Delta_1)i\gamma_{i+1,j-1,+}(t_2\sigma^z-i\Delta_2\sigma^y)\gamma_{i,j,-}\\
&-\mu_2(t_1-\Delta_1)i\gamma_{i,j,+}\sigma^z\gamma_{i+1,j,-}\\
&-\mu_2(t_1+\Delta_1)i\gamma_{i+1,j,+}\sigma^z\gamma_{i,j,-},\\
&-\mu_1i\gamma_{i,j,+}(t_2\sigma^z-i\Delta_2\sigma^y)\gamma_{i,j+1,-}\\
&-\mu_1i\gamma_{i,j+1,+}(t_2\sigma^z+i\Delta_2\sigma^y)\gamma_{i,j,-}\\
&-\mu_1\mu_2i\gamma_{i,j,+}\sigma^z\gamma_{i,j,-}.
\end{split}
\end{equation}
We execute the same similarity transformation as done in the MKC parallel case, which changes the Hamiltonian expression to,
\begin{equation}{\label{MKCperpMajorana}}
\begin{split}
H^c_{MKC,\perp} =& \frac{i}{2}\sum_{i,j}[-(t_1-\Delta_1)(t_2-\Delta_2)\tilde{\gamma}_{i,j,\uparrow,+}\tilde{\gamma}_{i+1,j+1,\downarrow,-}\\
&-(t_1+\Delta_1)(t_2+\Delta_2)\tilde{\gamma}_{i+1,j+1,\uparrow,+}\tilde{\gamma}_{i,j,\downarrow,-}\\
&-(t_1-\Delta_1)(t_2+\Delta_2)\tilde{\gamma}_{i,j,\uparrow,+}\tilde{\gamma}_{i+1,j-1,\downarrow,-}\\
&-(t_1+\Delta_1)(t_2-\Delta_2)\tilde{\gamma}_{i+1,j-1,\uparrow,+}\tilde{\gamma}_{i,j,\downarrow,-}\\
&-\mu_2(t_1-\Delta_1)\tilde{\gamma}_{i,j,\uparrow,+}\tilde{\gamma}_{i+1,j,\downarrow,-}\\
&-\mu_2(t_1+\Delta_1)\tilde{\gamma}_{i+1,j,\uparrow,+}\tilde{\gamma}_{i,j,\downarrow,-}\\
&-\mu_1(t_2-\Delta_2)\tilde{\gamma}_{i,j,\uparrow,+}\tilde{\gamma}_{i,j+1,\downarrow,-}\\
&-\mu_1(t_2+\Delta_2)\tilde{\gamma}_{i,j+1,\uparrow,+}\tilde{\gamma}_{i,j,\downarrow,-}
-\mu_1\mu_2\tilde{\gamma}_{i,j,\uparrow,+}\tilde{\gamma}_{i,j,\downarrow,-}]\\
& +\frac{i}{2}\sum_{i,j}[-(t_1-\Delta_1)(t_2+\Delta_2)\tilde{\gamma}_{i,j,\downarrow,+}\tilde{\gamma}_{i+1,j+1,\uparrow,-}\\
&-(t_1+\Delta_1)(t_2-\Delta_2)\tilde{\gamma}_{i+1,j+1,\downarrow,+}\tilde{\gamma}_{i,j,\uparrow,-}\\
&-(t_1-\Delta_1)(t_2-\Delta_2)\tilde{\gamma}_{i,j,\downarrow,+}\tilde{\gamma}_{i+1,j-1,\uparrow,-}\\
&-(t_1+\Delta_1)(t_2+\Delta_2)\tilde{\gamma}_{i+1,j-1,\downarrow,+}\tilde{\gamma}_{i,j,\uparrow,-}\\
&-\mu_2(t_1-\Delta_1)\tilde{\gamma}_{i,j,\downarrow,+}\tilde{\gamma}_{i+1,j,\uparrow,-}\\
&-\mu_2(t_1+\Delta_1)\tilde{\gamma}_{i+1,j,\downarrow,+}\tilde{\gamma}_{i,j,\uparrow,-}\\
&-\mu_1(t_2+\Delta_2)\tilde{\gamma}_{i,j,\downarrow,+}\tilde{\gamma}_{i,j+1,\uparrow,-}\\
&-\mu_1(t_2-\Delta_2)\tilde{\gamma}_{i,j+1,\downarrow,+}\tilde{\gamma}_{i,j,\uparrow,-}
-\mu_1\mu_2\tilde{\gamma}_{i,j,\downarrow,+}\tilde{\gamma}_{i,j,\uparrow,-}],\\
=& H_{1,\perp}+H_{2,\perp}.
\end{split}
\end{equation}

\begin{figure}
    \subfloat[$\mu_1=0$, $\mu_2=3$]{{\includegraphics[width=0.16\textwidth,trim={7cm 0 1.5cm 0},clip]{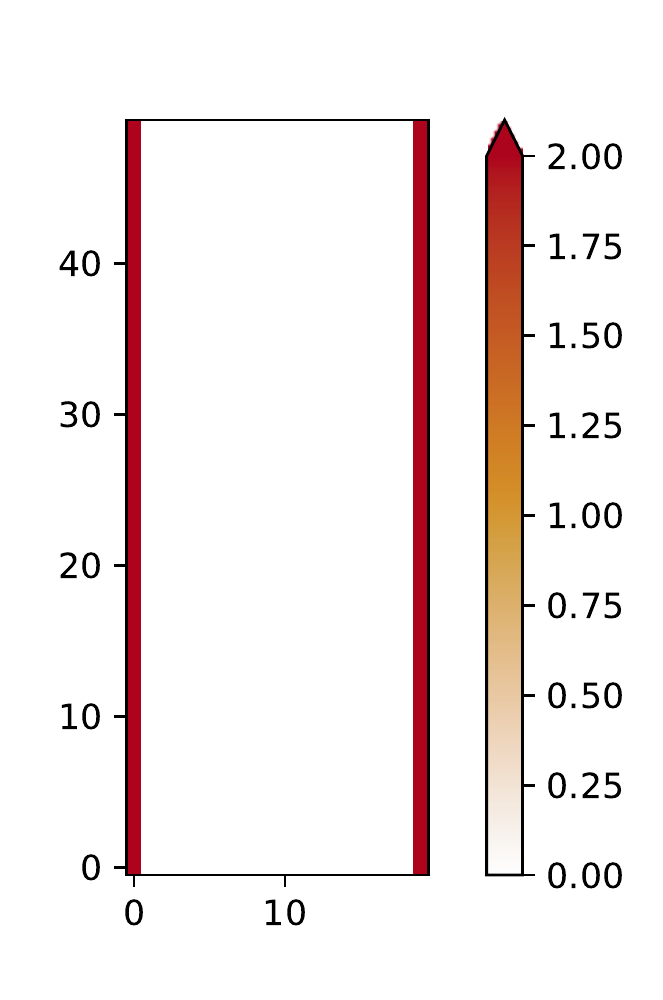}}}%
    \subfloat[$\mu_1=1$, $\mu_2=3$]{{\includegraphics[width=0.16\textwidth,trim={7cm 0 1.5cm 0},clip]{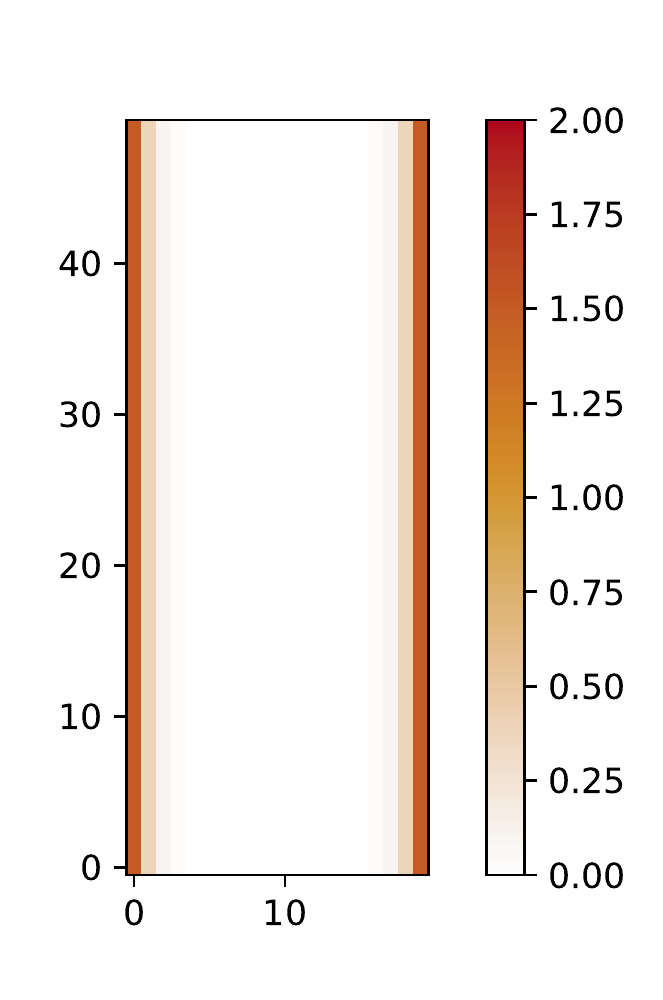}}}%
    \subfloat[$\mu_1=2$, $\mu_2=3$]{{\includegraphics[width=0.16\textwidth,trim={7cm 0 1.5cm 0},clip]{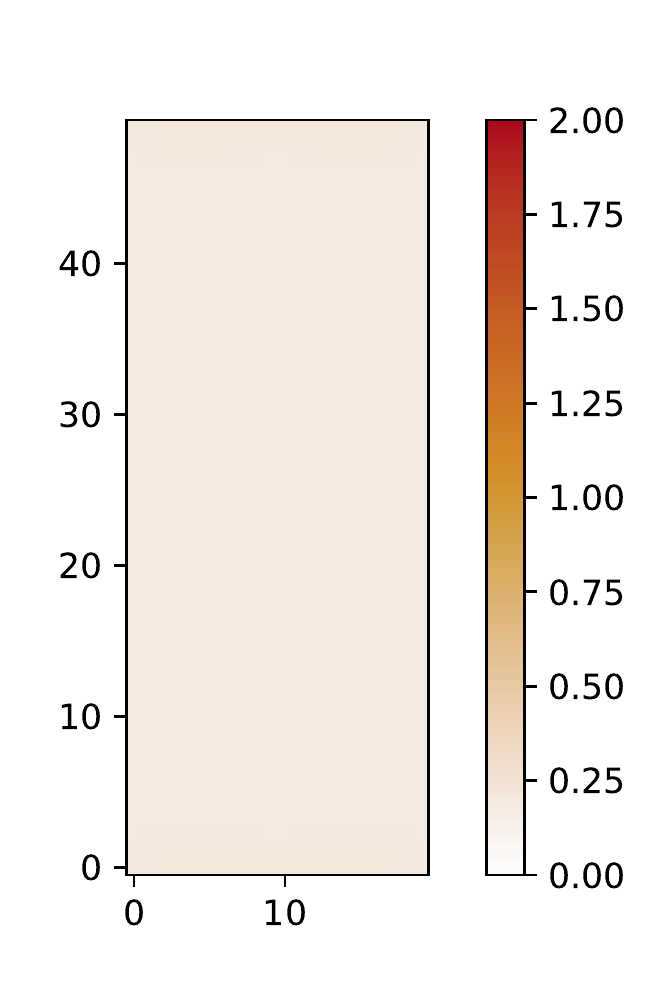}}}\\
    \subfloat[$\mu_1=3$, $\mu_2=0$]{{\includegraphics[width=0.16\textwidth,trim={7cm 0 1.5cm 0},clip]{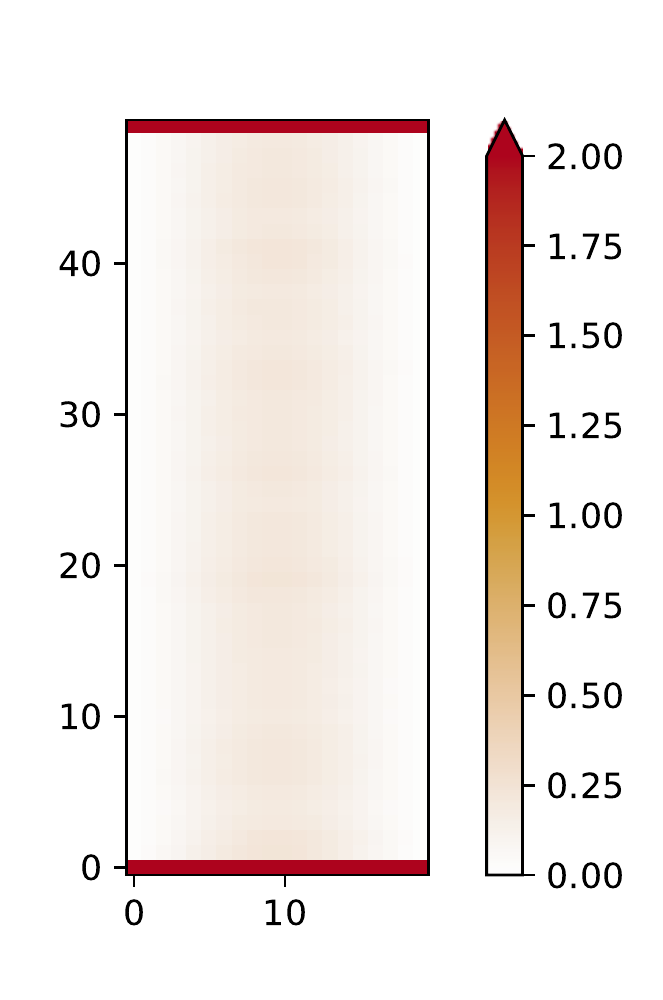}}}%
    \subfloat[$\mu_1=3$, $\mu_2=1$]{{\includegraphics[width=0.16\textwidth,trim={7cm 0 1.5cm 0},clip]{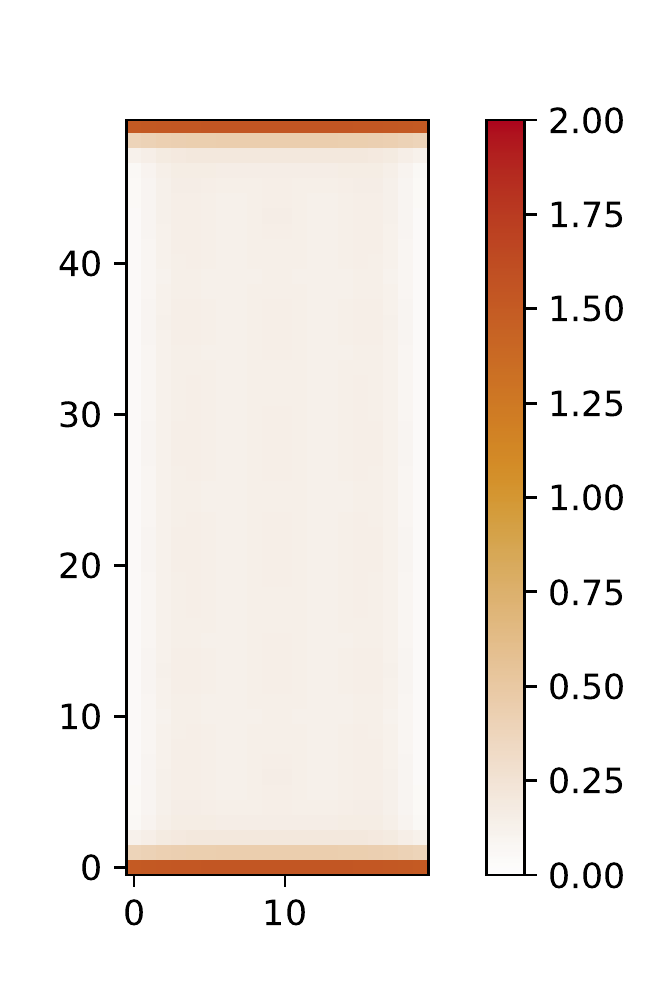}}}%
    \subfloat[$\mu_1=3$, $\mu_2=2$]{{\includegraphics[width=0.16\textwidth,trim={7cm 0 1.5cm 0},clip]{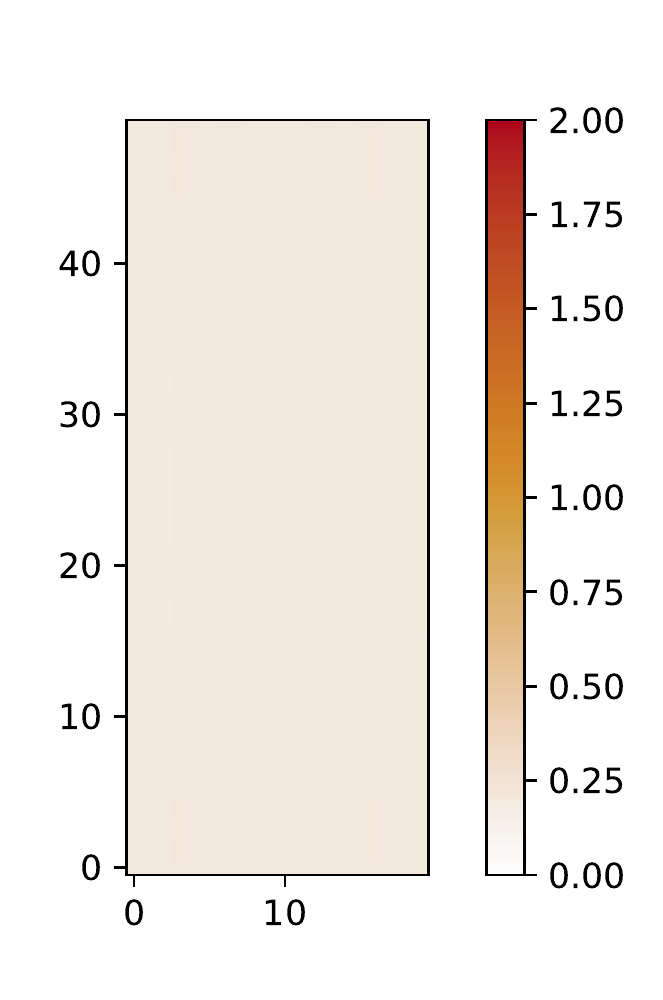}}}\\
    \subfloat[$\mu_1=0$, $\mu_2=0$]{{\includegraphics[width=0.16\textwidth,trim={7cm 0 1.5cm 0},clip]{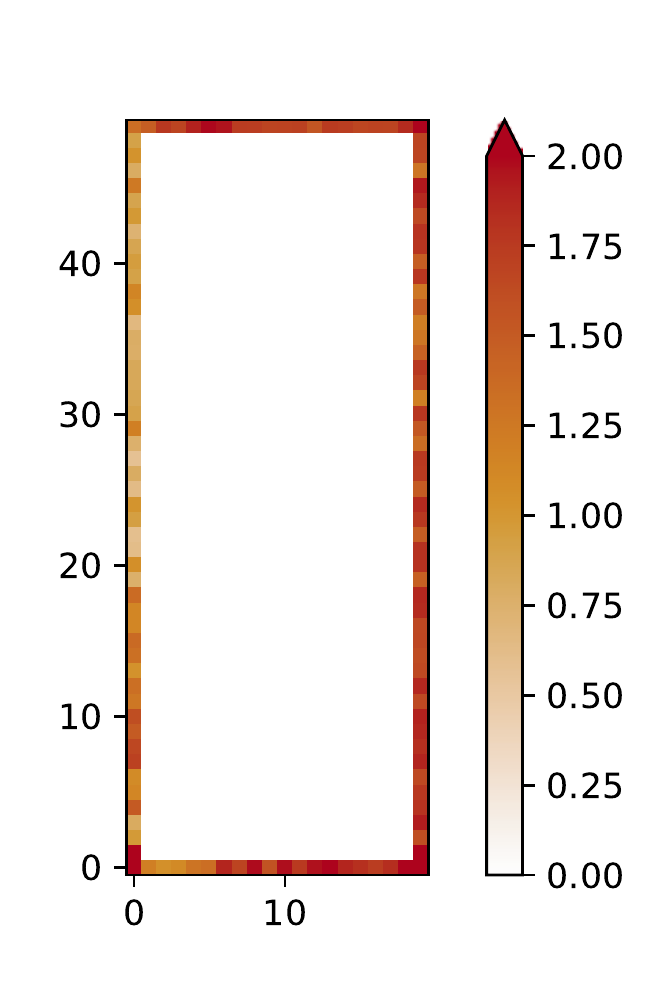}}}%
    \subfloat[$\mu_1=1$, $\mu_2=1$]{{\includegraphics[width=0.16\textwidth,trim={7cm 0 1.5cm 0},clip]{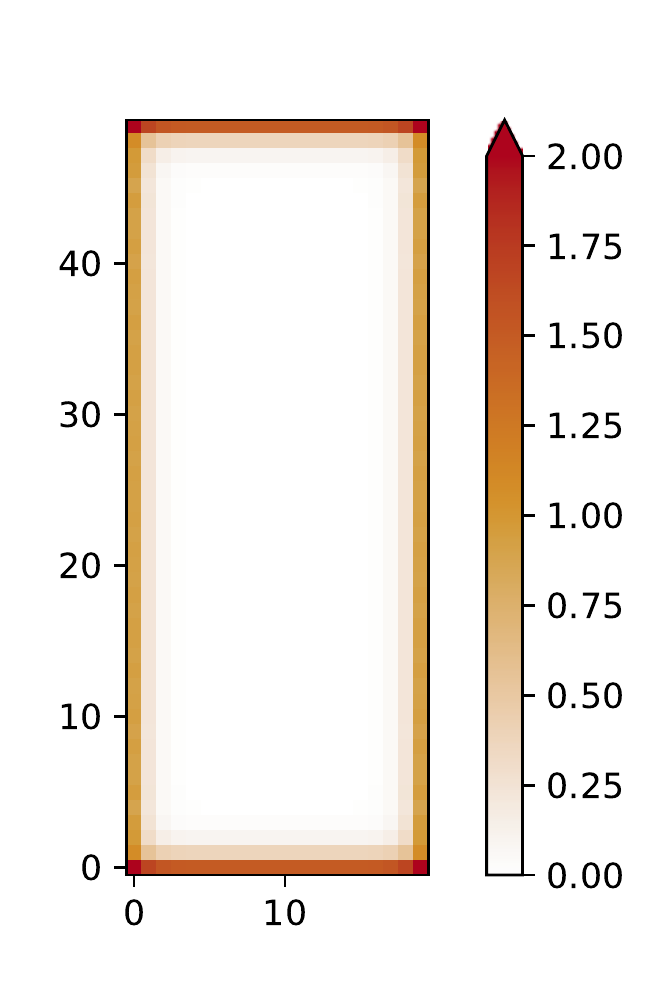}}}%
    \subfloat[$\mu_1=2$,$\mu_2=2$]{{\includegraphics[width=0.16\textwidth,trim={7cm 0 1.5cm 0},clip]{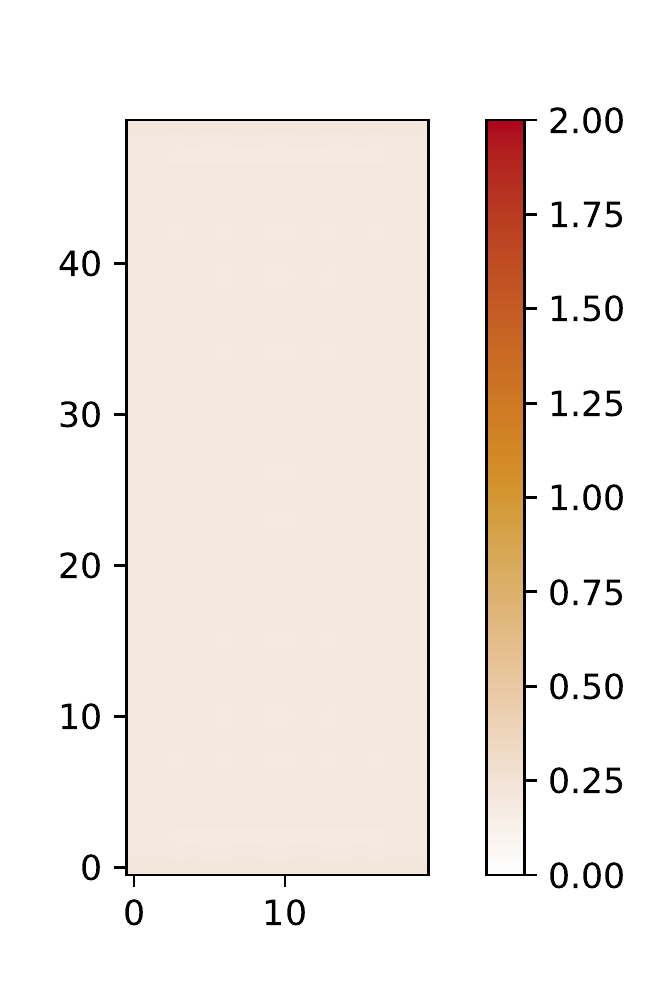}}}
    \caption{Density plots for zero energy Majorana modes for a finite $20\times 50$ slab of MKC perpendicular system at different values of $\mu_1$ and $\mu_2$. All the systems have $t_1=t_2=1$ and $\Delta_1=\Delta_2=1$.}
    \label{MKCperp_Density_plots}
\end{figure}

Based on the diagram shown in  Fig.~\ref{MKCperpschematic}, it is possible to deduce the possibility and placement of Majorana zero modes even if the system has finite length and width. We introduce all the interactions present with respect to one site in Fig.~\ref{MKCperpschematic}(a) and (b) for the component Hamiltonians $H_{\perp,1}$ and $H_{\perp,2}$ and then we prioritize the case for which $t_{1,2}=\Delta_{1,2}$ (Fig.~\ref{MKCperpschematic}(c) and (d)), where we find Majorana zero modes parent Hamiltonian for suitable $\mu_{1,2}$.
\begin{itemize}
\item \textit{Case 1}: We first look at the case when parent 1 is topological, with $\mu_1=0$, while parent 2 is trivial with $\mu_2>2t_2$. Fig.~\ref{MKCperpschematic}(f) and (g) show that both $H_{1,\perp}$ and $H_{2,\perp}$ have Majorana zero-modes running along both the edges parallel to the $y$-axis of the square lattice. Numerical simulation in Fig.~\ref{MKCperp_Density_plots}(a) agrees with this analytical calculation. The schematic diagram further illustrates that the states localized along each edge are two-fold degenerate, as each component Hamiltonian in Eqn.~\eqref{MKCperpMajorana} contributes a Majorana edge state.
\item \textit{Case 2}: Now, we consider parent 2 in the topological phase, with $\mu_2=0$, and parent 1 in the trivial phase by requiring that $\mu_1>2t_1$. From Fig.~\ref{MKCperpschematic}(e) and (f) for $H_{1,\perp}$ and $H_{2,\perp}$ respectively, one observes Majorana zero modes in the square lattice along the two edges parallel to the $x$-axis. Again, each Hamiltonian component contributes one Majorana state localized at each edge, yielding a two-fold degeneracy of the Majorana zero modes. Numerical simulations in Fig.~\ref{MKCperp_Density_plots}(d) are consistent with our analytical expressions.
\item \textit{Case 3}: Finally, we consider the case in which each parent is topologically non-trivial. This is illustrated in Fig.~\ref{MKCperpschematic}(i), (j) and  (k), (l) for $\mu_1=0,\mu_2<2t_2$ and $\mu_1<2t_1,\mu_2=0$, respectively. Notice the alternatively connected dashed and solid lines which indicates a number of decoupled Kitaev chains. The conditions $\mu_1<2t_1$ in Fig.~\ref{MKCperpschematic}(i) and (l) and $\mu_2<2t_2$ in Fig.~\ref{MKCperpschematic}(j) and (k) then naturally imply that each of the decoupled Kitaev chains are topologically non-trivial and hence have Majorana zero modes at the edges. The interesting fact to notice is however that the whole perimeter of the finite size system now has Majorana zero modes with a two-fold degeneracy (Each of $H_{1,\perp}$ and $H_{2,\perp}$ provide one MZM). This also agrees with our numerical simulation in Fig.~\ref{MKCperp_Density_plots}(g) and (h). In addition, the corners seem to host three degenerate Majoranas. This may indicate the presence of higher-order \cite{schindler2018higher} topological edge modes, but we defer this discussion to a later article.
\end{itemize}
We next discuss the topological invariants derived from the component Bloch Hamiltonians of the MKC perpendicular system in Eqns.~\ref{Hperp1Bloch} and \ref{Hperp2Bloch}.

\subsubsection{Topology of the perpendicular MKC characterized via chiral decomposition:}
It is possible to derive two separate Bloch Hamiltonians for each of the component Hamiltonians, $H_{\perp,1}$ and $H_{\perp,2}$ by comparing the form of the Hamiltonian in the Majorana basis for each of the component Hamiltonians to that of the 2-band 2d Kitaev chain with next-nearest neighbour interactions,
\begin{equation}
H_{\text{MKC},\perp}^c = \frac{1}{2}\sum_{k}\tilde{\mathbf{c}}^\dagger_{k,1}\mc{H}_{\perp,1}(k)\tilde{\mathbf{c}}_{k,1}+\frac{1}{2}\sum_{k}\tilde{\mathbf{c}}^\dagger_{k,2}\mc{H}_{\perp,2}(k)\tilde{\mathbf{c}}_{k,2},
\end{equation}
\begin{subequations}
\begin{equation}
\begin{split}
\mc{H}_{\perp,1}(k) =& -[2\mu_2t_1\cos k_x+2\mu_1t_2\cos k_y\\
&+2(t_1t_2+\Delta_1\Delta_2)\cos(k_x+k_y)\\
&+2(t_1t_2-\Delta_1\Delta_2)\cos(k_x-k_y)+\mu_1\mu_2]\sigma^z\\
&+[2\mu_2\Delta_1\sin k_x+2\mu_1\Delta_2\sin k_y\\
&+2(t_2\Delta_1+t_1\Delta_2)\sin(k_x+k_y)\\
&+2(t_2\Delta_1-t_1\Delta_2)\sin(k_x-k_y)]\sigma^y=\mathbf{d}_1(k)\cdot\boldsymbol{\sigma},
\end{split}
\label{Hperp1Bloch}
\end{equation}
\begin{equation}
\begin{split}
\mc{H}_{\perp,2}(k) =& -[2\mu_2t_1\cos k_x+2\mu_1t_2\cos k_y\\
&+2(t_1t_2-\Delta_1\Delta_2)\cos(k_x+k_y)\\
&+2(t_1t_2+\Delta_1\Delta_2)\cos(k_x-k_y)+\mu_1\mu_2]\sigma^z\\
&+[2\mu_2\Delta_1\sin k_x-2\mu_1\Delta_2\sin k_y\\
&+2(t_2\Delta_1-t_1\Delta_2)\sin(k_x+k_y)\\
&+2(t_2\Delta_1+t_1\Delta_2)\sin(k_x-k_y)]\sigma^y=\mathbf{d}_2(k)\cdot\boldsymbol{\sigma},
\end{split}
\label{Hperp2Bloch}
\end{equation}
\end{subequations}
where $\tilde{c}_{k,1}=(\tilde{c}_{k,\uparrow}, \tilde{c}^\dagger_{-k,\downarrow})^T$ and $\tilde{c}_{k,2}=(\tilde{c}_{k,\downarrow},\tilde{c}^\dagger_{-k,\uparrow})^T$.\\
It has been shown \cite{zhang2019majorana} that for 2d Kitaev chains, the topology is characterized by vortices due to the Bloch vector as one varies $k_x$ and $k_y$. But a Bloch vector field like representation in the 2d Brillouin Zone might not be suitable way to properly visualize these vortices. Rather we still stick to the winding number characterization for the MZMs and show that it is possible to figure out the topology as well as the number of the MZMs existing along a certain edge in OBC.
\begin{figure}[htb!]
\includegraphics[scale=0.6]{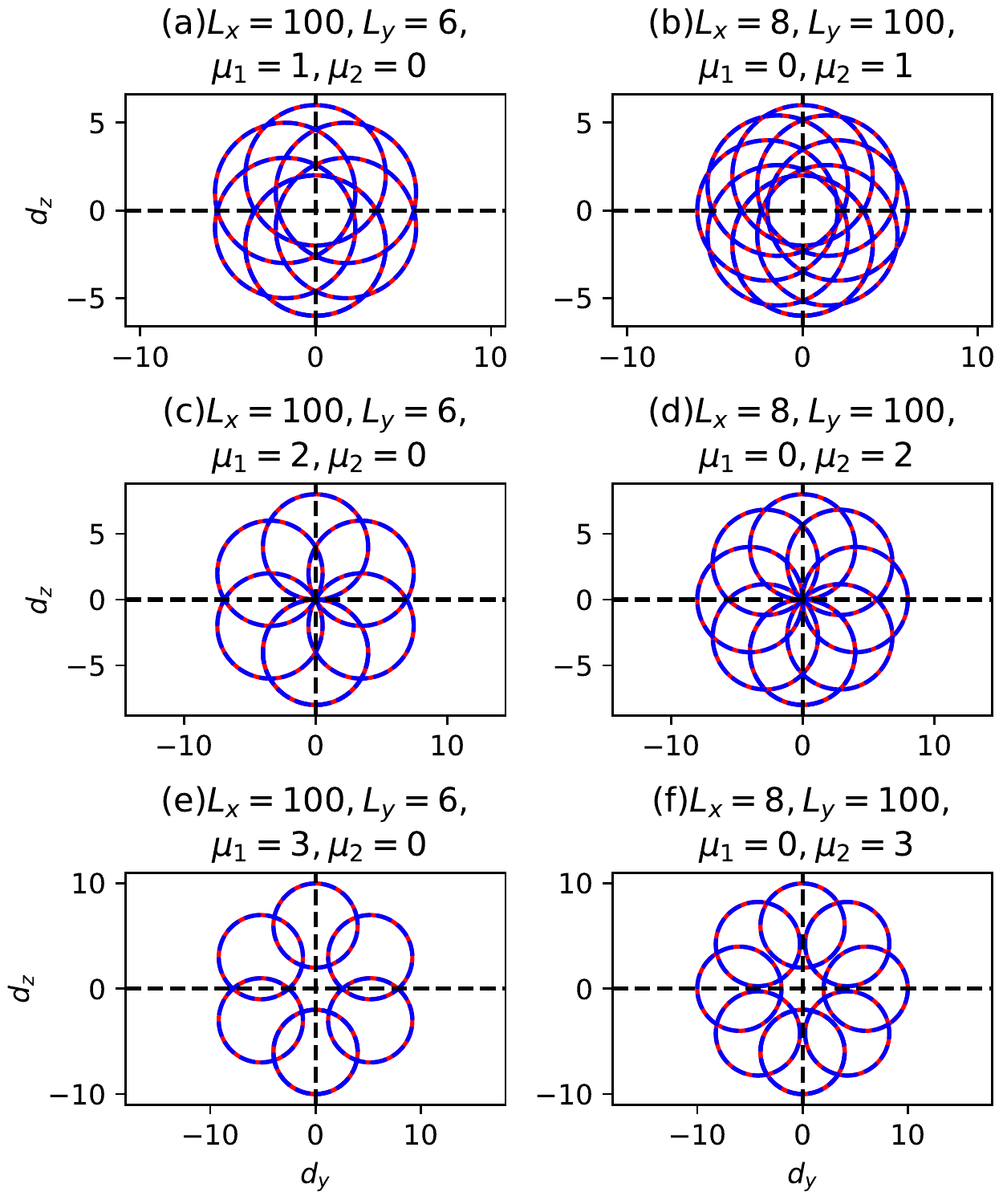}
\caption{Winding from the Bloch vectors $(d_{1,y},d_{1,z})$(red) and $(d_{2,y},d_{2,z})$(blue dashed) with PBC along both x and y directions. (a), (c), (e) are the closed curves due to PBC $L_x=100$ and $L_y=6$ where the 6 circles show the situation along the edge in the y-direction for (a) $\mu_1=1$, $\mu_2=0$ (MZMs present), (c) $\mu_1=2$, $\mu_2=0$ (critical), (e) $\mu_1=3$, $\mu_2=0$ (trivial along y edge) respectively. Similarly, (b), (d), (f) are the closed curves due to PBC $L_x=8$ and $L_y=100$ where the 8 circles show the situation alon the edge in the x-direction for (b) $\mu_1=0$, $\mu_2=1$(MZMs present), (d) $\mu_1=0$, $\mu_2=2$(critical), (f) $\mu_1=0$, $\mu_2=3$(trivial along x edge). The Bloch vectors $\mathbf{d}_1$ and $\mathbf{d}_2$ overlap so that the situation is similar for both the component Hamiltonians. All the cases assume $t_1=1=t_2=\Delta_1=\Delta_2$.}
\label{MKCperpBlochwinding}
\end{figure}
We will work with the matrices, $d_{1,y/z}(k_x,k_y)$ and $d_{2,y/z}(k_x,k_y)$ with PBC in both the x and y-directions so that the matrix element, $d_{1,2}(n,m)$ is given by $k_x=\frac{2\pi n}{L_x}$ and $k_y=\frac{2\pi m}{L_y}$ for $n\in\{0,...,L_x\}$ and $m\in\{0,...,L_y\}$, $L_x$ and $L_y$ being the number of sites in the x and y-directions respectively(we take $L_x+1$ or $L_y+1$ values for n and m respectively just to close the curve, only the first $L_x$ and $L_y$ values are considered for discussion). We then plot the n-th row of $d_{1,y}$ vs. n-th row of $d_{1,z}$ and similarly for $d_{2,y}$ and $d_{2,z}$. Since we have varying $k_x$ with given $k_y$ along a given row, we get the MZMs along the edge in the y-direction in the form of $L_y$ closed curves which enclose the origin if $\mu_1<2t_1$, touch the origin if $\mu_1=2t_1$ and do not contain the origin if $\mu_1>2t_1$. We show this, for the sake of clarity for $L_x=100$ and $L_y=6$ in Fig.~\ref{MKCperpBlochwinding} (a), (c) and (e) which corresponds to Fig.~\ref{MKCperpschematic}(g) and (h). Also for the case, $t_{1,2}=\Delta_{1,2}$ where the closed curve is a circle in the 2-band KC, here we see that the polygon created by joining the centers of the 6 circles also encloses the origin if $\mu_2<2t_2$, intersects the origin(only for $L_y$ even, otherwise may not exactly intersect if $L_y$ is odd) if $\mu_2=2t_2$ and does not enclose the origin if $\mu_2>2t_2$. From this one might imply that the windings of the MZMs in one direction are modulated by the winding in the perpendicular direction both in angle and radii. This does not however show the number of MZMs in the other direction - to provide an answer to this, one must plot the m-th column of $d_{1,y}$ vs. the m-th column of $d_{1,z}$ and similar for $d_{2,y}$ and $d_{2,z}$. Again we show the plot for $L_x=8$ and $L_y=100$ for the sake of clarity in Fig~\ref{MKCperpBlochwinding}(b), (d) and (f), which by comparing for varying $k_y$ and given $k_x$ shows the 7 closed curves encircling the origin corresponding to the 7 MZMs along the edge in the x-direction for $\mu_2<2t_2$ as shown schematically in Fig.~\ref{MKCperpschematic}(e) and (h).\\
The locus of the curves is shown at $t_1=\Delta_1$ for varying $k_x$ and constant $k_y$ as follows (detailed calculation in Supplementary materials \ref{WindingPerp}),
\begin{equation}
\begin{split}
&2t_1\sqrt{M_2^2+R_2^2}=(\cos\theta d_{1,y}+\sin\theta d_{1,z})^2 \\
&+\left(\cos\theta d_{1,z}-\sin\theta d_{1,y}+ \mu_1\sqrt{M_2^2+R_2^2}\right)^2,
\end{split}
\end{equation}
where we denote, $M_2=M_2+2t_2\cos k_y$, $R_2=2\Delta_2\sin k_y$ and $\tan\theta = \frac{R_2}{M_2}$. Essentially, $\theta$ here is the Bloch angle for the 2nd parent Hamiltonian. We observe that the winding curve is given as a circle in a rotated coordinate space and modulated by the dispersion at that $k_y$ value. This however does not affect the condition for non-zero winding number, which can still be written down as $|\mu_1|<2t_1$. The locus for the alternate case can be similarly calculated.\\
One must observe here the difference in the winding number characterization between the MKC parallel system and the MKC perpendicular system. For the MKC parallel system, Fig.~\ref{MKCpll_winding} has shown that the winding number flows between the two component Hamiltonians so that even if at least one of the parent systems is topological, the sum of the absolute value of winding derived from both the systems adds up to two. However, this flow is absent in the MKC perpendicular system. The winding curves of both the component Hamiltonians overlap, so that the component  Hamiltonians always have equal winding, given we are varying the momenta along a certain direction. The difference here, one can notice, is in the nature of the winding when one varies the $k_x$ direction compared to the $k_y$ direction, keeping of course, the perpendicular momenta constant for a given curve. Even Fig.~\ref{MKCperpschematic} agrees that we must get MZMs along the same edge for both the component Hamiltonians for the same set of parameter values.\\

\begin{figure}[htb!]
    \includegraphics[scale=0.6]{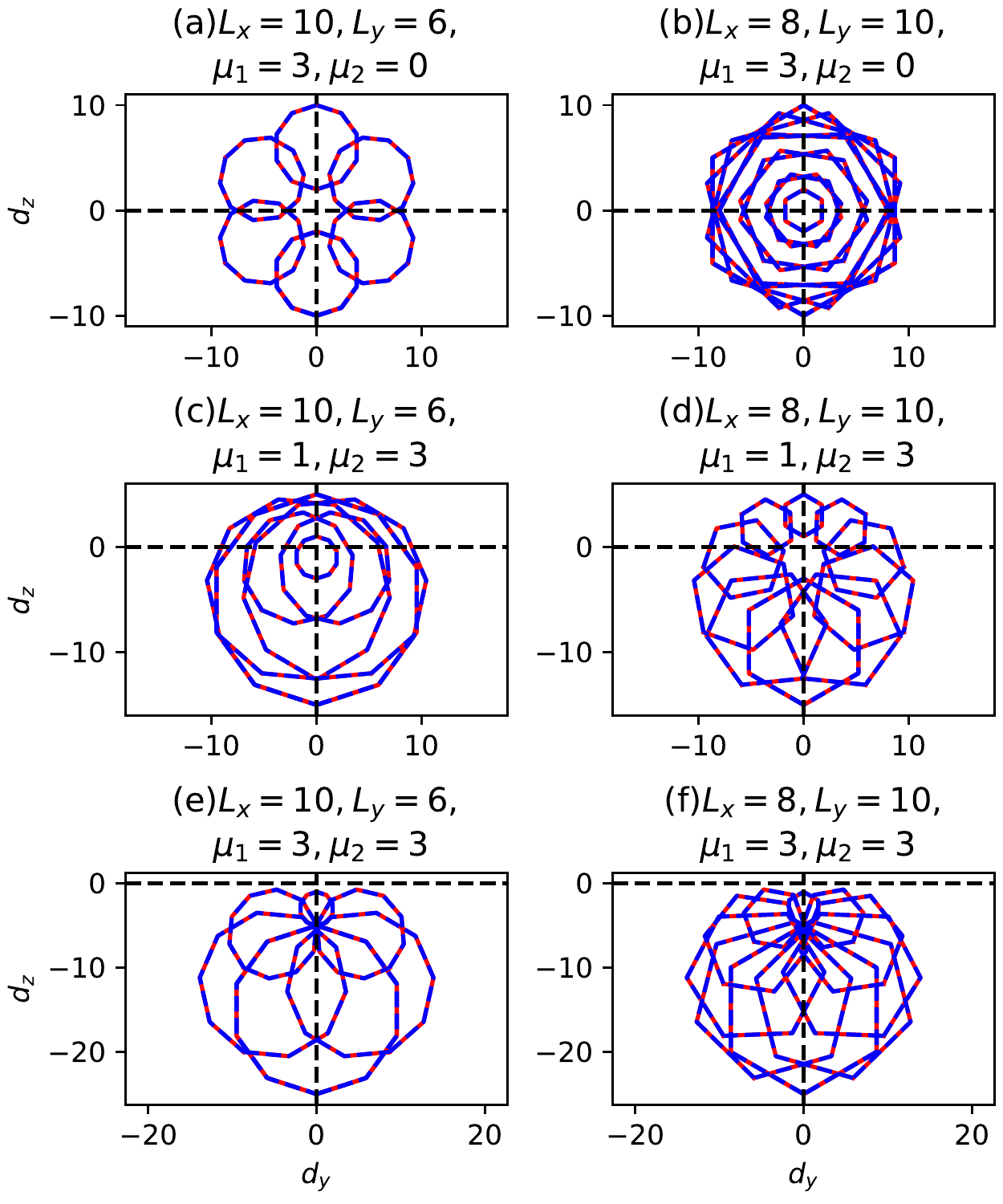}
    \caption{Winding from the Bloch vectors $(d_{1,y},d_{1,z})$(red) and $(d_{2,y},d_{2,z})$(blue dashed) with PBC along both x and y directions. We see more clearly the modulation of the winding curves in one direction due to its perpendicular part and the curves are actually polygons if both edges are smaller. But since the winding depends on the bulk this should not change the final outcome. (a), (c), (e) are the closed curves due to PBC $L_x=10$ and $L_y=6$ where the 6 circles show the situation along the edge in the y-direction for (a)$\mu_1=3,\mu_2=1$(MZMs along x, trivial along y), (c)$\mu_1=1,\mu_2=3$(MZMs along y trivial along x), (e)$\mu_1=3,\mu_2=3$(trivial along both edges) respectively. Similarly, (b), (d), (f) are the closed curves due to PBC $L_x=8$ and $L_y=10$ where the 8 circles show the situation alon the edge in the x-direction for (b)$\mu_1=3,\mu_2=1$(MZMs along x, trivial along y), (d)$\mu_1=1,\mu_2=3$(MZMs along y trivia along x), (f)$\mu_1=0,\mu_2=3$(trivial along both edges). The Bloch vectors $\mathbf{d}_1$ and $\mathbf{d}_2$ overlap so that the situation is similar for both the component Hamiltonians. All the cases assume $t_1=1=t_2=\Delta_1=\Delta_2$.}
    \label{MKCperpBlochwinding1}
\end{figure}

\subsubsection{Edge states of the MKC perpendicular system from the component Hamiltonians and entanglement:}
One can finally develop the explicit form of the edge state expressions obtained previously in Sec.~\ref{MKCperp_edges} with the edges from the component Bloch Hamiltonians in Eqns.~\ref{Hperp1Bloch} and \ref{Hperp2Bloch}. We have worked out the relations for zero energy obtained from $H_{\perp,1}$ and $H_{\perp,2}$, respectively, in Sec.~\ref{Sup_subsec_A2} of the supplementary materials. Here we assume OBC in both the x and y directions so that one can implement localization in both the directions as $k_x\rightarrow iq_x$ and $k_y\rightarrow iq_y$.
\begin{subequations}
\begin{equation}
\begin{split}
&[(2t_1\cosh q_x+\mu_1)\pm 2\Delta_1\sinh q_x]\\
&\times[(2t_2\cosh q_y+\mu_2)\pm 2\Delta_2\sinh q_y]=0,
\end{split}
\end{equation}
\begin{equation}
\begin{split}
&[(2t_1\cosh q_x+\mu_1)\pm 2\Delta_1\sinh q_x]\\
&\times[(2t_2\cosh q_y+\mu_2)\mp 2\Delta_2\sinh q_y]=0.
\end{split}
\end{equation}
\end{subequations}

\begin{figure}[tb]
    \includegraphics[scale=0.6]{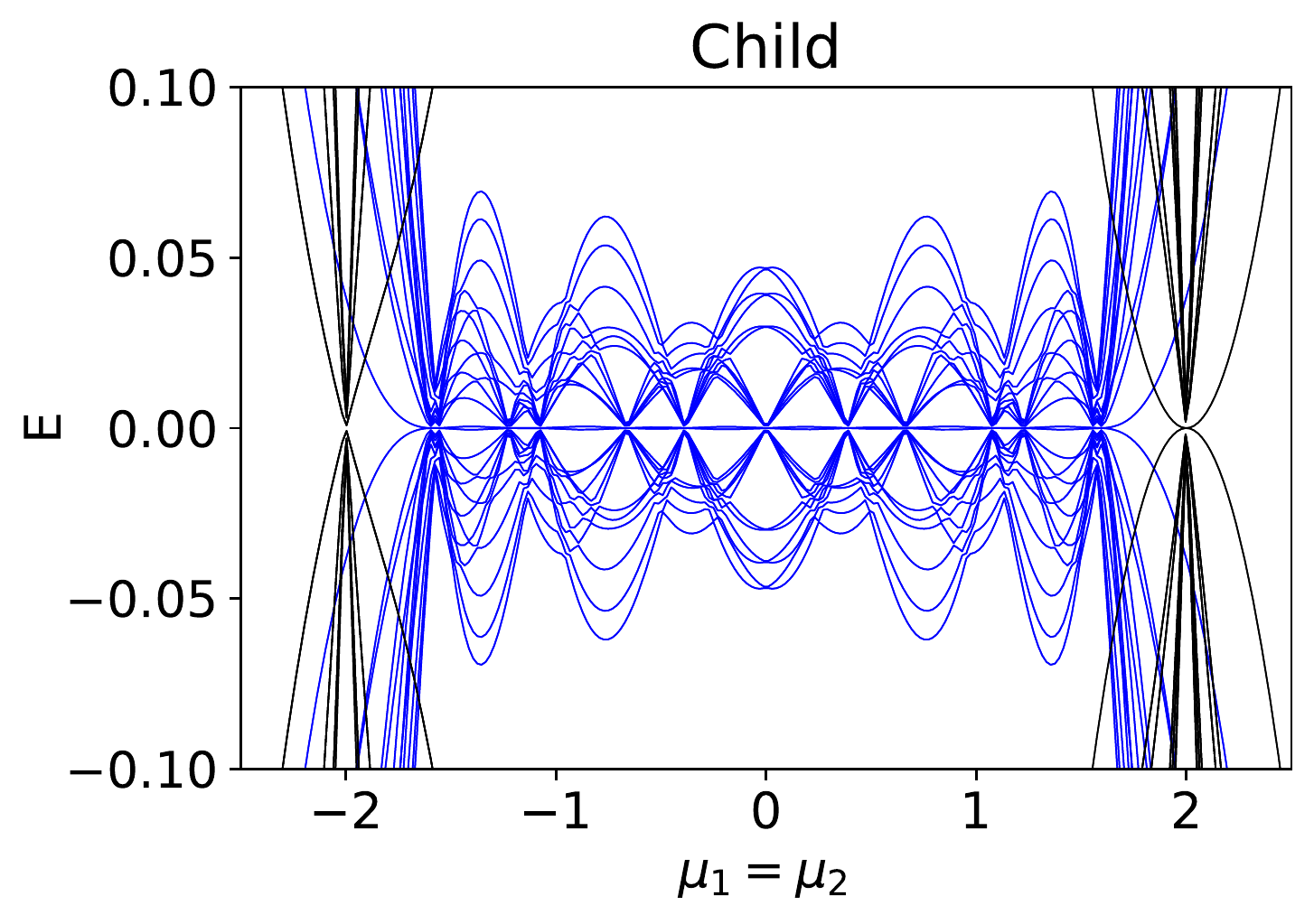}
    \caption{Spectrum E vs. $\mu_1=\mu_2$ for OBC(blue) and PBC(black) along both x and y directions with $L_x=6$ and $L_y=7$ for $t_1=t_2=1$ and $\Delta_1=\Delta_2=0.5$. We see 6 gapless points corresponding to OBC along x with 7-fold degeneracy and 7 gapless points corresponding to OBC along y with 6-fold degeneracy.}
    \label{MKCperpOBCxy}
\end{figure}

We assume the number of sites along the x-direction is $L_x$ and along the y-direction is $L_y$. Taking into account boundary conditions at $x=0,L_x+1$ for each $y$, and $y=0,L_y+1$ for each $x$, where the wavefunction needs to vanish irrespective of the other perpendicular axis site, we have, for $H_{\perp,1}$ for the sign $(+,+)$,
\begin{equation}
\Psi(j,l)\sim [p_{1,+}^j-p_{1,-}^j][s_{1,+}^l-s_{1,-}^l],
\end{equation}
and for $H_{\perp,2}$ for the sign $(+,-)$,
\begin{equation}
\Psi(j,l)\sim [p_{1,+}^j-p_{1,-}^j][s_{2,+}^l-s_{2,-}^l],
\end{equation}
where we have $p_{1,\pm}=\frac{-\mu_1\pm\sqrt{\mu_1^2-4(t_1^2-\Delta_1^2)}}{2(t_1+\Delta_1)}$, $p_{2,\pm}=\frac{-\mu_1\pm\sqrt{\mu_1^2-4(t_1^2-\Delta_1^2)}}{2(t_1-\Delta_1)}$, $s_{1,\pm}=\frac{-\mu_2\pm\sqrt{\mu_2^2-4(t_2^2-\Delta_2^2)}}{2(t_2+\Delta_2)}$, and $s_{2,\pm}=\frac{-\mu_2\pm\sqrt{\mu_2^2-4(t_2^2-\Delta_2^2)}}{2(t_2-\Delta_2)}$. The boundary conditions at $x=L_x+1$ and $y=L_y+1$ again imply,
\begin{equation}
\begin{split}
\mu_1&=2\sqrt{t_1^2-\Delta_1^2}\cos\frac{n_x\pi}{L_x+1},\quad n_x\in\{1,...,L_x\}\\
\mu_2&=2\sqrt{t_2^2-\Delta_2^2}\cos\frac{n_y\pi}{L_y+1},\quad n_y\in\{1,...,L_y\}.
\end{split}
\end{equation}
Then the plot of energy, $E$ vs. $\mu_1=\mu_2=\mu$ should include a total of $L_x\times L_y$ gapless points with the gapless points due to $\mu_1$ being $L_y$-degenerate(degenerate by the number of sites along the y-edge) and the gapless points due to $\mu_2$ being $L_x$-degenerate(degenrate by the number of sites along the x-axis) as we observe in Fig.~\ref{MKCperpOBCxy}.

\begin{figure}[htb]
    \subfloat[$\mu_1=0$, $\mu_2=3$, $t_1=t_2=1$,$\Delta_1=0.5$, $\Delta_2=1.0$]{\includegraphics[width=0.25\textwidth,trim={7cm 0 0 0},clip]{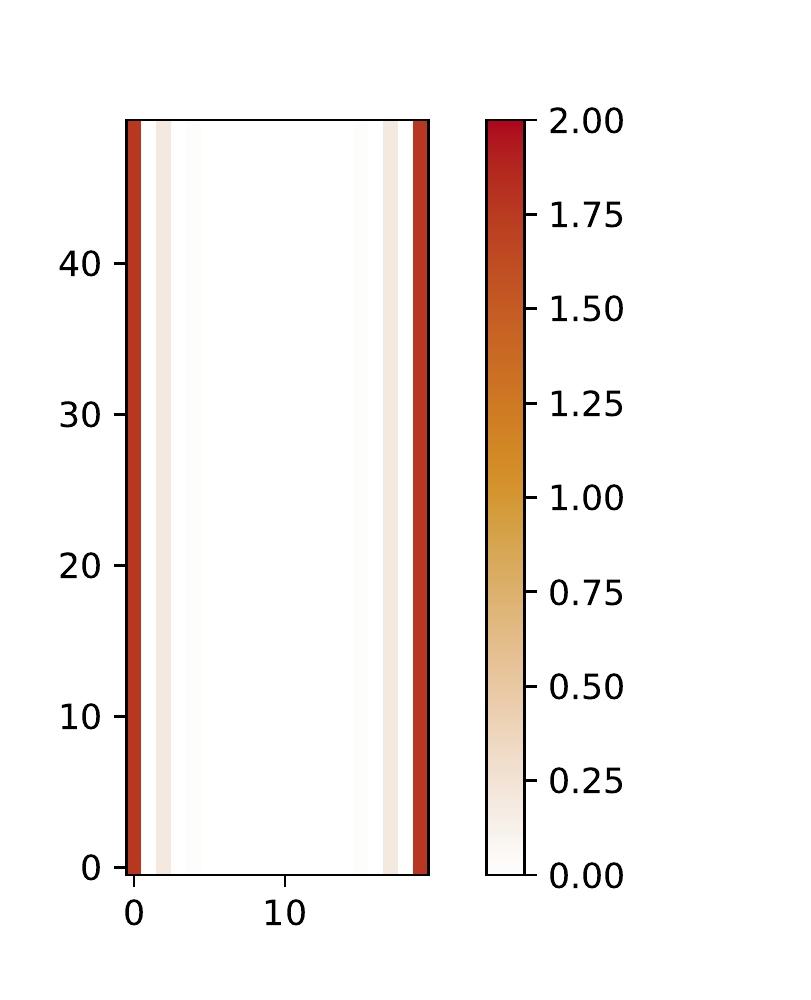}}%
    \subfloat[$\mu_1=0$, $\mu_2=3$, $t_1=t_2=1$, $\Delta_1=0.33$, $\Delta_2=1.0$]{\includegraphics[width=0.25\textwidth,trim={7cm 0 0 0},clip]{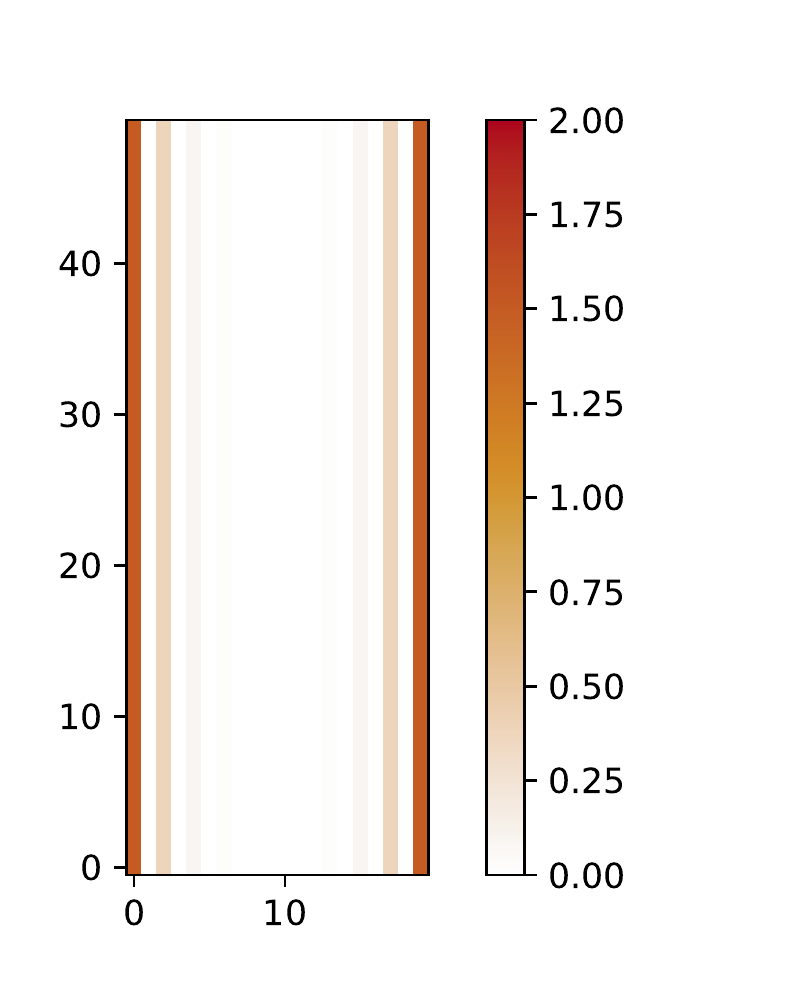}}\\
    \subfloat[$\mu_1=3$, $\mu_2=0$, $t_1=t_2=1$,$\Delta_1=1$, $\Delta_2=0.5$]{\includegraphics[width=0.25\textwidth,trim={7cm 0 0 0},clip]{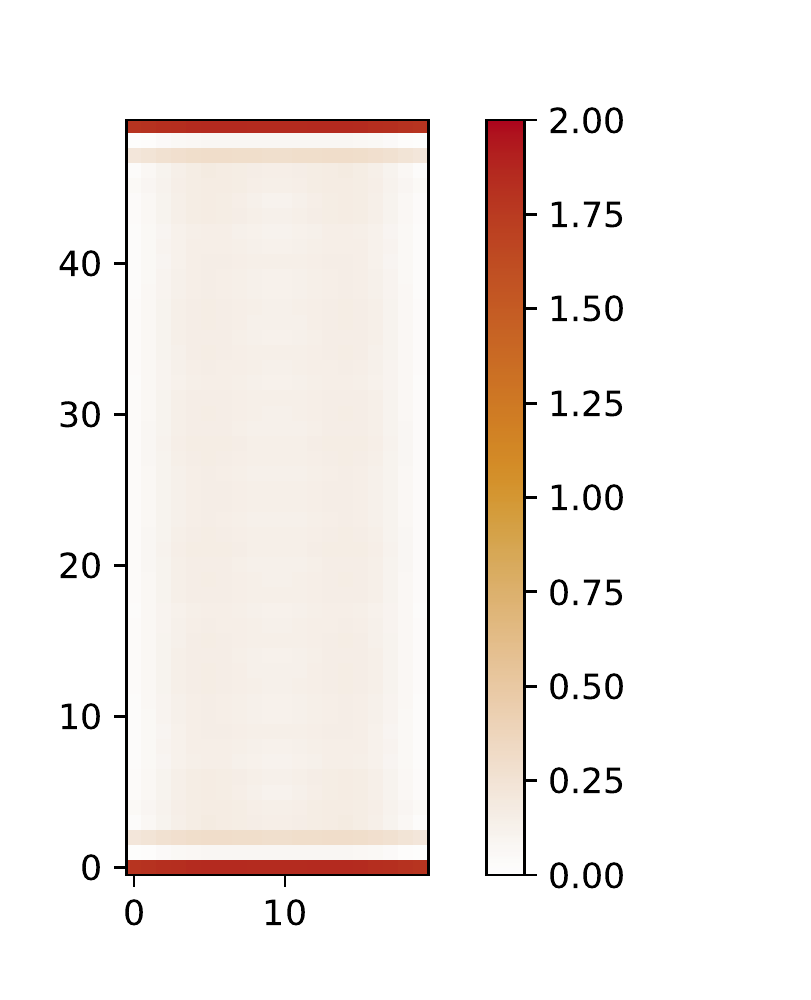}}%
    \subfloat[$\mu_1=3$, $\mu_2=0$, $t_1=t_2=3$, $\Delta_1=1$, $\Delta_2=0.33$]{\includegraphics[width=0.25\textwidth,trim={7cm 0 0 0},clip]{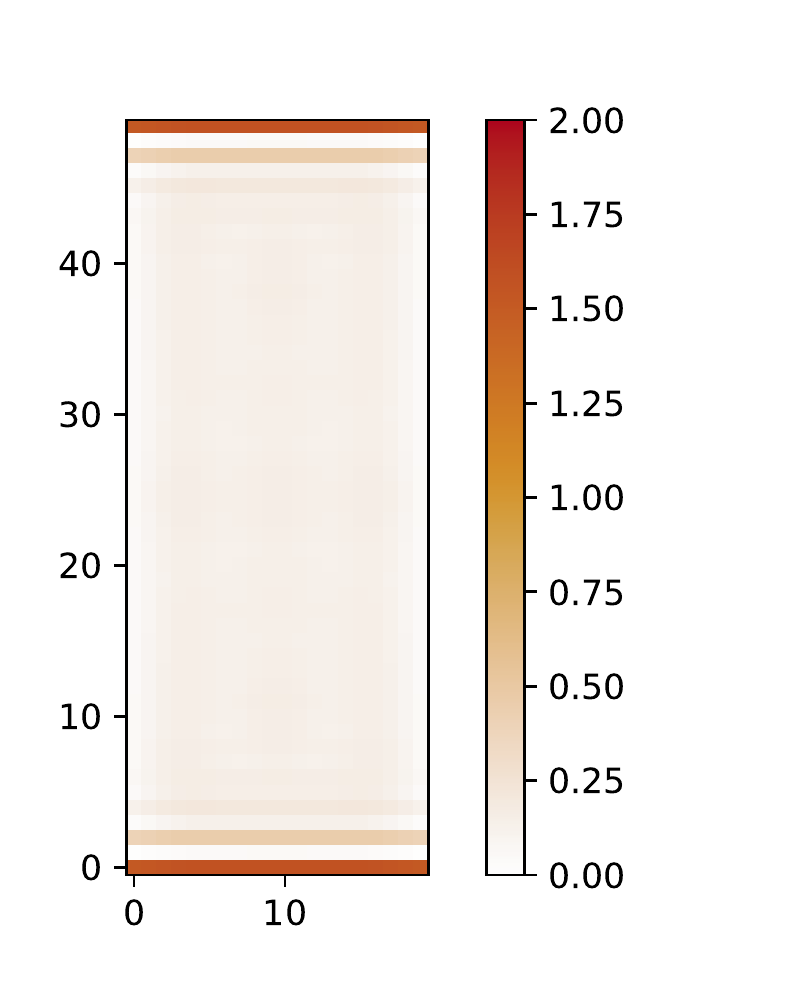}}
    \caption{Density plots for zero energy Majorana modes for a finite $20\times 50$ slab of MKC perpendicular system at different values of $\mu_1$,$\mu_2$ and $t_1$, $t_2$. All the systems have $\Delta_1=\Delta_2=1$.}
    \label{MKCperp_Density_diff_t}
\end{figure}

As in the case of the MKC parallel system, the Majorana zero modes (MZMs) can be shown to be entangled or product states based on the topological nature of the parent Hamiltonians or the boundary conditions one imposes. Unlike the MKC parallel system, the MKC perpendicular system has the added advantage that even if both the parents are topological, it is possible to change the entanglement by gluing together or not opening one of the edges. We again start with the localization, $k_x\rightarrow iq_x$ and $k_y\rightarrow iq_y$ along both the directions, so that $H_{\perp,1}$ and $H_{\perp,2}$ are given as,
\begin{subequations}
\begin{equation}
\begin{split}
\mc{H}_{\perp,1}(iq_x,iq_y)=&-[(\mu_1+2t_1\cosh q_x)(\mu_2+2t_2\cosh q_y)\\
&+4\Delta_1\Delta_2\sinh q_x\sinh q_y]\sigma^z\\
&+i[2\Delta_1\sinh q_x(\mu_2+2t_2\cosh q_y)\\
&+2\Delta_2\sinh q_y(\mu_1+2t_1\cosh q_x)]\sigma^y
\end{split}
\label{Hperploc1}
\end{equation}
\begin{equation}
\begin{split}
\mc{H}_{\perp,2}(iq_x,iq_y)=&-[(\mu_1+2t_1\cosh q_x)(\mu_2+2t_2\cosh q_y)\\
&-4\Delta_1\Delta_2\sinh q_x\sinh q_y]\sigma^z\\
&+i[2\Delta_1\sinh q_x(\mu_2+2t_2\cosh q_y)\\
&-2\Delta_2\sinh q_y(\mu_1+2t_1\cosh q_x)]\sigma^y
\end{split}
\label{Hperploc2}
\end{equation}
\end{subequations}
To get null-eigenvalues from the above expressions, we must satisfy the conditions,
\begin{equation}
\begin{split}
&[(2t_1\cosh q_x+\mu_1)\mp 2\Delta_1\sinh q_x]\\
&\times[(2t_2\cosh q_y+\mu_2)\mp 2\Delta_2\sinh q_y] = 0,
\end{split}
\label{Hperpcond1}
\end{equation}
for the component $H_{\perp,1}$ and,
\begin{equation}
\begin{split}
&[(2t_1\cosh q_x+\mu_1)\mp 2\Delta_1\sinh q_x]\\
&\times[(2t_2\cosh q_y+\mu_2)\pm 2\Delta_2\sinh q_y] = 0,
\end{split}
\label{Hperpcond2}
\end{equation}

\begin{table*}[ht]
\begin{tabular}{|c|c|c|c|c|c|c|}
\hline
\multicolumn{3}{|c|}{Parent 1} & \multicolumn{3}{c|}{Parent 2} & \multirow{2}{*}{MZM Eigenvectors} \\
\cline{1-6}
Phase & $\frac{\text{sgn}(t_1)}{\text{sgn}(\Delta_1)}$ & x-BC & Phase & $\frac{\text{sgn}(t_1)}{\text{sgn}(\Delta_1)}$ & y-BC & \\
\hline
\multirow{4}{*}{topo} & + & OBC & \multirow{4}{*}{topo} & + & OBC & $\{\frac{1}{\sqrt{2}}(\ket{00}-\ket{11}),\ket{01},\ket{10}\}$\\
 & + & OBC & & - & OBC & $\{\frac{1}{\sqrt{2}}(\ket{01}-\ket{10}),\ket{00},\ket{11}\}$\\
 & - & OBC & & + & OBC & $\{\frac{1}{\sqrt{2}}(\ket{01}+\ket{10}),\ket{00},\ket{11}\}$\\
 & - & OBC & & - & OBC & $\{\frac{1}{\sqrt{2}}(\ket{00}+\ket{11}),\ket{01},\ket{10}\}$\\
\hline
\multirow{4}{*}{topo} & + & OBC & \multirow{4}{*}{topo} & +,- & PBC & $\{\frac{1}{\sqrt{2}}(\ket{00}-\ket{11}),\frac{1}{\sqrt{2}}(\ket{01}-\ket{10})\}$\\
 & - & OBC & & +,- & PBC & $\{\frac{1}{\sqrt{2}}(\ket{00}+\ket{11}),\frac{1}{\sqrt{2}}(\ket{01}+\ket{10})\}$\\
 & +,- & PBC & & + & OBC & $\{\frac{1}{\sqrt{2}}(\ket{00}-\ket{11}),\frac{1}{\sqrt{2}}(\ket{01}+\ket{10})\}$\\
 & +,- & PBC & & - & OBC & $\{\frac{1}{\sqrt{2}}(\ket{00}+\ket{11}),\frac{1}{\sqrt{2}}(\ket{01}-\ket{10})\}$\\
\hline
\multirow{2}{*}{topo} & + & OBC & \multirow{2}{*}{triv} & & & $\{\frac{1}{\sqrt{2}}(\ket{00}-\ket{11}),\frac{1}{\sqrt{2}}(\ket{01}-\ket{10})\}$\\
 & - & OBC & &  & & $\{\frac{1}{\sqrt{2}}(\ket{00}+\ket{11}),\frac{1}{\sqrt{2}}(\ket{01}+\ket{10})\}$\\
\hline
\multirow{2}{*}{triv} & & & \multirow{2}{*}{topo} & + & OBC & $\{\frac{1}{\sqrt{2}}(\ket{00}-\ket{11}),\frac{1}{\sqrt{2}}(\ket{01}+\ket{10})\}$\\
 & & &  & - & OBC & $\{\frac{1}{\sqrt{2}}(\ket{00}+\ket{11}),\frac{1}{\sqrt{2}}(\ket{01}-\ket{10})\}$\\
\hline
\end{tabular}
\caption{Eigen-vectors of the MKC perpendicular system for different topological characterizations of the two parent systems, ratio of signs of $t_i$ and $\Delta_i$, $i\in\{1,2\}$, and boundary conditions.}
\label{MMZMpll}
\end{table*}
for the component $H_{\perp,2}$. The schematic diagram Fig.~\ref{MKCperpschematic} shows that the edges where the MZMs are localized depend on the topological nature of the parents. But if one of the directions remain unopened or with periodic boundary conditions, one will not observe the MZMs although the relevant parent is topological. Let us consider the condition Eqn.~\ref{Hperpcond1} for $\text{sgn}(t_i)=\text{sgn}(\Delta_i)$, $i\in\{1,2\}$,
\begin{equation}
\begin{split}
&[(2t_1\cosh q_x+\mu_1)-2\Delta_1\sinh q_x]\\
&\times [(2t_2\cosh q_y+\mu_2)-2\Delta_2\sinh q_y]=0.
\end{split}
\end{equation}
If there exists \textit{OBC along both x and y directions and both the parents are topological}, we have, $2t_1\cosh q_x+\mu_1=2\Delta_1\sinh q_x$ and $2t_2\cosh q_y+\mu_2=2\Delta_2\sinh q_y$, which when substituted into Eqns.~\ref{Hperploc1} and \ref{Hperploc2} shows that $\mc{H}_{\perp,2}$ vanishes. We are actually working in the full basis of the MKC perpendicular system, given by the four degrees of freedom, $(\tilde{c}_{\boldsymbol{k},\uparrow},\tilde{c}_{\boldsymbol{k},\downarrow},\tilde{c}^\dagger_{-\boldsymbol{k},\uparrow},\tilde{c}^\dagger_{-\boldsymbol{k},\downarrow})^T$. which combines the degrees of freedom of the two components. In this basis, the null eigenvectors derived from $\mc{H}_{\perp,1}(iq_x,iq_y)$ are given as,
\begin{equation}
\ket{\Psi}_{MZM}=\{\frac{1}{\sqrt{2}}(\ket{00}-\ket{11}),\ket{01},\ket{10}\},
\end{equation}
where $\ket{0}=(1,0)^T$ and $\ket{1}=(0,1)^T$.\\
Now, say if \textit{parent 1 is topological  while parent 2 is trivial while we retain OBC in both x and y directions}, we must only satisfy the condition, $2t_1\cosh q_x+\mu_1=2\Delta_1\sinh q_x$. Substituting the identity into Eqn.~ \ref{Hperploc1} and \ref{Hperploc2}, the null eigen-vectors in the full basis with four degrees of freedom is shown to be,
\begin{equation}
\ket{\Psi}_{MZM}=\{\frac{1}{\sqrt{2}}(\ket{00}-\ket{11}),\frac{1}{\sqrt{2}}(\ket{01}-\ket{10})\}.
\end{equation}
One must note that the above eigen-vectors are also valid if the \textit{y-direction is unopened or in PBC so that the topological nature of the second parent does not matter}. Then only the condition $2t_1\cosh q_x+\mu_1=2\Delta_1\sinh q$ holds. Detailed calculations can be found in Supplementary section \ref{Sup_subsec_A2}. The extra part here compared to the MKC parallel system is that one can control the entanglement between maximally entangled Bell states and product states, not only via the topological nature of the parents but also by the boundary conditions along the two directions. We provide a small table (Table II) showing all the possible MZM eigen-vectors under various parent topology and boundary conditions. The table lists only the MMZM eigenvector at edges $x=0$ and $y=0$. The eigenvectors at the other edges can be found by changing, $\frac{\text{sqn}(t_i)}{\text{sgn}(\Delta_i)}$ from $+$ to $-$ and vice-versa for both the parents. We will recover a total of four eigenvectors with two common eigenvectors for both signs when both parents are topological. Thus, the MZMs obtained in this case have a similar entanglement structure as the \textit{Multiplicative Majorana Zero Modes(MMZMs)} in the parallel case so that one may refer to the MZMs in the MKC perpendicular system as \textit{Multiplicative Majorana Zero Modes(MMZMs)} as well.

\subsubsection{Parallel quantum gates without braiding:}
The MMZMs of the perpendicular MKC system can also be  entangled states and separable two-qubit states via variation of the system parameters at a given parity. In this case, however, there is the potential to perform parallel gate operations: since the number of MMZM pairs the perpendicular system has is proportional to the perimeter of the system (when there are open boundary conditions in each direction), it is possible to carry out CNOT operations simultaneously on a large number of MMZM pairs. The full potential for universal quantum computation schemes by manipulating multiplicative topological phases in combination with this potential for parallelized gate operations warrants further investigation, but this is beyond the scope of this work.

\section{Discussion and Conclusion}
In this work, we introduce the concept of a multiplicative Majorana zero-mode(MMZM), a zero-energy, symmetry-protected tensor product state or maximally-entangled Bell state composed of one or more \textit{unpaired} Majorana zero-modes. We find that the recently-introduced multiplicative topological phases~\cite{cook2022mult} realize such zero-modes through bulk-boundary correspondence, specifically considering a canonical Hamiltonian for realizing such multiplicative topological phases consisting of a symmetry-protected tensor product of two Kitaev chain Hamiltonians. While considerable important work currently focuses on smoking-gun experimental confirmation of unpaired Majorana zero-modes and individual topological qubits in experiment, it remains important to identify practical platforms for scalable topological quantum computers. Results discussed here are relevant to realizing such scalable systems of many topological qubits, given that multiplicative Majorana zero-modes are individual states composed of multiple symmetry-protected unpaired Majorana zero-modes. Additionally, results here indicate there are opportunities for controlled introduction of entanglement between degrees of freedom derived from both parents in the Majorana eigenvectors, potentially useful for performing gate operations of topological quantum computation schemes.

We demonstrate the richness of multiplicative topological phases by constructing one-dimensional \textit{but also two-dimensional} multiplicative Kitaev chain models capable of realizing myriad topologically non-trivial phases. These models consist of either two parent Kitaev chain Hamiltonians that depend on the same momentum component, or perpendicular momentum components, combined in a symmetry-protected, tensor product construction. We lay the groundwork for studying these systems by characterizing bulk topology and corresponding topologically-protected boundary states, focusing on the dependence of the resultant multiplicative topological phases on the topology of the parents.

We characterize the bulk of multiplicative Kitaev chains first by demonstrating that eigenvalues of the bulk spectrum are products of the eigenvalues of the parent Kitaev chain bulk spectra, indicating topological phases of the child are stable up to gap-closing of either parent. We also explore characterization of multiplicative topology in the bulk, and find that Wilson loop spectra successfully characterize some multiplicative topological phases, but can also indicate trivial topology in the case when each parent is topologically non-trivial. We show, however, that it is possible to decompose the MKC into chiral subsectors to more fully characterize the topology under certain conditions. This exploits the fact that the degrees of freedom of these Hamiltonians are symmetry-constrained, locking together into pseudospins yielding winding numbers that successfully characterize all topologically non-trivial states realized through different combinations of trivial and non-trivial parents considered here. Fully characterizing multiplicative topological phases, however, is an important issue to explore in future works.

Topologically-protected boundary states possible for the multiplicative Kitaev chain Hamiltonians are varied.  We consider child Hamiltonians, which can be block-diagonalized into chiral subsectors. Based on the topology of the parents, the MMZMs of the child may either possess a tensor product or maximally-entangled Bell state structure. We characterize topology of the child chiral subsectors in the bulk by computing winding numbers for the parallel case, which seem to possess an algebra as one might infer from addition of angular momentum. We find a relationship between the winding numbers of the child chiral subsectors in the case of two parallel parent Kitaev chains. Schematically, from real space Hamiltonian expressions, we show that for suitable parametric conditions, MMZMs are localized at the outermost and second outermost sites for 1d (parallel) case or along two or four edges for the 2d (perpendicular) case.

Similarly, we illustrate a winding number calculation for the perpendicular case which accurately reflects the number of MMZMs and the edge along which they are localized. A quantization condition for the existence of topologically-protected boundary modes in finite size MKC systems has also been obtained, and we have shown that it agrees with our numerical results for one of the simpler cases. More complicated cases may still be studied, such as one example in Sec. \ref{Sup_subsec_A1} of the Supplementary Materials. This shows that a topologically-protected, multiplicative Majorana zero-mode of the child MKC, in both the parallel and the perpendicular case, is not just a tensor product of parent Hamiltonian states in general. Instead, they can more generally possess emergent properties evident in their localization, entanglement and topological robustness.

Future work will explore topological characterization in systems with lower symmetry, for which the multiplicative Majorana zero-modes are expected to take more general forms, as well as control of the entanglement properties, which hold great promise for developing more robust and versatile topological quantum computation schemes. This could include further study of the potential for braiding schemes, with the degenerate manifold of zero-energy states for the case of each parent topologically non-trivial being a particularly interesting case for such future study, as well as further study of the potential for alternatives to braiding schemes for topologically-protected quantum computation.

\textbf{Acknowledgements} - We gratefully acknowledge helpful discussions with J.~E.~Moore, I.~A.~Day and R.~Calderon.

\textbf{Correspondence} - Correspondence and requests for materials should be addressed to A.M.C. (email: cooka@pks.mpg.de).

\bibliography{ref.bib}

\begin{thebibliography}{36}%
\makeatletter
\providecommand \@ifxundefined [1]{%
 \@ifx{#1\undefined}
}%
\providecommand \@ifnum [1]{%
 \ifnum #1\expandafter \@firstoftwo
 \else \expandafter \@secondoftwo
 \fi
}%
\providecommand \@ifx [1]{%
 \ifx #1\expandafter \@firstoftwo
 \else \expandafter \@secondoftwo
 \fi
}%
\providecommand \natexlab [1]{#1}%
\providecommand \enquote  [1]{``#1''}%
\providecommand \bibnamefont  [1]{#1}%
\providecommand \bibfnamefont [1]{#1}%
\providecommand \citenamefont [1]{#1}%
\providecommand \href@noop [0]{\@secondoftwo}%
\providecommand \href [0]{\begingroup \@sanitize@url \@href}%
\providecommand \@href[1]{\@@startlink{#1}\@@href}%
\providecommand \@@href[1]{\endgroup#1\@@endlink}%
\providecommand \@sanitize@url [0]{\catcode `\\12\catcode `\$12\catcode
  `\&12\catcode `\#12\catcode `\^12\catcode `\_12\catcode `\%12\relax}%
\providecommand \@@startlink[1]{}%
\providecommand \@@endlink[0]{}%
\providecommand \url  [0]{\begingroup\@sanitize@url \@url }%
\providecommand \@url [1]{\endgroup\@href {#1}{\urlprefix }}%
\providecommand \urlprefix  [0]{URL }%
\providecommand \Eprint [0]{\href }%
\providecommand \doibase [0]{http://dx.doi.org/}%
\providecommand \selectlanguage [0]{\@gobble}%
\providecommand \bibinfo  [0]{\@secondoftwo}%
\providecommand \bibfield  [0]{\@secondoftwo}%
\providecommand \translation [1]{[#1]}%
\providecommand \BibitemOpen [0]{}%
\providecommand \bibitemStop [0]{}%
\providecommand \bibitemNoStop [0]{.\EOS\space}%
\providecommand \EOS [0]{\spacefactor3000\relax}%
\providecommand \BibitemShut  [1]{\csname bibitem#1\endcsname}%
\let\auto@bib@innerbib\@empty
\bibitem [{\citenamefont {Kitaev}(2003)}]{kitaev2003fault}%
  \BibitemOpen
  \bibfield  {author} {\bibinfo {author} {\bibfnamefont {A~Yu}\ \bibnamefont
  {Kitaev}},\ }\bibfield  {title} {\enquote {\bibinfo {title} {Fault-tolerant
  quantum computation by anyons},}\ }\href@noop {} {\bibfield  {journal}
  {\bibinfo  {journal} {Annals of Physics}\ }\textbf {\bibinfo {volume}
  {303}},\ \bibinfo {pages} {2--30} (\bibinfo {year} {2003})}\BibitemShut
  {NoStop}%
\bibitem [{\citenamefont {Aasen}\ \emph {et~al.}(2016)\citenamefont {Aasen},
  \citenamefont {Hell}, \citenamefont {Mishmash}, \citenamefont {Higginbotham},
  \citenamefont {Danon}, \citenamefont {Leijnse}, \citenamefont {Jespersen},
  \citenamefont {Folk}, \citenamefont {Marcus}, \citenamefont {Flensberg} \emph
  {et~al.}}]{aasen2016milestones}%
  \BibitemOpen
  \bibfield  {author} {\bibinfo {author} {\bibfnamefont {David}\ \bibnamefont
  {Aasen}}, \bibinfo {author} {\bibfnamefont {Michael}\ \bibnamefont {Hell}},
  \bibinfo {author} {\bibfnamefont {Ryan~V}\ \bibnamefont {Mishmash}}, \bibinfo
  {author} {\bibfnamefont {Andrew}\ \bibnamefont {Higginbotham}}, \bibinfo
  {author} {\bibfnamefont {Jeroen}\ \bibnamefont {Danon}}, \bibinfo {author}
  {\bibfnamefont {Martin}\ \bibnamefont {Leijnse}}, \bibinfo {author}
  {\bibfnamefont {Thomas~S}\ \bibnamefont {Jespersen}}, \bibinfo {author}
  {\bibfnamefont {Joshua~A}\ \bibnamefont {Folk}}, \bibinfo {author}
  {\bibfnamefont {Charles~M}\ \bibnamefont {Marcus}}, \bibinfo {author}
  {\bibfnamefont {Karsten}\ \bibnamefont {Flensberg}},  \emph {et~al.},\
  }\bibfield  {title} {\enquote {\bibinfo {title} {Milestones toward
  majorana-based quantum computing},}\ }\href@noop {} {\bibfield  {journal}
  {\bibinfo  {journal} {Physical Review X}\ }\textbf {\bibinfo {volume} {6}},\
  \bibinfo {pages} {031016} (\bibinfo {year} {2016})}\BibitemShut {NoStop}%
\bibitem [{\citenamefont {Str{\"u}bi}\ \emph {et~al.}(2011)\citenamefont
  {Str{\"u}bi}, \citenamefont {Belzig}, \citenamefont {Choi},\ and\
  \citenamefont {Bruder}}]{strubi2011interferometric}%
  \BibitemOpen
  \bibfield  {author} {\bibinfo {author} {\bibfnamefont {Gr{\'e}gory}\
  \bibnamefont {Str{\"u}bi}}, \bibinfo {author} {\bibfnamefont {Wolfgang}\
  \bibnamefont {Belzig}}, \bibinfo {author} {\bibfnamefont {Mahn-Soo}\
  \bibnamefont {Choi}}, \ and\ \bibinfo {author} {\bibfnamefont {Christoph}\
  \bibnamefont {Bruder}},\ }\bibfield  {title} {\enquote {\bibinfo {title}
  {Interferometric and noise signatures of majorana fermion edge states in
  transport experiments},}\ }\href@noop {} {\bibfield  {journal} {\bibinfo
  {journal} {Physical review letters}\ }\textbf {\bibinfo {volume} {107}},\
  \bibinfo {pages} {136403} (\bibinfo {year} {2011})}\BibitemShut {NoStop}%
\bibitem [{\citenamefont {J{\"a}ck}\ \emph {et~al.}(2019)\citenamefont
  {J{\"a}ck}, \citenamefont {Xie}, \citenamefont {Li}, \citenamefont {Jeon},
  \citenamefont {Bernevig},\ and\ \citenamefont
  {Yazdani}}]{jack2019observation}%
  \BibitemOpen
  \bibfield  {author} {\bibinfo {author} {\bibfnamefont {Berthold}\
  \bibnamefont {J{\"a}ck}}, \bibinfo {author} {\bibfnamefont {Yonglong}\
  \bibnamefont {Xie}}, \bibinfo {author} {\bibfnamefont {Jian}\ \bibnamefont
  {Li}}, \bibinfo {author} {\bibfnamefont {Sangjun}\ \bibnamefont {Jeon}},
  \bibinfo {author} {\bibfnamefont {B~Andrei}\ \bibnamefont {Bernevig}}, \ and\
  \bibinfo {author} {\bibfnamefont {Ali}\ \bibnamefont {Yazdani}},\ }\bibfield
  {title} {\enquote {\bibinfo {title} {Observation of a majorana zero mode in a
  topologically protected edge channel},}\ }\href@noop {} {\bibfield  {journal}
  {\bibinfo  {journal} {Science}\ }\textbf {\bibinfo {volume} {364}},\ \bibinfo
  {pages} {1255--1259} (\bibinfo {year} {2019})}\BibitemShut {NoStop}%
\bibitem [{\citenamefont {Karzig}\ \emph {et~al.}(2017)\citenamefont {Karzig},
  \citenamefont {Knapp}, \citenamefont {Lutchyn}, \citenamefont {Bonderson},
  \citenamefont {Hastings}, \citenamefont {Nayak}, \citenamefont {Alicea},
  \citenamefont {Flensberg}, \citenamefont {Plugge}, \citenamefont {Oreg} \emph
  {et~al.}}]{karzig2017scalable}%
  \BibitemOpen
  \bibfield  {author} {\bibinfo {author} {\bibfnamefont {Torsten}\ \bibnamefont
  {Karzig}}, \bibinfo {author} {\bibfnamefont {Christina}\ \bibnamefont
  {Knapp}}, \bibinfo {author} {\bibfnamefont {Roman~M}\ \bibnamefont
  {Lutchyn}}, \bibinfo {author} {\bibfnamefont {Parsa}\ \bibnamefont
  {Bonderson}}, \bibinfo {author} {\bibfnamefont {Matthew~B}\ \bibnamefont
  {Hastings}}, \bibinfo {author} {\bibfnamefont {Chetan}\ \bibnamefont
  {Nayak}}, \bibinfo {author} {\bibfnamefont {Jason}\ \bibnamefont {Alicea}},
  \bibinfo {author} {\bibfnamefont {Karsten}\ \bibnamefont {Flensberg}},
  \bibinfo {author} {\bibfnamefont {Stephan}\ \bibnamefont {Plugge}}, \bibinfo
  {author} {\bibfnamefont {Yuval}\ \bibnamefont {Oreg}},  \emph {et~al.},\
  }\bibfield  {title} {\enquote {\bibinfo {title} {Scalable designs for
  quasiparticle-poisoning-protected topological quantum computation with
  majorana zero modes},}\ }\href@noop {} {\bibfield  {journal} {\bibinfo
  {journal} {Physical Review B}\ }\textbf {\bibinfo {volume} {95}},\ \bibinfo
  {pages} {235305} (\bibinfo {year} {2017})}\BibitemShut {NoStop}%
\bibitem [{\citenamefont {Lian}\ \emph {et~al.}(2018)\citenamefont {Lian},
  \citenamefont {Sun}, \citenamefont {Vaezi}, \citenamefont {Qi},\ and\
  \citenamefont {Zhang}}]{lian2018topological}%
  \BibitemOpen
  \bibfield  {author} {\bibinfo {author} {\bibfnamefont {Biao}\ \bibnamefont
  {Lian}}, \bibinfo {author} {\bibfnamefont {Xiao-Qi}\ \bibnamefont {Sun}},
  \bibinfo {author} {\bibfnamefont {Abolhassan}\ \bibnamefont {Vaezi}},
  \bibinfo {author} {\bibfnamefont {Xiao-Liang}\ \bibnamefont {Qi}}, \ and\
  \bibinfo {author} {\bibfnamefont {Shou-Cheng}\ \bibnamefont {Zhang}},\
  }\bibfield  {title} {\enquote {\bibinfo {title} {Topological quantum
  computation based on chiral majorana fermions},}\ }\href@noop {} {\bibfield
  {journal} {\bibinfo  {journal} {Proceedings of the National Academy of
  Sciences}\ }\textbf {\bibinfo {volume} {115}},\ \bibinfo {pages}
  {10938--10942} (\bibinfo {year} {2018})}\BibitemShut {NoStop}%
\bibitem [{\citenamefont {Plugge}\ \emph {et~al.}(2017)\citenamefont {Plugge},
  \citenamefont {Rasmussen}, \citenamefont {Egger},\ and\ \citenamefont
  {Flensberg}}]{plugge2017majorana}%
  \BibitemOpen
  \bibfield  {author} {\bibinfo {author} {\bibfnamefont {Stephan}\ \bibnamefont
  {Plugge}}, \bibinfo {author} {\bibfnamefont {Asbj{\o}rn}\ \bibnamefont
  {Rasmussen}}, \bibinfo {author} {\bibfnamefont {Reinhold}\ \bibnamefont
  {Egger}}, \ and\ \bibinfo {author} {\bibfnamefont {Karsten}\ \bibnamefont
  {Flensberg}},\ }\bibfield  {title} {\enquote {\bibinfo {title} {Majorana box
  qubits},}\ }\href@noop {} {\bibfield  {journal} {\bibinfo  {journal} {New
  Journal of Physics}\ }\textbf {\bibinfo {volume} {19}},\ \bibinfo {pages}
  {012001} (\bibinfo {year} {2017})}\BibitemShut {NoStop}%
\bibitem [{\citenamefont {Leijnse}\ and\ \citenamefont
  {Flensberg}(2012)}]{leijnse2012parity}%
  \BibitemOpen
  \bibfield  {author} {\bibinfo {author} {\bibfnamefont {Martin}\ \bibnamefont
  {Leijnse}}\ and\ \bibinfo {author} {\bibfnamefont {Karsten}\ \bibnamefont
  {Flensberg}},\ }\bibfield  {title} {\enquote {\bibinfo {title} {Parity qubits
  and poor man's majorana bound states in double quantum dots},}\ }\href@noop
  {} {\bibfield  {journal} {\bibinfo  {journal} {Physical Review B}\ }\textbf
  {\bibinfo {volume} {86}},\ \bibinfo {pages} {134528} (\bibinfo {year}
  {2012})}\BibitemShut {NoStop}%
\bibitem [{\citenamefont {Calzona}\ \emph {et~al.}(2020)\citenamefont
  {Calzona}, \citenamefont {Bauer},\ and\ \citenamefont
  {Trauzettel}}]{calzona_2020}%
  \BibitemOpen
  \bibfield  {author} {\bibinfo {author} {\bibfnamefont {Alessio}\ \bibnamefont
  {Calzona}}, \bibinfo {author} {\bibfnamefont {Nicolas~P.}\ \bibnamefont
  {Bauer}}, \ and\ \bibinfo {author} {\bibfnamefont {Björn}\ \bibnamefont
  {Trauzettel}},\ }\bibfield  {title} {\enquote {\bibinfo {title} {{Holonomic
  implementation of CNOT gate on topological Majorana qubits}},}\ }\href
  {\doibase 10.21468/SciPostPhysCore.3.2.014} {\bibfield  {journal} {\bibinfo
  {journal} {SciPost Phys. Core}\ }\textbf {\bibinfo {volume} {3}},\ \bibinfo
  {pages} {14} (\bibinfo {year} {2020})}\BibitemShut {NoStop}%
\bibitem [{\citenamefont {Cook}\ and\ \citenamefont
  {Moore}(2022)}]{cook2022mult}%
  \BibitemOpen
  \bibfield  {author} {\bibinfo {author} {\bibfnamefont {Ashley~M.}\
  \bibnamefont {Cook}}\ and\ \bibinfo {author} {\bibfnamefont {Joel~E.}\
  \bibnamefont {Moore}},\ }\bibfield  {title} {\enquote {\bibinfo {title}
  {Multiplicative topological phases},}\ }\href {\doibase
  10.1038/s42005-022-01022-x} {\bibfield  {journal} {\bibinfo  {journal}
  {Communications Physics}\ }\textbf {\bibinfo {volume} {5}},\ \bibinfo {pages}
  {262} (\bibinfo {year} {2022})}\BibitemShut {NoStop}%
\bibitem [{\citenamefont {Kitaev}(2001)}]{Kitaev_2001}%
  \BibitemOpen
  \bibfield  {author} {\bibinfo {author} {\bibfnamefont {A~Yu}\ \bibnamefont
  {Kitaev}},\ }\bibfield  {title} {\enquote {\bibinfo {title} {Unpaired
  majorana fermions in quantum wires},}\ }\href {\doibase
  10.1070/1063-7869/44/10s/s29} {\bibfield  {journal} {\bibinfo  {journal}
  {Physics-Uspekhi}\ }\textbf {\bibinfo {volume} {44}},\ \bibinfo {pages}
  {131--136} (\bibinfo {year} {2001})}\BibitemShut {NoStop}%
\bibitem [{\citenamefont {Pientka}\ \emph {et~al.}(2017)\citenamefont
  {Pientka}, \citenamefont {Keselman}, \citenamefont {Berg}, \citenamefont
  {Yacoby}, \citenamefont {Stern},\ and\ \citenamefont
  {Halperin}}]{2deg_halperin}%
  \BibitemOpen
  \bibfield  {author} {\bibinfo {author} {\bibfnamefont {Falko}\ \bibnamefont
  {Pientka}}, \bibinfo {author} {\bibfnamefont {Anna}\ \bibnamefont
  {Keselman}}, \bibinfo {author} {\bibfnamefont {Erez}\ \bibnamefont {Berg}},
  \bibinfo {author} {\bibfnamefont {Amir}\ \bibnamefont {Yacoby}}, \bibinfo
  {author} {\bibfnamefont {Ady}\ \bibnamefont {Stern}}, \ and\ \bibinfo
  {author} {\bibfnamefont {Bertrand~I.}\ \bibnamefont {Halperin}},\ }\bibfield
  {title} {\enquote {\bibinfo {title} {Topological superconductivity in a
  planar josephson junction},}\ }\href {\doibase 10.1103/PhysRevX.7.021032}
  {\bibfield  {journal} {\bibinfo  {journal} {Phys. Rev. X}\ }\textbf {\bibinfo
  {volume} {7}},\ \bibinfo {pages} {021032} (\bibinfo {year}
  {2017})}\BibitemShut {NoStop}%
\bibitem [{\citenamefont {Hell}\ \emph {et~al.}(2017)\citenamefont {Hell},
  \citenamefont {Leijnse},\ and\ \citenamefont {Flensberg}}]{2deg_karsten}%
  \BibitemOpen
  \bibfield  {author} {\bibinfo {author} {\bibfnamefont {Michael}\ \bibnamefont
  {Hell}}, \bibinfo {author} {\bibfnamefont {Martin}\ \bibnamefont {Leijnse}},
  \ and\ \bibinfo {author} {\bibfnamefont {Karsten}\ \bibnamefont
  {Flensberg}},\ }\bibfield  {title} {\enquote {\bibinfo {title}
  {Two-dimensional platform for networks of majorana bound states},}\ }\href
  {\doibase 10.1103/PhysRevLett.118.107701} {\bibfield  {journal} {\bibinfo
  {journal} {Phys. Rev. Lett.}\ }\textbf {\bibinfo {volume} {118}},\ \bibinfo
  {pages} {107701} (\bibinfo {year} {2017})}\BibitemShut {NoStop}%
\bibitem [{\citenamefont {Lutchyn}\ \emph {et~al.}(2018)\citenamefont
  {Lutchyn}, \citenamefont {Bakkers}, \citenamefont {Kouwenhoven},
  \citenamefont {Krogstrup}, \citenamefont {Marcus},\ and\ \citenamefont
  {Oreg}}]{Lutchyn2018}%
  \BibitemOpen
  \bibfield  {author} {\bibinfo {author} {\bibfnamefont {R.~M.}\ \bibnamefont
  {Lutchyn}}, \bibinfo {author} {\bibfnamefont {E.~P. A.~M.}\ \bibnamefont
  {Bakkers}}, \bibinfo {author} {\bibfnamefont {L.~P.}\ \bibnamefont
  {Kouwenhoven}}, \bibinfo {author} {\bibfnamefont {P.}~\bibnamefont
  {Krogstrup}}, \bibinfo {author} {\bibfnamefont {C.~M.}\ \bibnamefont
  {Marcus}}, \ and\ \bibinfo {author} {\bibfnamefont {Y.}~\bibnamefont
  {Oreg}},\ }\bibfield  {title} {\enquote {\bibinfo {title} {Majorana zero
  modes in superconductor--semiconductor heterostructures},}\ }\href@noop {}
  {\bibfield  {journal} {\bibinfo  {journal} {Nature Reviews Materials}\
  }\textbf {\bibinfo {volume} {3}},\ \bibinfo {pages} {52--68} (\bibinfo {year}
  {2018})}\BibitemShut {NoStop}%
\bibitem [{\citenamefont {Mourik}\ \emph {et~al.}(2012)\citenamefont {Mourik},
  \citenamefont {Zuo}, \citenamefont {Frolov}, \citenamefont {Plissard},
  \citenamefont {Bakkers},\ and\ \citenamefont
  {Kouwenhoven}}]{science_majoranas}%
  \BibitemOpen
  \bibfield  {author} {\bibinfo {author} {\bibfnamefont {V.}~\bibnamefont
  {Mourik}}, \bibinfo {author} {\bibfnamefont {K.}~\bibnamefont {Zuo}},
  \bibinfo {author} {\bibfnamefont {S.~M.}\ \bibnamefont {Frolov}}, \bibinfo
  {author} {\bibfnamefont {S.~R.}\ \bibnamefont {Plissard}}, \bibinfo {author}
  {\bibfnamefont {E.~P. A.~M.}\ \bibnamefont {Bakkers}}, \ and\ \bibinfo
  {author} {\bibfnamefont {L.~P.}\ \bibnamefont {Kouwenhoven}},\ }\bibfield
  {title} {\enquote {\bibinfo {title} {Signatures of majorana fermions in
  hybrid superconductor-semiconductor nanowire devices},}\ }\href {\doibase
  10.1126/science.1222360} {\bibfield  {journal} {\bibinfo  {journal}
  {Science}\ }\textbf {\bibinfo {volume} {336}},\ \bibinfo {pages} {1003--1007}
  (\bibinfo {year} {2012})}\BibitemShut {NoStop}%
\bibitem [{\citenamefont {Nadj-Perge}\ \emph {et~al.}(2013)\citenamefont
  {Nadj-Perge}, \citenamefont {Drozdov}, \citenamefont {Bernevig},\ and\
  \citenamefont {Yazdani}}]{ali_proposal}%
  \BibitemOpen
  \bibfield  {author} {\bibinfo {author} {\bibfnamefont {S.}~\bibnamefont
  {Nadj-Perge}}, \bibinfo {author} {\bibfnamefont {I.~K.}\ \bibnamefont
  {Drozdov}}, \bibinfo {author} {\bibfnamefont {B.~A.}\ \bibnamefont
  {Bernevig}}, \ and\ \bibinfo {author} {\bibfnamefont {Ali}\ \bibnamefont
  {Yazdani}},\ }\bibfield  {title} {\enquote {\bibinfo {title} {Proposal for
  realizing majorana fermions in chains of magnetic atoms on a
  superconductor},}\ }\href {\doibase 10.1103/PhysRevB.88.020407} {\bibfield
  {journal} {\bibinfo  {journal} {Phys. Rev. B}\ }\textbf {\bibinfo {volume}
  {88}},\ \bibinfo {pages} {020407} (\bibinfo {year} {2013})}\BibitemShut
  {NoStop}%
\bibitem [{\citenamefont {Alicea}(2010)}]{jason_alicea_proposal}%
  \BibitemOpen
  \bibfield  {author} {\bibinfo {author} {\bibfnamefont {Jason}\ \bibnamefont
  {Alicea}},\ }\bibfield  {title} {\enquote {\bibinfo {title} {Majorana
  fermions in a tunable semiconductor device},}\ }\href {\doibase
  10.1103/PhysRevB.81.125318} {\bibfield  {journal} {\bibinfo  {journal} {Phys.
  Rev. B}\ }\textbf {\bibinfo {volume} {81}},\ \bibinfo {pages} {125318}
  (\bibinfo {year} {2010})}\BibitemShut {NoStop}%
\bibitem [{\citenamefont {Alicea}\ \emph {et~al.}(2011)\citenamefont {Alicea},
  \citenamefont {Oreg}, \citenamefont {Refael}, \citenamefont {von Oppen},\
  and\ \citenamefont {Fisher}}]{Alicea2011}%
  \BibitemOpen
  \bibfield  {author} {\bibinfo {author} {\bibfnamefont {Jason}\ \bibnamefont
  {Alicea}}, \bibinfo {author} {\bibfnamefont {Yuval}\ \bibnamefont {Oreg}},
  \bibinfo {author} {\bibfnamefont {Gil}\ \bibnamefont {Refael}}, \bibinfo
  {author} {\bibfnamefont {Felix}\ \bibnamefont {von Oppen}}, \ and\ \bibinfo
  {author} {\bibfnamefont {Matthew P.~A.}\ \bibnamefont {Fisher}},\ }\bibfield
  {title} {\enquote {\bibinfo {title} {Non-abelian statistics and topological
  quantum information processing in 1d wire networks},}\ }\href {\doibase
  10.1038/nphys1915} {\bibfield  {journal} {\bibinfo  {journal} {Nature
  Physics}\ }\textbf {\bibinfo {volume} {7}},\ \bibinfo {pages} {412--417}
  (\bibinfo {year} {2011})}\BibitemShut {NoStop}%
\bibitem [{\citenamefont {Alicea}(2012)}]{Alicea_2012}%
  \BibitemOpen
  \bibfield  {author} {\bibinfo {author} {\bibfnamefont {Jason}\ \bibnamefont
  {Alicea}},\ }\bibfield  {title} {\enquote {\bibinfo {title} {New directions
  in the pursuit of majorana fermions in solid state systems},}\ }\href
  {\doibase 10.1088/0034-4885/75/7/076501} {\bibfield  {journal} {\bibinfo
  {journal} {Reports on Progress in Physics}\ }\textbf {\bibinfo {volume}
  {75}},\ \bibinfo {pages} {076501} (\bibinfo {year} {2012})}\BibitemShut
  {NoStop}%
\bibitem [{\citenamefont {Beenakker}(2013)}]{Beenakker_review}%
  \BibitemOpen
  \bibfield  {author} {\bibinfo {author} {\bibfnamefont {C.W.J.}\ \bibnamefont
  {Beenakker}},\ }\bibfield  {title} {\enquote {\bibinfo {title} {Search for
  majorana fermions in superconductors},}\ }\href {\doibase
  10.1146/annurev-conmatphys-030212-184337} {\bibfield  {journal} {\bibinfo
  {journal} {Annual Review of Condensed Matter Physics}\ }\textbf {\bibinfo
  {volume} {4}},\ \bibinfo {pages} {113--136} (\bibinfo {year} {2013})},\
  \Eprint
  {http://arxiv.org/abs/https://doi.org/10.1146/annurev-conmatphys-030212-184337}
  {https://doi.org/10.1146/annurev-conmatphys-030212-184337} \BibitemShut
  {NoStop}%
\bibitem [{\citenamefont {Nayak}\ \emph {et~al.}(2008)\citenamefont {Nayak},
  \citenamefont {Simon}, \citenamefont {Stern}, \citenamefont {Freedman},\ and\
  \citenamefont {Das~Sarma}}]{sankar_das_sarma_review}%
  \BibitemOpen
  \bibfield  {author} {\bibinfo {author} {\bibfnamefont {Chetan}\ \bibnamefont
  {Nayak}}, \bibinfo {author} {\bibfnamefont {Steven~H.}\ \bibnamefont
  {Simon}}, \bibinfo {author} {\bibfnamefont {Ady}\ \bibnamefont {Stern}},
  \bibinfo {author} {\bibfnamefont {Michael}\ \bibnamefont {Freedman}}, \ and\
  \bibinfo {author} {\bibfnamefont {Sankar}\ \bibnamefont {Das~Sarma}},\
  }\bibfield  {title} {\enquote {\bibinfo {title} {Non-abelian anyons and
  topological quantum computation},}\ }\href {\doibase
  10.1103/RevModPhys.80.1083} {\bibfield  {journal} {\bibinfo  {journal} {Rev.
  Mod. Phys.}\ }\textbf {\bibinfo {volume} {80}},\ \bibinfo {pages}
  {1083--1159} (\bibinfo {year} {2008})}\BibitemShut {NoStop}%
\bibitem [{\citenamefont {Sau}\ \emph {et~al.}(2010)\citenamefont {Sau},
  \citenamefont {Lutchyn}, \citenamefont {Tewari},\ and\ \citenamefont
  {Das~Sarma}}]{j_sau_2010}%
  \BibitemOpen
  \bibfield  {author} {\bibinfo {author} {\bibfnamefont {Jay~D.}\ \bibnamefont
  {Sau}}, \bibinfo {author} {\bibfnamefont {Roman~M.}\ \bibnamefont {Lutchyn}},
  \bibinfo {author} {\bibfnamefont {Sumanta}\ \bibnamefont {Tewari}}, \ and\
  \bibinfo {author} {\bibfnamefont {S.}~\bibnamefont {Das~Sarma}},\ }\bibfield
  {title} {\enquote {\bibinfo {title} {Generic new platform for topological
  quantum computation using semiconductor heterostructures},}\ }\href {\doibase
  10.1103/PhysRevLett.104.040502} {\bibfield  {journal} {\bibinfo  {journal}
  {Phys. Rev. Lett.}\ }\textbf {\bibinfo {volume} {104}},\ \bibinfo {pages}
  {040502} (\bibinfo {year} {2010})}\BibitemShut {NoStop}%
\bibitem [{\citenamefont {Chiu}\ \emph {et~al.}(2016)\citenamefont {Chiu},
  \citenamefont {Teo}, \citenamefont {Schnyder},\ and\ \citenamefont
  {Ryu}}]{chiu2016classification}%
  \BibitemOpen
  \bibfield  {author} {\bibinfo {author} {\bibfnamefont {Ching-Kai}\
  \bibnamefont {Chiu}}, \bibinfo {author} {\bibfnamefont {Jeffrey~CY}\
  \bibnamefont {Teo}}, \bibinfo {author} {\bibfnamefont {Andreas~P}\
  \bibnamefont {Schnyder}}, \ and\ \bibinfo {author} {\bibfnamefont {Shinsei}\
  \bibnamefont {Ryu}},\ }\bibfield  {title} {\enquote {\bibinfo {title}
  {Classification of topological quantum matter with symmetries},}\ }\href@noop
  {} {\bibfield  {journal} {\bibinfo  {journal} {Reviews of Modern Physics}\
  }\textbf {\bibinfo {volume} {88}},\ \bibinfo {pages} {035005} (\bibinfo
  {year} {2016})}\BibitemShut {NoStop}%
\bibitem [{\citenamefont {Groth}\ \emph {et~al.}(2014)\citenamefont {Groth},
  \citenamefont {Wimmer}, \citenamefont {Akhmerov},\ and\ \citenamefont
  {Waintal}}]{Groth_2014}%
  \BibitemOpen
  \bibfield  {author} {\bibinfo {author} {\bibfnamefont {Christoph~W}\
  \bibnamefont {Groth}}, \bibinfo {author} {\bibfnamefont {Michael}\
  \bibnamefont {Wimmer}}, \bibinfo {author} {\bibfnamefont {Anton~R}\
  \bibnamefont {Akhmerov}}, \ and\ \bibinfo {author} {\bibfnamefont {Xavier}\
  \bibnamefont {Waintal}},\ }\bibfield  {title} {\enquote {\bibinfo {title}
  {Kwant: a software package for quantum transport},}\ }\href {\doibase
  10.1088/1367-2630/16/6/063065} {\bibfield  {journal} {\bibinfo  {journal}
  {New Journal of Physics}\ }\textbf {\bibinfo {volume} {16}},\ \bibinfo
  {pages} {063065} (\bibinfo {year} {2014})}\BibitemShut {NoStop}%
\bibitem [{\citenamefont {Varjas}\ \emph {et~al.}(2018)\citenamefont {Varjas},
  \citenamefont {Rosdahl},\ and\ \citenamefont {Akhmerov}}]{Varjas_2018}%
  \BibitemOpen
  \bibfield  {author} {\bibinfo {author} {\bibfnamefont {D{\'{a}}niel}\
  \bibnamefont {Varjas}}, \bibinfo {author} {\bibfnamefont {T{\'{o}}mas~Ö}\
  \bibnamefont {Rosdahl}}, \ and\ \bibinfo {author} {\bibfnamefont {Anton~R}\
  \bibnamefont {Akhmerov}},\ }\bibfield  {title} {\enquote {\bibinfo {title}
  {Qsymm: algorithmic symmetry finding and symmetric hamiltonian generation},}\
  }\href {\doibase 10.1088/1367-2630/aadf67} {\bibfield  {journal} {\bibinfo
  {journal} {New Journal of Physics}\ }\textbf {\bibinfo {volume} {20}},\
  \bibinfo {pages} {093026} (\bibinfo {year} {2018})}\BibitemShut {NoStop}%
\bibitem [{\citenamefont {Alexandradinata}\ \emph {et~al.}(2014)\citenamefont
  {Alexandradinata}, \citenamefont {Dai},\ and\ \citenamefont
  {Bernevig}}]{wilson_loop_paper}%
  \BibitemOpen
  \bibfield  {author} {\bibinfo {author} {\bibfnamefont {A.}~\bibnamefont
  {Alexandradinata}}, \bibinfo {author} {\bibfnamefont {Xi}~\bibnamefont
  {Dai}}, \ and\ \bibinfo {author} {\bibfnamefont {B.~Andrei}\ \bibnamefont
  {Bernevig}},\ }\bibfield  {title} {\enquote {\bibinfo {title} {Wilson-loop
  characterization of inversion-symmetric topological insulators},}\ }\href
  {\doibase 10.1103/PhysRevB.89.155114} {\bibfield  {journal} {\bibinfo
  {journal} {Phys. Rev. B}\ }\textbf {\bibinfo {volume} {89}},\ \bibinfo
  {pages} {155114} (\bibinfo {year} {2014})}\BibitemShut {NoStop}%
\bibitem [{\citenamefont {Leumer}\ \emph {et~al.}(2020)\citenamefont {Leumer},
  \citenamefont {Marganska}, \citenamefont {Muralidharan},\ and\ \citenamefont
  {Grifoni}}]{Leumer_2020}%
  \BibitemOpen
  \bibfield  {author} {\bibinfo {author} {\bibfnamefont {Nico}\ \bibnamefont
  {Leumer}}, \bibinfo {author} {\bibfnamefont {Magdalena}\ \bibnamefont
  {Marganska}}, \bibinfo {author} {\bibfnamefont {Bhaskaran}\ \bibnamefont
  {Muralidharan}}, \ and\ \bibinfo {author} {\bibfnamefont {Milena}\
  \bibnamefont {Grifoni}},\ }\bibfield  {title} {\enquote {\bibinfo {title}
  {Exact eigenvectors and eigenvalues of the finite kitaev chain and its
  topological properties},}\ }\href {\doibase 10.1088/1361-648x/ab8bf9}
  {\bibfield  {journal} {\bibinfo  {journal} {Journal of Physics: Condensed
  Matter}\ }\textbf {\bibinfo {volume} {32}},\ \bibinfo {pages} {445502}
  (\bibinfo {year} {2020})}\BibitemShut {NoStop}%
\bibitem [{\citenamefont {Cook}\ and\ \citenamefont
  {Nielsen}(2022)}]{cook2022}%
  \BibitemOpen
  \bibfield  {author} {\bibinfo {author} {\bibfnamefont {A.~M.}\ \bibnamefont
  {Cook}}\ and\ \bibinfo {author} {\bibfnamefont {A.~E.~B.}\ \bibnamefont
  {Nielsen}},\ }\bibfield  {title} {\enquote {\bibinfo {title} {Finite-size
  topology},}\ }\href@noop {} {\bibfield  {journal} {\bibinfo  {journal}
  {submitted}\ } (\bibinfo {year} {2022})}\BibitemShut {NoStop}%
\bibitem [{\citenamefont {Leumer}(2021)}]{leumer2021spectral}%
  \BibitemOpen
  \bibfield  {author} {\bibinfo {author} {\bibfnamefont {Nico~Gerhard}\
  \bibnamefont {Leumer}},\ }\emph {\bibinfo {title} {Spectral and transport
  signatures of 1d topological superconductors of finite size in the sub-and
  supra-gap regime: An analytical study}},\ \href@noop {} {Ph.D. thesis}
  (\bibinfo {year} {2021})\BibitemShut {NoStop}%
\bibitem [{\citenamefont {Schindler}\ \emph {et~al.}(2018)\citenamefont
  {Schindler}, \citenamefont {Cook}, \citenamefont {Vergniory}, \citenamefont
  {Wang}, \citenamefont {Parkin}, \citenamefont {Bernevig},\ and\ \citenamefont
  {Neupert}}]{schindler2018higher}%
  \BibitemOpen
  \bibfield  {author} {\bibinfo {author} {\bibfnamefont {Frank}\ \bibnamefont
  {Schindler}}, \bibinfo {author} {\bibfnamefont {Ashley~M}\ \bibnamefont
  {Cook}}, \bibinfo {author} {\bibfnamefont {Maia~G}\ \bibnamefont
  {Vergniory}}, \bibinfo {author} {\bibfnamefont {Zhijun}\ \bibnamefont
  {Wang}}, \bibinfo {author} {\bibfnamefont {Stuart~SP}\ \bibnamefont
  {Parkin}}, \bibinfo {author} {\bibfnamefont {B~Andrei}\ \bibnamefont
  {Bernevig}}, \ and\ \bibinfo {author} {\bibfnamefont {Titus}\ \bibnamefont
  {Neupert}},\ }\bibfield  {title} {\enquote {\bibinfo {title} {Higher-order
  topological insulators},}\ }\href@noop {} {\bibfield  {journal} {\bibinfo
  {journal} {Science advances}\ }\textbf {\bibinfo {volume} {4}},\ \bibinfo
  {pages} {eaat0346} (\bibinfo {year} {2018})}\BibitemShut {NoStop}%
\bibitem [{\citenamefont {Zhang}\ \emph {et~al.}(2019)\citenamefont {Zhang},
  \citenamefont {Wang},\ and\ \citenamefont {Song}}]{zhang2019majorana}%
  \BibitemOpen
  \bibfield  {author} {\bibinfo {author} {\bibfnamefont {KL}~\bibnamefont
  {Zhang}}, \bibinfo {author} {\bibfnamefont {Peng}\ \bibnamefont {Wang}}, \
  and\ \bibinfo {author} {\bibfnamefont {Zhi}\ \bibnamefont {Song}},\
  }\bibfield  {title} {\enquote {\bibinfo {title} {Majorana flat band edge
  modes of topological gapless phase in 2d kitaev square lattice},}\
  }\href@noop {} {\bibfield  {journal} {\bibinfo  {journal} {Scientific
  reports}\ }\textbf {\bibinfo {volume} {9}},\ \bibinfo {pages} {1--9}
  (\bibinfo {year} {2019})}\BibitemShut {NoStop}%
\bibitem [{\citenamefont {King-Smith}\ and\ \citenamefont
  {Vanderbilt}(1993)}]{theory_polarization_vanderbilt}%
  \BibitemOpen
  \bibfield  {author} {\bibinfo {author} {\bibfnamefont {R.~D.}\ \bibnamefont
  {King-Smith}}\ and\ \bibinfo {author} {\bibfnamefont {David}\ \bibnamefont
  {Vanderbilt}},\ }\bibfield  {title} {\enquote {\bibinfo {title} {Theory of
  polarization of crystalline solids},}\ }\href {\doibase
  10.1103/PhysRevB.47.1651} {\bibfield  {journal} {\bibinfo  {journal} {Phys.
  Rev. B}\ }\textbf {\bibinfo {volume} {47}},\ \bibinfo {pages} {1651--1654}
  (\bibinfo {year} {1993})}\BibitemShut {NoStop}%
\bibitem [{\citenamefont {Vanderbilt}\ and\ \citenamefont
  {King-Smith}(1993)}]{electric_pol_vanderbilt}%
  \BibitemOpen
  \bibfield  {author} {\bibinfo {author} {\bibfnamefont {David}\ \bibnamefont
  {Vanderbilt}}\ and\ \bibinfo {author} {\bibfnamefont {R.~D.}\ \bibnamefont
  {King-Smith}},\ }\bibfield  {title} {\enquote {\bibinfo {title} {Electric
  polarization as a bulk quantity and its relation to surface charge},}\ }\href
  {\doibase 10.1103/PhysRevB.48.4442} {\bibfield  {journal} {\bibinfo
  {journal} {Phys. Rev. B}\ }\textbf {\bibinfo {volume} {48}},\ \bibinfo
  {pages} {4442--4455} (\bibinfo {year} {1993})}\BibitemShut {NoStop}%
\bibitem [{\citenamefont {Alexandradinata}\ \emph {et~al.}(2016)\citenamefont
  {Alexandradinata}, \citenamefont {Wang},\ and\ \citenamefont
  {Bernevig}}]{alexandradinata2016gcohomology}%
  \BibitemOpen
  \bibfield  {author} {\bibinfo {author} {\bibfnamefont {A.}~\bibnamefont
  {Alexandradinata}}, \bibinfo {author} {\bibfnamefont {Zhijun}\ \bibnamefont
  {Wang}}, \ and\ \bibinfo {author} {\bibfnamefont {B.~Andrei}\ \bibnamefont
  {Bernevig}},\ }\bibfield  {title} {\enquote {\bibinfo {title} {Topological
  insulators from group cohomology},}\ }\href {\doibase
  10.1103/PhysRevX.6.021008} {\bibfield  {journal} {\bibinfo  {journal} {Phys.
  Rev. X}\ }\textbf {\bibinfo {volume} {6}},\ \bibinfo {pages} {021008}
  (\bibinfo {year} {2016})}\BibitemShut {NoStop}%
\bibitem [{\citenamefont {Asb{\'o}th}\ \emph {et~al.}(2016)\citenamefont
  {Asb{\'o}th}, \citenamefont {Oroszl{\'a}ny},\ and\ \citenamefont
  {P{\'a}lyi}}]{asboth2016short}%
  \BibitemOpen
  \bibfield  {author} {\bibinfo {author} {\bibfnamefont {J{\'a}nos~K}\
  \bibnamefont {Asb{\'o}th}}, \bibinfo {author} {\bibfnamefont
  {L{\'a}szl{\'o}}\ \bibnamefont {Oroszl{\'a}ny}}, \ and\ \bibinfo {author}
  {\bibfnamefont {Andr{\'a}s}\ \bibnamefont {P{\'a}lyi}},\ }\bibfield  {title}
  {\enquote {\bibinfo {title} {A short course on topological insulators},}\
  }\href@noop {} {\bibfield  {journal} {\bibinfo  {journal} {Lecture notes in
  physics}\ }\textbf {\bibinfo {volume} {919}},\ \bibinfo {pages} {166}
  (\bibinfo {year} {2016})}\BibitemShut {NoStop}%
\bibitem [{\citenamefont {Wieder}\ \emph {et~al.}(2018)\citenamefont {Wieder},
  \citenamefont {Bradlyn}, \citenamefont {Wang}, \citenamefont {Cano},
  \citenamefont {Kim}, \citenamefont {Kim}, \citenamefont {Rappe},
  \citenamefont {Kane},\ and\ \citenamefont {Bernevig}}]{wieder2018wallpaper}%
  \BibitemOpen
  \bibfield  {author} {\bibinfo {author} {\bibfnamefont {Benjamin~J}\
  \bibnamefont {Wieder}}, \bibinfo {author} {\bibfnamefont {Barry}\
  \bibnamefont {Bradlyn}}, \bibinfo {author} {\bibfnamefont {Zhijun}\
  \bibnamefont {Wang}}, \bibinfo {author} {\bibfnamefont {Jennifer}\
  \bibnamefont {Cano}}, \bibinfo {author} {\bibfnamefont {Youngkuk}\
  \bibnamefont {Kim}}, \bibinfo {author} {\bibfnamefont {Hyeong-Seok~D}\
  \bibnamefont {Kim}}, \bibinfo {author} {\bibfnamefont {Andrew~M}\
  \bibnamefont {Rappe}}, \bibinfo {author} {\bibfnamefont {CL}~\bibnamefont
  {Kane}}, \ and\ \bibinfo {author} {\bibfnamefont {B~Andrei}\ \bibnamefont
  {Bernevig}},\ }\bibfield  {title} {\enquote {\bibinfo {title} {Wallpaper
  fermions and the nonsymmorphic dirac insulator},}\ }\href@noop {} {\bibfield
  {journal} {\bibinfo  {journal} {Science}\ }\textbf {\bibinfo {volume}
  {361}},\ \bibinfo {pages} {246--251} (\bibinfo {year} {2018})}\BibitemShut
  {NoStop}%
\end{thebibliography}%

\cleardoublepage


\makeatletter
\renewcommand{\theequation}{S\arabic{equation}}
\renewcommand{\thefigure}{S\arabic{figure}}
\renewcommand{\thesection}{S\arabic{section}}
\setcounter{equation}{0}
\setcounter{section}{0}
\onecolumngrid
\begin{center}
  \textbf{\large Supplemental material for ``Multiplicative Majorana zero-modes''}\\[.2cm]
  Adipta Pal$^{1,2}$, Joe H. Winter$^{1,2,3}$, and Ashley M. Cook$^{1,2,*}$\\[.1cm]
  {\itshape ${}^1$Max Planck Institute for Chemical Physics of Solids, Nöthnitzer Strasse 40, 01187 Dresden, Germany\\
  ${}^2$Max Planck Institute for the Physics of Complex Systems, Nöthnitzer Strasse 38, 01187 Dresden, Germany}\\
  ${}^3$SUPA, School of Physics and Astronomy, University of St.\ Andrews, North Haugh, St.\ Andrews KY16 9SS, UK \\
  ${}^*$Electronic address: cooka@pks.mpg.de\\
\end{center}

\section{Calculations for finite size KC and MKC:}\label{Sup_sec_A}
The KC Bloch Hamiltonian, $H_{KC}(k)=-(2t\cos k+\mu)\tau^z+2\Delta\sin k$ anti-commutes with the Chirality operator $\Pi=\tau^x$ so that we have eigen-states $\psi$ and $\tau^x\psi$ with energy $E$ and $-E$ respectively. In the position basis, the KC Hamiltonian may be expressed as,

\begin{equation}
    \begin{aligned}
    \mathbb{H}_{KC} = \frac{1}{2}[-\mu\tau^z\otimes \mathbb{I}-t\tau^z\otimes\mathbb{M}_+-i\Delta\tau^y\otimes\mathbb{M}_-],
    \end{aligned}
\end{equation}

where $\mathbb{M}_+=\delta_{i+1,j}+\delta_{i,j+1}$ and $\mathbb{M}_-=\delta_{i,j+1}-\delta_{i+1,j}$. We therefore perform a chiral decomposition of the KC Hamiltonian. With the similarity transformation, $S=\frac{1}{\sqrt{2}}(I-i\tau^y)\otimes \mathbb{I}$,

\begin{equation}
    \begin{aligned}
    \tilde{\mathbb{H}}_{KC} = \mathbb{S}\mathbb{H}_{KC}\mathbb{S}^\dagger = \tau^+\tilde{\mathbb{H}}_{KC,R}+\tau^-\tilde{\mathbb{H}}_{KC,L},
    \end{aligned}
\end{equation}

where $\tau^{\pm} = \frac{1}{2}(\tau^x\pm i\tau^y)$ and

\begin{equation}
\begin{aligned}
\tilde{\mathbb{H}}_{KC,R} =& \frac{1}{2}(-\mu\mathbb{I}-t\mathbb{M}_+-\Delta\mathbb{M}_-),\\
\tilde{\mathbb{H}}_{KC,L} =& \frac{1}{2}(-\mu\mathbb{I}-t\mathbb{M}_++\Delta\mathbb{M}_-).
\end{aligned}
\end{equation}
Alternatively, one may perform the chiral decomposition via the transformation, $k\rightarrow iq$. This corresponds to localization, so that we search for the eigenvectors $\Phi$ of,
\begin{equation}
H_{KC}(iq) = -(2t\cosh q+\mu)\tau^z+2i\Delta\sinh q\tau^y,
\end{equation}
so that $H_{KC}(iq)\Phi=0$ which should lead us to the expressions for the Majorana zero modes at the edges. Since $H_{KC}(iq)$ is of the form $\mathbf{d}(iq) \cdot\boldsymbol{\sigma}$, the condition for zero eigenvalues is $|\mathbf{d}(iq)|=0$, which gives rise to two conditions,
\begin{equation}
2t\cosh q+\mu = \pm 2\Delta\sinh q.
\end{equation}
Substituting these conditions into $H_{KC}(iq)$, we want null vectors of $\tau^z\mp i\tau^y$, which are given as
$$
\Phi\sim
\begin{pmatrix}
1\\
\pm 1
\end{pmatrix}
.
$$
The function form for the zero modes requires, however that we solve for $e^{-q}$ which are a natural conversion from the plane waves to a localized wavefunction, $e^{ikx}\rightarrow e^{-qx}$. We illustrate for one of the conditions,
$$
\begin{aligned}
&2t\cosh q+\mu-2\Delta\sinh q = 0,\\
\implies & (t+\Delta)e^{-2q}+\mu e^{-q}+(t-\Delta)=0,\\
\implies & e^{-q_{\pm}} = \frac{-\mu\pm\sqrt{\mu^2-4(t^2-\Delta^2)}}{2(t+\Delta)}.
\end{aligned}
$$
This condition is equivalent to solving for the zero energy eigenfunction for $\tilde{\mathbb{H}}_{KC,L}$ with the ansatz, $s^j\sim (e^{-q})^j$
For the two-band Kitaev chain,the functional form of the Majorana zero energy states then must be of the form,

\begin{equation}
\begin{aligned}
\Psi(j) = \alpha s_+^j+\beta s_-^j,
\end{aligned}
\end{equation}

where $s_\pm = e^{-q_\pm}$. A finite chain with only $N$ sites implies that the boundary conditions $\Psi(0)=0=\Psi(N+1)$ must hold. This means,

\begin{equation}
\begin{aligned}
&\alpha+\beta = 0,\\
&\alpha s_+^{N+1}+\beta s_-^{N+1} = 0.
\end{aligned}
\end{equation}

It is not possible to satisfy both the conditions unless the wave function is oscillatory, which implies $|\mu|<2\sqrt{t^2+\Delta^2}$, which leads to the following equation,

\begin{equation}
\begin{aligned}
&\alpha \bigg{(}s_+^{N+1}-s_-^{N+1}\bigg{)}=0,\\
\implies & \bigg{(}-\frac{\mu}{2(t+\Delta)}+i\frac{\sqrt{4(t^2-\Delta^2)-\mu^2}}{2(t+\Delta)}\bigg{)}^{N+1}\\
&-\bigg{(}-\frac{\mu}{2(t+\Delta)}-i\frac{\sqrt{4(t^2-\Delta^2)-\mu^2}}{2(t+\Delta)}\bigg{)}^{N+1} = 0,\\
\implies & R^{N+1}(e^{i(N+1)\theta}-e^{-i(N+1)\theta}) = 0,\\
\implies & \sin((N+1)\theta) = 0,\\
\implies & \cos\theta = \cos\frac{n\pi}{N+1},
\end{aligned}
\end{equation}

where, $R = \sqrt{\frac{t-\Delta}{t+\Delta}}$ and $\cos\theta = \frac{\mu}{2\sqrt{t^2-\Delta^2}}$. Then, one can have oscillatory zero energy Majorana modes only at,

\begin{equation}
\begin{aligned}
\mu = 2\sqrt{t^2-\Delta^2}\cos\frac{n\pi}{N+1}, \quad (n=1,...,N).
\end{aligned}
\end{equation}

\subsection{MKC parallel zero energy modes:}\label{Sup_subsec_A1}
One can similarly work out the null vectors and zero mode functional form for the MKC parallel system,
\begin{equation}
\begin{split}
H_{MKC,||}(k) =& [-(2t_1\cos k+\mu_1)\tau^z+2\Delta_1\sin k \tau^y]\otimes [(2t_2\cos k+\mu_2)\sigma^z+2\Delta_2\sin k\sigma^y],\\
=& \mathbf{d}_1(k)\cdot\boldsymbol{\tau}\otimes \mathbf{d}_2(k)\cdot \boldsymbol{\sigma}\cdot
\end{split}
\end{equation}
Carrying out the transformation for the localization, $k\rightarrow iq$,
\begin{equation}
H_{MKC,||}(iq) = [-(2t_1\cosh q+\mu_1)\tau^z+2i\Delta_1\sinh q \tau^y]\otimes [(2t_2\cosh q+\mu_2)\sigma^z+2i\Delta_2\sinh q\sigma^y]
\end{equation}
we will have zero eigenvalues if we have $|\mathbf{d}_1(iq)|\times |\mathbf{d}_2(iq)|=0$, as evident from the tensor product structure. We then have the conditions,
\begin{equation}
((2t_1\cosh q+\mu_1)^2-4\Delta_1^2\sinh^2 q)((2t_2\cosh q+\mu_2)^2-4\Delta_2^2\sinh^2 q)=0.
\end{equation}
From the four conditions due to different sign combinations, it is easy to infer that we get the following eigenvectors,
\begin{equation}
\Phi=\frac{1}{2}
\begin{pmatrix}
1\\
\pm 1
\end{pmatrix}
\otimes
\begin{pmatrix}
1\\
\pm 1
\end{pmatrix}
.
\end{equation}
The problem with this approach is that for our composite system, there is another possibility for the eigenvectors, namely the Bell states which also conserve the respective parities of the full system. It is therefore better to consider consequences of the four constraints on the two component Hamiltonians, $H_{||,1}$ and $H_{||,2}$. The component Bloch Hamiltonians,
\begin{equation}
\begin{split}
\mc{H}_{||,1}(k) =& -[2(\mu_1t_2+\mu_2t_1)\cos k+2(t_1t_2+\Delta_1\Delta_2)\cos 2k+\mu_1\mu_2+2t_1t_2-2\Delta_1\Delta_2]\sigma^z\\
&+[2(\mu_2\Delta_1+\mu_1\Delta_2)\sin k+2(t_2\Delta_1+t_1\Delta_2)\sin 2k]\sigma^y=\mathbf{d}_1(k)\cdot\boldsymbol{\sigma},
\end{split}
\label{Hpll1_Bloch}
\end{equation}
\begin{equation}
\begin{split}
\mc{H}_{||,2}(k) =& -[2(\mu_1t_2+\mu_2t_1)\cos k+2(t_1t_2-\Delta_1\Delta_2)\cos 2k+\mu_1\mu_2+2t_1t_2+2\Delta_1\Delta_2]\sigma^z\\
&+[2(\mu_2\Delta_1-\mu_1\Delta_2)\sin k+2(t_2\Delta_1-t_1\Delta_2)\sin 2k]\sigma^y=\mathbf{d}_2(k)\cdot\boldsymbol{\sigma}.
\end{split}
\label{Hpll2_Bloch}
\end{equation}
After localization, $k\rightarrow iq$ the condition for null eigenvalues yield,
\begin{subequations}
\begin{equation}
\begin{split}
&[2(\mu_1t_2+\mu_2t_1)\cosh q+2(t_1t_2+\Delta_1\Delta_2)\cosh 2q+\mu_1\mu_2+2t_1t_2-2\Delta_1\Delta_2]\\
&=\pm[2(\mu_2\Delta_1+\mu_1\Delta_2)\sinh q+2(t_2\Delta_1+t_1\Delta_2)\sinh 2q],\\
\implies & [(2t_1\cosh q+\mu_1)\mp 2\Delta_1\sinh q][(2t_2\cosh q+\mu_2)\mp 2\Delta_2\sinh q]=0,
\end{split}
\end{equation}
\begin{equation}
\begin{split}
&[2(\mu_1t_2+\mu_2t_1)\cosh q+2(t_1t_2-\Delta_1\Delta_2)\cosh 2q+\mu_1\mu_2+2t_1t_2+2\Delta_1\Delta_2]\\
&=\pm[2(\mu_2\Delta_1-\mu_1\Delta_2)\sinh q+2(t_2\Delta_1-t_1\Delta_2)\sinh 2q],\\
\implies & [(2t_1\cosh q+\mu_1)\mp 2\Delta_1\sinh q][(2t_2\cosh q+\mu_2)\pm 2\Delta_2\sinh q]=0.
\end{split}
\end{equation}
\end{subequations}
After localization, $k\rightarrow iq$, the respective component Bloch Hamiltonians are,
\begin{subequations}
\begin{equation}
\begin{split}
\mc{H}_{||,1}(iq) =& -[(2t_1\cosh q+\mu_1)(2t_2\cosh q+\mu_2)+4\Delta_1\Delta_2\sinh^2q]\sigma^z\\
&+[2\Delta_1\sinh q(2t_2\cosh q+\mu_2)+2\Delta_2\sinh q(2t_1\cosh q+\mu_1)]\sigma^y,
\end{split}
\end{equation}
\begin{equation}
\begin{split}
\mc{H}_{||,2}(iq) =&-[(2t_1\cosh q+\mu_1)(2t_2\cosh q+\mu_2)-4\Delta_1\Delta_2\sinh^2q]\sigma^z\\
&+[2\Delta_1\sinh q(2t_2\cosh q+\mu_2)-2\Delta_2\sinh q(2t_1\cosh q+\mu_1)]\sigma^y.
\end{split}
\end{equation}
\end{subequations}
Depending on whether the parents are topological or trivial, we have two cases. We show here for one of the conditions,
\begin{equation}
[(2t_1\cosh q+\mu_1)-2\Delta_1\sinh q][(2t_2\cosh q+\mu_2)-2\Delta_2\sinh q]=0.
\end{equation}

The other conditions follow similarly. The basis of the full system is $\tilde{c}_k=(\tilde{c}_{k,\uparrow},\tilde{c}_{k,\downarrow},\tilde{c}^\dagger_{-k,\uparrow},\tilde{c}^\dagger_{-k,\downarrow})^T$. We therefore combine the bases $\tilde{c}_{k,1}=(\tilde{c}_{k,\uparrow},\tilde{c}^\dagger_{-k,\downarrow})^T$ for $\mathcal{H}_{||,1}$ and $\tilde{c}_{k,2}=(\tilde{c}_{k,\downarrow},\tilde{c}^\dagger_{-k,\uparrow})^T$ for $\mathcal{H}_{||,2}$ and search for the null eigenvectors of the $4\times 4$ matrix derived from the full basis.
\begin{itemize}
\item \textit{Case 1}: If both the parents are topological, we have $(2t_1\cosh q+\mu_1)=2\Delta_1\sinh q$ and $(2t_2\cosh q+\mu_2)=2\Delta_2\sinh q$ in separate situations except if the parameters of both the parents are proportional to each other, i.e., $\big{|}\frac{\mu_1}{t_1}\big{|}=\big{|}\frac{\mu_2}{t_2}\big{|}$ and $\big{|}\frac{t_1}{\Delta_1}\big{|}=\big{|}\frac{t_2}{\Delta_2}\big{|}$. If such cases, say $\frac{\mu_1}{t_1}=\frac{\mu_2}{t_2}$ and $\frac{t_1}{\Delta_1}=\frac{t_2}{\Delta_2}$, we get $+$ signs on the right-hand side (rhs) for the first lines in Eqns.~(S16a) and (S16b) but also the $-$ sign for Eqn.~(S16b). Just substituting for $2t_1\cosh q+\mu_1$ and $2t_2\cosh q+\mu_2$ into Eqn.~(S17a) and (S17b) implies that the incidence of both $+$ and $-$ sign on the rhs of Eqn.~(S16a) is equivalent to getting $\mc{H}(iq)=0$. Therefore, our MZM eigenvectors in this case must be null eigenvectors of the matrix,
\begin{equation}
\begin{pmatrix}
d_{1,z}(iq) & 0 & 0 & d_{1,z}(iq)\\
0 & 0 & 0 & 0\\
0 & 0 & 0 & 0\\
-d_{1,z}(iq) & 0 & 0 & -d_{1,z}(iq)
\end{pmatrix}
,
\end{equation}
which are given as, $\{\frac{1}{\sqrt{2}}(\ket{00}-\ket{11}),\ket{10},\ket{01}\}$. If the parents, on the other hand, do not have such related parameters, we substitute the conditions $(2t_1\cosh q+\mu_1)=2\Delta_1\sinh q$ and $(2t_2\cosh q+\mu_2)=2\Delta_2\sinh q$ one by one, so that our MZM eigenvectors are eigenvectors of the matrix,
\begin{equation}
\begin{pmatrix}
d_{1,z}(iq) & 0 & 0 & d_{1,z}(iq)\\
0 & d_{2,z}(iq) & d_{2,z}(iq) & 0\\
0 & -d_{2,z}(iq) & -d_{2,z}(iq) & 0\\
-d_{1,z}(iq) & 0 & 0 & -d_{1,z}(iq)\\
\end{pmatrix}
,
\end{equation}
which are given as, $\{\frac{1}{\sqrt{2}}(\ket{00}-\ket{11}),\frac{1}{\sqrt{2}}(\ket{01}-\ket{10})\}$.
\item \textit{Case 2}: If one of the parents, say parent 1, is topological and parent 2 is trivial, we have only $2t_1\cosh q+\mu_1=2\Delta_1\sinh q$. The MMZM eigenvectors must then be null eigenvectors of the matrix,
\begin{equation}
\begin{pmatrix}
d_{1,z}(iq) & 0 & 0 & d_{1,z}(iq)\\
0 & d_{2,z}(iq) & d_{2,z}(iq) & 0\\
0 & -d_{2,z}(iq) & -d_{2,z}(iq) & 0\\
-d_{1,z}(iq) & 0 & 0 & -d_{1,z}(iq)\\
\end{pmatrix}
,
\end{equation}
which are given as, $\{\frac{1}{\sqrt{2}}(\ket{00}-\ket{11}),\frac{1}{\sqrt{2}}(\ket{01}-\ket{10})\}$.
\end{itemize}
For both cases, we have used $\ket{0}=(1,0)^T$ and $\ket{1}=(0,1)^T$.
as the Majorana eigenvectors. Then for the chosen signs, $(+,+)$ on the two rhs, we get the following non-zero null eigenvectors,
\begin{equation}
(+,+): \ket{\Psi} = \frac{1}{\sqrt{2}}(\ket{00}-\ket{11}),\frac{1}{\sqrt{2}}(\ket{01}-\ket{10}),
\end{equation}
where  These are the maximally-entangled Bell states. We have provided all of the Bell state combinations that arise due to the different chosen signs in the main text.\\
Now we look into the functional form for our Bell state MMZMs. We get four solutions for $e^{-q}$ from Eq. ~(S17),
\begin{equation}
e^{-q} = \frac{-\mu_1\pm\sqrt{\mu_1^2-4(t_1^2-\Delta_1^2)}}{2(t_1+\Delta_1)},\frac{-\mu_2\pm\sqrt{\mu_2^2-4(t_2^2-\Delta_2^2)}}{2(t_2+\Delta_2)}.
\end{equation}
The functional form for the Majorana edge modes, $\Psi(j)$, must then be a linear combination of $e^{-qj}$, where $j$ corresponds to the discrete lattice site index. This is, of course, subject to the following boundary conditions for a chain length, $N$,
\begin{equation}
\Psi(0)=0, \quad \Psi(N+1)=0, \quad \Psi(-1)=0, \quad \Psi(N+2)=0.
\end{equation}
The last two conditions arise because we have considered next-nearest neighbour interactions. They can be derived if one considers the recurrence relation arising out of the chiral decomposition of the Bloch Hamiltonian, as we have previously shown for the two-band Kitaev chain. Since the chain is finite in length, we must have complex roots of $e^{-q}$ for oscillating solutions, so that one may write,
$e^{-q}=\frac{-\mu_l\pm i\sqrt{4(t_l^2-\Delta_l^2)-\mu_l^2}}{2(t_l+\Delta_l)}=R_le^{\pm i\theta_l}$, $l\in \{1,2\}$. We then write down the following ansatz,
\begin{equation}
\Psi(j) = A_1R_1^je^{ij\theta_1}+A_2R_1^je^{-ij\theta_1}+B_1R_2^je^{ij\theta_2}+B_2R_2^je^{-ij\theta_2},
\label{genwavefunc}
\end{equation}
whereby the boundary conditions are given as follows,
\begin{equation}
\begin{split}
&A_1+A_2+B_1+B_2=0,\\
&
\begin{pmatrix}
R_1^{-1}\cos\theta_1-R_2^{-1}\cos\theta_2 & -R_1^{-1}\sin\theta_1 & -R_2^{-1}\sin\theta_2\\
R_1^{N+1}\cos((N+1)\theta_1)-R_2^{N+1}\cos((N+1)\theta_2) & R_1^{N+1}\sin((N+1)\theta_1) & R_2^{N+1}\sin((N+1)\theta_2)\\
R_1^{N+2}\cos((N+2)\theta_1)-R_2^{N+2}\cos((N+2)\theta_2) & R_1^{N+2}\sin((N+2)\theta_1) & R_2^{N+2}\sin((N+2)\theta_2)
\end{pmatrix}
\begin{pmatrix}
A_1+A_2\\
i(A_1-A_2)\\
i(B_1-B_2)
\end{pmatrix}
=0.
\end{split}
\end{equation}
Equating the determinant for the above matrix to zero provides the quantization condition,
\begin{equation}
\frac{R_1^{2(N+2)}+R_2^{2(N+2)}-2R_1^{N+2}R_2^{N+2}\cos(2(N+2)\theta_+)}{R_1^2+R_2^2-2R_1R_2\cos 2\theta_+}=\frac{R_1^{2(N+2)}+R_2^{2(N+2)}-2R_1^{N+2}R_2^{N+2}\cos(2(N+2)\theta_-)}{R_1^2+R_2^2-2R_1R_2\cos 2\theta_-},
\end{equation}
where $\theta_\pm = \frac{\theta_1\pm\theta_2}{2}$. We check if this quantization condition holds true by applying it to the case $\mu_1=\mu_2$, $t_1=-t_2$ and $\Delta_1=\Delta_2$. Here one can simply calculate that $R_2e^{i\theta_2}=-R_1e^{i\theta_1}$. Substituting into the above equation, we get,
\begin{equation}
1-(-1)^{N+2}\cos 2(N+2)\theta_1=(1-(-1)^{N+2})\cos^2\theta_1.
\end{equation}
We calculate separately for $N$ odd and $N$ even,
\begin{equation}
\begin{split}
&(N=odd)\quad 2\sin(N+3)\theta_1\sin(N+1)\theta_1=0,\implies \theta_1=\frac{n\pi}{N+1}, \frac{m\pi}{N+3}\quad n\in\{1,...,N+1\},m\in\{1,...,N+3\}\\
&(N=even)\quad 2\sin^2(N+2)\theta_1=0,\implies \theta_1 = \frac{n\pi}{N+2}\quad n\in\{1,...,N+2\}.
\end{split}
\label{unitmodcond}
\end{equation}
We see that the result derived via a schematic approach and the one from this quantization condition both indicate that we should observe two chains of even length based on which the Majorana points must be calculated.\\
We calculate the eigen-function for the above case using the physical basis provided by the schematic approach. Consider first that we have an MKC parallel system with $N=2L$ sites with $t_1=-t_2=-t$ and $\Delta_1=\Delta_2=\Delta$. For the above case, the MZM wavefunction for each of the two L-sized KC is given as,
\begin{equation}
\psi(j) = \bigg{[}\frac{t-\Delta}{t+\Delta}\bigg{]}^j\sin\bigg{(}\frac{n\pi j}{L+1}\bigg{)}, \quad n\in \{1,...,L\}.
\end{equation}
In terms of site index $l$ in the MKC parallel system, we have two eigen-functions, and taking into account that only nearest neighbour interactions are present, the wave-function of one of the constituent KC maintains a zero value in the site-index belonging to the other KC in the full MKC parallel lattice. Therefore, site index $l$ for the full MKC parallel lattice is related to the site index $j$ for the first constituent KC as $2j=l$ and for the second KC as $2j-1=l$.
\begin{subequations}
\begin{align}
\Psi_1(l) =&
\begin{cases}
&\big{[}\frac{t-\Delta}{t+\Delta}\big{]}^{\frac{l}{2}}\sin\big{(}\frac{n\pi l}{2L+2}\big{)} = \big{[}\frac{t-\Delta}{t+\Delta}\big{]}^{\frac{l}{2}}\sin\big{(}\frac{n\pi l}{N+2}\big{)},\quad  \text{if} \quad l=\text{even},\\
& 0,\quad \text{if} \quad l=\text{odd}.
\end{cases}\\
\Psi_2(l) =&
\begin{cases}
& 0,\quad \text{if} \quad l=\text{even},\\
&\big{[}\frac{t-\Delta}{t+\Delta}\big{]}^{\frac{l+1}{2}}\sin\big{(}\frac{n\pi (l+1)}{2L+2}\big{)} = \big{[}\frac{t-\Delta}{t+\Delta}\big{]}^{\frac{l+1}{2}}\sin\big{(}\frac{n\pi (l+1)}{N+2}\big{)},\quad  \text{if} \quad l=\text{odd}.
\end{cases}
\end{align}
\end{subequations}
In terms of the general wavefunction, Eqn.~\ref{genwavefunc} of the MZMs for the MKC parallel system, this can be represented as $A_1=B_1$ and $A_2=B_2$ with $A_1=-A_2$, for $\Psi_1(l)$,
\begin{equation}\label{MZMstatepi}
\Psi_1(l) \sim R_1^l(1+(-1)^l)e^{il\theta_1}-R_1^l(1+(-1)^l)e^{-il\theta_1},
\end{equation}
where $\theta_1=\frac{n\pi}{N+2}$, $\{n=1,...,N+2\}$ from the quantization condition we proved a while back. For $\Psi_2(l)$, a similar calculation can be done or it can simply be read as a shifted $\Psi_1$, so that $\Psi_2(l)=\Psi_1(l+1)$ so that it is shifted one step to the left. Similarly, for MKC parallel lattice of size $N=2L+1$, one can show from the schematic diagram that the following holds,
\begin{subequations}
\begin{align}
\Psi_1(l) =&
\begin{cases}
&\big{[}\frac{t-\Delta}{t+\Delta}\big{]}^{\frac{l}{2}}\sin\big{(}\frac{n\pi l}{2L+2}\big{)} = \big{[}\frac{t-\Delta}{t+\Delta}\big{]}^{\frac{l}{2}}\sin\big{(}\frac{n\pi l}{N+1}\big{)},\quad  \text{if} \quad l=\text{even},\\
& 0,\quad \text{if} \quad l=\text{odd}.
\end{cases}\\
\Psi_2(l) =&
\begin{cases}
& 0,\quad \text{if} \quad l=\text{even},\\
&\big{[}\frac{t-\Delta}{t+\Delta}\big{]}^{\frac{l+1}{2}}\sin\big{(}\frac{n\pi (l+1)}{2L+4}\big{)} = \big{[}\frac{t-\Delta}{t+\Delta}\big{]}^{\frac{l+1}{2}}\sin\big{(}\frac{n\pi (l+1)}{N+3}\big{)},\quad  \text{if} \quad l=\text{odd}.
\end{cases}
\end{align}
\end{subequations}
Again, we get the same expression as in Eqn.~\ref{MZMstatepi} for $\Psi_1(l)$ with $\theta_1=\frac{n\pi}{N+1}$ and $\Psi_2(l)=\Psi_1(l+1)$ but with $\theta_1=\frac{n\pi}{N+3}$. \\
Let us now extrapolate to the case when $R_1=R_2$. This is possible if $\frac{t_1}{\Delta_1}=\pm\frac{t_2}{\Delta_2}$. Consider now $R_2e^{i\theta_2}=R_1e^{i\delta}e^{i\theta_1}$, which implies $\theta_2-\theta_1=\delta$. This is a generalization of the previous case, where we had $\delta=\pi$. The quantization condition in this case is,
\begin{equation}
\frac{1+(e^{i\delta})^{2(N+2)}-2(e^{i\delta})^{N+2}\cos(2(N+1)\theta_1)}{1+e^{2i\delta}-2e^{i\delta}\cos2\theta_1}=\frac{1+(e^{i\delta})^{2(N+2)}-2(e^{i\delta})^{N+2}}{1+e^{2i\delta}-2e^{i\delta_1}}.
\label{deltaquant}
\end{equation}
Let us consider the next simplest case i.e., the phase difference is the third root of unity, denoted as $\delta=e^{i\frac{2\pi}{3}}=\omega$. Following our previous analysis, we must consider three kinds of system sizes, $N=3L,3L+1$ and $3L+2$,
\begin{itemize}
\item \textit{Case 1:}(\textbf{N=3L}) From the Eqn.~\ref{deltaquant}, we get,
\begin{equation}
\begin{split}
&\frac{1+\omega-2\omega\cos(2(N+2)\theta_1)}{1+\omega^2-2\omega\cos 2\theta_1}=\omega,\implies 4\omega^2\sin((N+1)\theta_1)\sin((N+3)\theta_1)=0,\\ \implies &\theta_1=\frac{n\pi}{N+1},\frac{m\pi}{N+3}, \quad n\in\{1,...,N\},m\in\{1,...,N+2\}.
\end{split}
\end{equation}
\item \textit{Case 2:}(\textbf{N=3L+1}) From Eqn.~\ref{deltaquant}, we get,
\begin{equation}
\begin{split}
&\frac{2-2\cos(2(N+2)\theta_1)}{1+\omega^2-2\omega\cos 2\theta_1}=0,\implies  4\sin^2((N+2)\theta_1)=0,\\
&\implies \theta_1=\frac{n\pi}{N+2}, \quad n\in\{1,...,N+1\}.
\end{split}
\end{equation}
\item \textit{Case 3:}(\textbf{N=3L+2}) From Eqn.~\ref{deltaquant}, we have,
\begin{equation}
\begin{split}
&\frac{1+\omega^2-2\omega\cos((N+2)\theta_1)}{1+\omega^2-2\omega\cos 2\theta_1}=1, \implies 4\omega\sin((N+1)\theta_1)\sin((N+3)\theta_1)=0,\\
\implies & \theta_1 = \frac{n\pi}{N+1},\frac{m\pi}{N+3},\quad n\in\{1,...,N\}, m\in\{1,...,N+2\}.
\end{split}
\end{equation}
\end{itemize}

\subsection{MKC perpendicular zero energy modes:}\label{Sup_subsec_A2}
Here, we compute the zero energy modes for the MKC perpendicular system given by the Hamiltonian,
\begin{equation}
\begin{split}
H_{MKC,\perp}(k_x,k_y) =& [-(2t_1\cos k_x+\mu_1)\tau^z+2\Delta_1\sin k_x \tau^y]\otimes [(2t_2\cos k_y+\mu_2)\sigma^z+2\Delta_2\sin k_y\sigma^y],\\
=& \mathbf{d}_1(k_x)\cdot\boldsymbol{\tau}\otimes \mathbf{d}_2(k_y)\cdot\boldsymbol{\sigma}.
\end{split}
\end{equation}
Localizing the Hamiltonian in the $\hat{x}$-direction via $k_x\rightarrow iq_x$,
\begin{equation}
H_{MKC,\perp}(iq_x,k_y) = [-(2t_1\cosh q_x+\mu_1)\tau^z+2i\Delta_1\sinh q_x \tau^y]\otimes [(2t_2\cos k_y+\mu_2)\sigma^z+2\Delta_2\sin k_y\sigma^y],
\end{equation}
the null condition must be,
\begin{equation}
\begin{split}
&((2t_1\cosh q_x+\mu_1)^2-4\Delta_1^2\sinh^2 q_x)((2t_2\cos k_y+\mu_2)^2+4\Delta_2^2\sin^2k_y)=0,\\
\implies & 2t_1\cosh q_x+\mu_1=\pm 2\Delta_1\sinh q_x.
\end{split}
\label{MKCperp_OBCx}
\end{equation}
As we show for the KC case, the edge states must be of the form, $\Psi(j,k_y)\sim (e^{-q_{x,+}j}-e^{-q_{x,-}j})e^{ik_yy}\Phi$, where $\Phi$ is derived from the null vectors of $H_{MKC,\perp}(iq_x,k_y)$. Denoting $M_2=2t_2\cos k_y+\mu_2$ and $R_2=2\Delta_2\sin k_y$, the two signs in Eqn. \eqref{MKCperp_OBCx} imply,
\begin{equation}
\Phi_\pm = \frac{1}{2\sqrt{M_2(M_2\pm \sqrt{M_2^2+R_2^2})}}
\begin{pmatrix}
1\\
\pm 1
\end{pmatrix}
\otimes
\begin{pmatrix}
M_2\pm\sqrt{M_2^2+R_2^2}\\
iR_2
\end{pmatrix}.
\end{equation}
We obtain two solutions for $e^{-q_x}$ from each of the two signs in Eqn. \eqref{MKCperp_OBCx},
\begin{equation}
+: e^{-q_x} = \frac{-\mu_1\pm\sqrt{\mu_1^2-4(t_1^2-\Delta_1^2)}}{2(t_1+\Delta_1)},\quad -: e^{-q_x} = \frac{-\mu_1\pm\sqrt{\mu_1^2-4(t_1^2-\Delta_1^2)}}{2(t_1-\Delta_1)}.
\end{equation}
Similarly, localizing only along the y-direction, $k_y\rightarrow iq_y$,
\begin{equation}
H_{MKC,\perp}(k_x,iq_y) = [-(2t_1\cos k_x+\mu_1)\tau^z+2\Delta_1\sin k_x \tau^y]\otimes [(2t_2\cosh q_y+\mu_2)\sigma^z+2i\Delta_2\sinh q_y\sigma^y],
\end{equation}
the null condition stands as,
\begin{equation}
\begin{split}
&((2t_1\cos k_x+\mu_1)^2+4\Delta_1^2\sin^2 k_x)((2t_2\cosh q_y+\mu_2)^2-4\Delta_2^2\sinh^2q_y)=0,\\
\implies & 2t_2\cosh q_y+\mu_2 = \pm 2\Delta_2\sinh q_y.
\end{split}
\label{MKCperp_OBCy}
\end{equation}
Similar to the previous case, the edge states must be of the form, $\Psi(k_x,j)\sim e^{ik_xx}(e^{-q_{y,+}j}-e^{-q_{y,-}j})\Phi$. Denoting $M_1=-(2t_1\cos k_x+\mu_1)$ and $R_1=2\Delta_1\sin k_x$, for the two signs of Eqn. \eqref{MKCperp_OBCy} respectively, the null vector is given as,
\begin{equation}
\Phi_\pm = \frac{1}{2\sqrt{M_1(M_1\pm\sqrt{M_1^2+R_1^2})}}
\begin{pmatrix}
M_1\pm\sqrt{M_1^2+R_1^2}\\
iR_1
\end{pmatrix}
\otimes
\begin{pmatrix}
1\\
\mp 1
\end{pmatrix}
.
\end{equation}
We obtain two solutions for $e^{-q_y}$ from each of the two signs in Eqn. \eqref{MKCperp_OBCy},
\begin{equation}
+: e^{-q_y} = \frac{-\mu_2\pm\sqrt{\mu_2^2-4(t_2^2-\Delta_2^2)}}{2(t_2+\Delta_2)},\quad -: e^{-q_y} = \frac{-\mu_2\pm\sqrt{\mu_2^2-4(t_2^2-\Delta_2^2)}}{2(t_2-\Delta_2)}.
\end{equation}
Again, if we localize along both the $\hat{x}$- and $\hat{y}$-directions, $k_x\rightarrow iq_x$, $k_y\rightarrow iq_y$,
\begin{equation}
H_{MKC,\perp}(iq_x,iq_y) = [-(2t_1\cosh q_x+\mu_1)\tau^z+2i\Delta_1\sinh q_x \tau^y]\otimes [(2t_2\cosh q_y+\mu_2)\sigma^z+2i\Delta_2\sinh q_y\sigma^y],
\end{equation}
the null condition is realized as,
\begin{equation}
((2t_1\cosh q_x+\mu_1)^2-4\Delta_1^2\sinh^2 q_x)((2t_2\cosh q_y+\mu_2)^2-4\Delta_2^2\sinh^2q_y)=0.
\end{equation}
This relation does not provide the conditions which occur simultaneously, for which we need to consider the component Bloch Hamiltonians for the MKC perpendicular,
\begin{equation}
\begin{split}
\mc{H}_{\perp,1}(k) =& -[2\mu_2t_1\cos k_x+2\mu_1t_2\cos k_y+2(t_1t_2+\Delta_1\Delta_2)\cos(k_x+k_y)+2(t_1t_2-\Delta_1\Delta_2)\cos(k_x-k_y)+\mu_1\mu_2]\sigma^z\\
&+[2\mu_2\Delta_1\sin k_x+2\mu_1\Delta_2\sin k_y+2(t_2\Delta_1+t_1\Delta_2)\sin(k_x+k_y)+2(t_2\Delta_1-t_1\Delta_2)\sin(k_x-k_y)]\sigma^y\\
&=\mathbf{d}_1(k)\cdot\boldsymbol{\sigma},
\end{split}
\label{Hperp1_Bloch}
\end{equation}
\begin{equation}
\begin{split}
\mc{H}_{\perp,2}(k) =& -[2\mu_2t_1\cos k_x+2\mu_1t_2\cos k_y+2(t_1t_2-\Delta_1\Delta_2)\cos(k_x+k_y)+2(t_1t_2+\Delta_1\Delta_2)\cos(k_x-k_y)+\mu_1\mu_2]\sigma^z\\
&+[2\mu_2\Delta_1\sin k_x-2\mu_1\Delta_2\sin k_y+2(t_2\Delta_1-t_1\Delta_2)\sin(k_x+k_y)+2(t_2\Delta_1+t_1\Delta_2)\sin(k_x-k_y)]\sigma^y\\
&=\mathbf{d}_2(k)\cdot\boldsymbol{\sigma},
\end{split}
\label{Hperp2_Bloch}
\end{equation}
Performing localization $k_x\rightarrow iq_x$ and $k_y\rightarrow iq_y$ then provides the following conditions,
\begin{subequations}
\begin{equation}
\begin{split}
&[2\mu_2t_1\cosh q_x+2\mu_1t_2\cosh q_y+2(t_1t_2+\Delta_1\Delta_2)\cosh(q_x+q_y)+2(t_1t_2-\Delta_1\Delta_2)\cosh(q_x-q_y)+\mu_1\mu_2]\\
&=\pm [2\mu_2\Delta_1\sinh q_x+2\mu_1\Delta_2\sinh q_y+2(t_2\Delta_1+t_1\Delta_2)\sinh(q_x+q_y)+2(t_2\Delta_1-t_1\Delta_2)\sinh(q_x-q_y)],\\
\implies &[(2t_1\cosh q_x+\mu_1)\pm 2\Delta_1\sinh q_x][(2t_2\cosh q_y+\mu_2)\pm 2\Delta_2\sinh q_y]=0,
\end{split}
\end{equation}
\begin{equation}
\begin{split}
&[2\mu_2t_1\cosh q_x+2\mu_1t_2\cosh q_y+2(t_1t_2-\Delta_1\Delta_2)\cosh(q_x+q_y)+2(t_1t_2+\Delta_1\Delta_2)\cosh(q_x-q_y)+\mu_1\mu_2]\\
&=\pm [2\mu_2\Delta_1\sinh q_x-2\mu_1\Delta_2\sinh q_y+2(t_2\Delta_1-t_1\Delta_2)\sinh(q_x+q_y)+2(t_2\Delta_1+t_1\Delta_2)\sinh(q_x-q_y)],\\
\implies &[(2t_1\cosh q_x+\mu_1)\pm 2\Delta_1\sinh q_x][(2t_2\cosh q_y+\mu_2)\mp 2\Delta_2\sinh q_y]=0,
\end{split}
\end{equation}
\end{subequations}
Each condition for each component Hamiltonian gives rise to two solutions for $e^{-q_x}$ and two for $e^{-q_y}$. The Bloch Hamiltonians after localization are given as,
\begin{subequations}
\begin{equation}
\begin{split}
\mc{H}_{\perp,1}(iq_x,iq_y)=& -[(2t_1\cosh q_x+\mu_1)(2t_2\cosh q_y+\mu_2)+4\Delta_1\Delta_2\sinh q_x\sinh q_y]\sigma^z\\
&+[2\Delta_1\sinh q_x(2t_2\cosh q_y+\mu_2)+2\Delta_2\sinh q_y(2t_1\cosh q_x+\mu_1)]\sigma^y,
\end{split}
\end{equation}
\begin{equation}
\begin{split}
\mc{H}_{\perp,2}(iq_x,iq_y)=& -[(2t_1\cosh q_x+\mu_1)(2t_2\cosh q_y+\mu_2)-4\Delta_1\Delta_2\sinh q_x\sinh q_y]\sigma^z\\
&+[2\Delta_1\sinh q_x(2t_2\cosh q_y+\mu_2)-2\Delta_2\sinh q_y(2t_1\cosh q_x+\mu_1)]\sigma^y.
\end{split}
\end{equation}
\end{subequations}
Similar to the parallel system, even here we must work with the basis for the full system, $(\tilde{c}_{\boldsymbol{k},\uparrow},c\tilde{c}_{\boldsymbol{k},\downarrow},c^{\dagger}_{-\boldsymbol{k},\uparrow},c^\dagger_{-\boldsymbol{k},\downarrow})^T$. The eigenvectors for the MZMs then depend on whether the parents are topological or trivial, and which boundary conditions are open,  so that we have two cases,
\begin{itemize}
\item \textit{Case 1}: If both the x and y boundary conditions are open, and both the parents are topological, let us have the following condition fulfilled,
\begin{equation}
[(2t_1\cosh q_x+\mu_1)-2\Delta_1\sinh q_x][(2t_2\cosh q_y+\mu_2)-2\Delta_2\sinh q_y]=0.
\end{equation}
Here both the factors are zero since both the parents are topological, so that $2t_1\cosh q_x+\mu_1=2\Delta_1\sinh q_x$ and $2t_2\cosh q_y+\mu_2=2\Delta_2\sinh q_y$. Substituting into Eqns.~(S47a) and (S47b), we see that $\mc{H}_{\perp,2}(iq_x,iq_y)=0$ and then the MZM eigenvectors must be null vectors of the matrix,
\begin{equation}
\begin{pmatrix}
d_{1,z}(iq_x,iq_y) & 0 & 0 & d_{1,z}(iq_x,iq_y)\\
0 & 0 & 0 & 0\\
0 & 0 & 0 & 0\\
-d_{1,z}(iq_x,iq_y) & 0 & 0 & -d_{1,z}(iq_x,iq_y)
\end{pmatrix}
.
\end{equation}
The MZM eigenvectors are then given as $\{\frac{1}{\sqrt{2}}(\ket{00}-\ket{11}),\ket{01},\ket{10}\}$.
\item\textit{Case 2}: Lets suppose that only the x boundary condition is open or only parent 1 is topological. One can show that we must facilitate the following condition,
\begin{equation}
2t_1\cosh q_x+\mu_1=2\Delta_1\sinh q_x.
\end{equation}
The MZM eigenvectors are then null vectors for the following matrix,
\begin{equation}
\begin{pmatrix}
d_{1,z}(iq_x,iq_y/k_y) & 0 & 0 & d_{1,z}(iq_x,iq_y/k_y)\\
0 & d_{2,z}(iq_x,iq_y/k_y) & d_{2,z}(1q_x,iq_y/k_y) & 0\\
0 & -d_{2,z}(iq_x,iq_y/k_y) & -d_{2,z}(1q_x,iq_y/k_y) & 0\\
-d_{1,z}(iq_x,iq_y/k_y) & 0 & 0 & -d_{1,z}(iq_x,iq_y/k_y)
\end{pmatrix}
,
\end{equation}
which are given as, $\{\frac{1}{\sqrt{2}}(\ket{00}-\ket{11}),\frac{1}{\sqrt{2}}(\ket{01}-\ket{10})\}$.
\end{itemize}

\section{Wilson loop}
\label{app:wilson_loop}

The Wilson loop \cite{wilson_loop_paper} is a unitary operator defined over a closed path as:

\begin{equation}
    \begin{aligned}
    \mathcal{W} = \overline{\exp}{\Big [ i \int_{BZ} d \bm{k} \cdot \bm{A}(\bm{k}) \Big ]}.
    \end{aligned}
\end{equation}

Here $\bm{A} = A_x \hat{k}_x + A_y \hat{k}_y + A_z \hat{k}_z$ is the non-Abelian Berry connection:

\begin{equation}
    \begin{aligned}
    \bm{A}_{mn}(\bm k) = i\bra{u_m(\bm k)} \nabla_{\bm k} \ket{u_n(\bm k)},
    \end{aligned}
\end{equation}

a Hermitian operator as $A_{mn}=A^\ast_{nm}$ in this convention. In the definition above $\ket{u_n(\bm k)}$ are eigenvectors of a Bloch Hamiltonian satisfying $H(\bm k) \ket{u_{n}(\bm k)} = E_{n}(\bm k)  \ket{u_{n}(\bm k)}$ and $1 \leq m, n \leq M$, with $M$ being the number of occupied bands.

The eigenvalues of the Wilson operator defined over a closed path, a Wilson loop, are Gauge independent and unitary. Therefore they can be expressed as $e^{i2\pi \nu_i}$, with $\nu_i$ corresponding to the Wannier centers \cite{theory_polarization_vanderbilt, electric_pol_vanderbilt}.

To numerically compute the Wilson loop and avoid complications related to a lack of Gauge fixing, we compute the Wilson loop over a discrete and closed path in momentum space divided into $R+1$ segments \cite{wilson_loop_paper}:

\begin{equation}
    \begin{aligned}
    \mathcal{W}_{mn} = \bra{u_{m}(\bm k_0)} \lim_{R \to \infty} \prod_{i=R}^1  P(\bm k_i) \ket{u_n(\bm k_0)},
    \end{aligned}
\end{equation}

where $P(\bm k) = \sum_{n'=1}^M \ket{u_{n'}(\bm k)} \bra{u_{n'}(\bm k)}$ is the projector operator in the occupied subspace as $1 \leq n, m \leq M$, with $M$ the number of occupied bands.

\subsection{Properties of the Wilson loop spectrum of a child Hamiltonian}

We analytically derive here the numerical results presented in Fig.~\ref{fig:Wannier-centers-parallel}, where the Wannier centers of a child Hamiltonian correspond to the addition of the parents' Wannier centers of charge.

The child Hamiltonian for the parallel multiplicative chain is given by Eq.~\ref{KCparallel}, repeated here for convenience,

\begin{equation}
\begin{aligned}
H^c_{MKC,||}(k) =& [-(2t_1\cos k+\mu_1)\tau^z+2\Delta_1\sin k\tau^y]\\
&\otimes [(2t_2\cos k+\mu_2)\sigma^z+2\Delta_2\sin k\sigma^y].
\end{aligned}
\end{equation}

Given the tensor product construction of the Hamiltonian, its eigenvectors can be represented using the parents eigendecomposition,

\begin{equation}
\begin{aligned}
\ket{u_1(k)} = \ket{v_{1+}(k)} \otimes \ket{v_{2-}(k)} ; \quad H^c_{MKC,||}(k)\ket{u_1(k)} = \epsilon^{(1)}_+ \epsilon^{(2)}_- \ket{u_1(k)}\\
\ket{u_2(k)} = \ket{v_{1-}(k)} \otimes \ket{v_{2+}(k)} ; \quad H^c_{MKC,||}(k)\ket{u_2(k)} = \epsilon^{(1)}_- \epsilon^{(2)}_+ \ket{u_2(k)} \\
\ket{u_3(k)} = \ket{v_{1-}(k)} \otimes \ket{v_{2-}(k)} ; \quad H^c_{MKC,||}(k)\ket{u_3(k)} = \epsilon^{(1)}_- \epsilon^{(2)}_- \ket{u_3(k)}\\
\ket{u_4(k)} = \ket{v_{1+}(k)} \otimes \ket{v_{2+}(k)} ; \quad H^c_{MKC,||}(k)\ket{u_4(k)} = \epsilon^{(1)}_+ \epsilon^{(2)}_+ \ket{u_4(k)}
\end{aligned}
\end{equation}

where $\ket{v_{1 \pm}(k)}$, are the first parent's eigenvectors with corresponding eigenvalues $\epsilon^{(1)}_{\pm}$, and $\ket{v_{2 \pm}(k)}$ are the eigenvectors of the second parent, with corresponding eigenvalues $\epsilon^{(2)}_{\pm}$, after mapping $t_2 \rightarrow -t_2$ and $\mu_2 \rightarrow -\mu_2$. As $\epsilon^{(1)}_{\pm}=\epsilon^{(2)}_{\pm}$, we obtain doubly degenerate bands and identify $\ket{u_1(k)}, \ket{u_2(k)}$ as the eigenvectors in the occupied subspace at half-filling. This results in a $2 \times 2$ matrix representation for the non-Abelian Berry connection,

\begin{equation}
    \begin{aligned}
    A(k) &= i
    \begin{pmatrix}
    \bra{v_{1-}(k)}\partial_k \ket{v_{1-}(k)} + \bra{v_{2+}(k)}\partial_k \ket{v_{2+}(k)}& 0\\
    0 & \bra{v_{1+}(k)}\partial_k \ket{v_{1+}(k)} + \bra{v_{2-}(k)}\partial_k \ket{v_{2-}(k)}
    \end{pmatrix} \\
    &=i
    \begin{pmatrix}
    \bra{v_{1-}(k)}\partial_k \ket{v_{1-}(k)} & 0\\
    0 & \bra{v_{1+}(k)}\partial_k \ket{v_{1+}(k)}
    \end{pmatrix}
    +i
    \begin{pmatrix}
    \bra{v_{2+}(k)} \partial_k \ket{v_{2+}(k)}& 0\\
    0 & \bra{v_{2-}(k)} \partial_k \ket{v_{2-}(k)}
    \end{pmatrix}
     \\
\implies A &= A_1 + A_2
    \end{aligned}
\end{equation}

where $A_1$ and $A_2$ are not the Berry connections for each parent, as they are constructed from the full space and not just the occupied subspace. Importantly, $A_1$ and $A_2$ are Hermitian and diagonal.

\begin{equation}
    \begin{aligned}
    \mathcal{W} &=
    \overline{\exp}{\Big [ i \int_{BZ} dk A(k) \Big ]} \\
    &= \overline{\exp}{\Big [ i \int_{BZ} dk A_1(k) +i \int_{BZ} dk A_2(k)\Big ]} \\
    &= \overline{\exp} \Big [ i \int_{BZ} dk A_1(k) \Big ] \quad \overline{\exp}{\Big [ i \int_{BZ} dk A_2(k)\Big ]} \\
    \implies \mathcal{W}&= \mathcal{W}_1 \mathcal{W}_2
    \end{aligned}
    \label{eq:wilson_loop_equivalence}
\end{equation}

In order to separate the exponential, we have used the fact that as $A_1$ and $A_2$ are diagonal matrices, they satisfy $[A_1, A_2]=0$. It is worth noting that the hermiticity of $A_1$ and $A_2$ makes $\mathcal{W}_1$ and $\mathcal{W}_2$ unitary operators with unitary eigenvalues. Moreover, due to $A_1$ and $A_2$ being diagonal we conclude $\mathcal{W}_1$, $\mathcal{W}_2$, and consequently $\mathcal{W}$, are diagonal too.

As stated earlier, the Wilson loop $\mathcal{W}$ is a unitary operator, and its  $i$\textsuperscript{th} eigenvalue can therefore be represented by $\exp(i2\pi \nu_i)$. Using Eq.~\ref{eq:wilson_loop_equivalence}, we establish a direct relation between the eigenvalues of $\mathcal{W}$ and the ones of $\mathcal{W}_1$ and $\mathcal{W}_2$. We first denote the $i$\textsuperscript{th} eigenvalue of $\mathcal{W}_1$ and $\mathcal{W}_2$ as $\exp(i2\pi \nu^{(1)}_i)$ and $\exp(i2\pi \nu^{(2)}_i)$, respectively. Noting that $\nu_1^{(1)}=\nu_2^{(1)}$ and $\nu_1^{(2)}=\nu_2^{(2)}$, $\exp(i2\pi \nu_i)$ may then be expressed as:

\begin{equation}
    \begin{aligned}
    \implies
    \exp(i2\pi \nu_i) &=  \exp(i2\pi \nu^{(1)}) \exp(i2\pi \nu^{(2)}) \\
     \implies \nu_i &= \nu^{(1)} + \nu^{(2)} \mod 1
    \end{aligned}
\end{equation}

\begin{equation}
    \begin{aligned}
        \nu_i &= \nu^{(1)} + \nu^{(2)} \mod 1
    \end{aligned}
\end{equation}

This result demonstrates that a child obtained from two topological parents ($\nu^{(1)}= \nu^{(2)}=0.5$) is not distinguished from a child of trivial parents ($\nu^{(1)}= \nu^{(2)}=0$) by means of a Wilson loop spectrum, as $\nu_i=0$ for $i \in \{1, 2\}$ in each case.

We can also examine another formulation for the Wilson loop, in terms of projectors onto occupied states, which is widely-used for numerical calculations \cite{alexandradinata2016gcohomology, asboth2016short, wieder2018wallpaper}, and arrive at the same conclusion.

\begin{equation}
    \begin{aligned}
    \mathcal{W}_{mn} = \bra{u_{m}(\bm k_0)} \lim_{R \to \infty} \prod_{i=R}^1  P(\bm k_i) \ket{u_n(\bm k_0)},
    \end{aligned}
    \label{WLProj}
\end{equation}

At half-filling the projector operator corresponds to

\begin{equation}
    \begin{aligned}
P(k_i)
&= \ket{u_{1}(k)} \bra{u_1(k)} + \ket{u_{2}(k)} \bra{u_2(k)} \\
&= \ket{v_{1-}(k)} \bra{v_{1-}(k)} \otimes \ket{v_{2+}(k)} \bra{v_{2+}(k)} + \ket{v_{1+}(k)} \bra{v_{1+}(k)} \otimes \ket{v_{2-}(k)} \bra{v_{2-}(k)}
    \end{aligned}
\end{equation}

Consequently, after a discretization of the BZ into $R+1$ segments such that the wavefunctions vary smoothly enough,

\begin{equation}
    \begin{aligned}
\lim_{R \to \infty} \prod_{i=R}^{1} P(k_i)
&= \lim_{R \to \infty} \prod_{i=R}^{1} \Big [ \ket{v_{1-}(k)} \bra{v_{1-}(k)} \otimes \ket{v_{2+}(k)} \bra{v_{2+}(k)} + \ket{v_{1+}(k)} \bra{v_{1+}(k)} \otimes \ket{v_{2-}(k)} \bra{v_{2-}(k)}  \Big ] \\
&= \lim_{R \to \infty} \prod_{i=R}^{1} \Big [ \ket{v_{1-}(k)} \bra{v_{1-}(k)} \otimes \ket{v_{2+}(k)} \bra{v_{2+}(k)} \Big ] + \prod_{i=R}^{1} \Big [ \ket{v_{1+}(k)} \bra{v_{1+}(k)} \otimes \ket{v_{2-}(k)} \bra{v_{2-}(k)} \Big ] \\
&= \lim_{R \to \infty} \prod_{i=R}^{1}  \ket{v_{1-}(k)} \bra{v_{1-}(k)} \otimes \prod_{i=R}^{1}  \ket{v_{2+}(k)} \bra{v_{2+}(k)}  + \prod_{i=R}^{1} \ket{v_{1+}(k)} \bra{v_{1+}(k)} \otimes \prod_{i=R}^{1}  \ket{v_{2-}(k)} \bra{v_{2-}(k)}  \\
&= \lim_{R \to \infty} \prod_{i=R}^{1}  P_{1-}(k_i) \otimes \prod_{i=R}^{1}  P_{2+}(k_i)  + \prod_{i=R}^{1}  P_{1+}(k_i) \otimes \prod_{i=R}^{1}  P_{2-}(k_i)
    \end{aligned}
\end{equation}

\begin{equation}
    \begin{aligned}
    \implies \mathcal{W}_{mn} = \bra{u_{m}(k_0)} \lim_{R \to \infty} \prod_{i=R}^{1}  P_{1-}(k_i) \otimes \prod_{i=R}^{1}  P_{2+}(k_i)   \ket{u_n(k_0)} +
    \bra{u_{m}(k_0)} \lim_{R \to \infty} \prod_{i=R}^{1}  P_{1+}(k_i) \otimes \prod_{i=R}^{1}  P_{2-}(k_i)  \ket{u_n(k_0)}
    \end{aligned}
\end{equation}

where
\begin{equation}
    \begin{aligned}
    \mathcal{W}_{11} &=
    \bra{v_{1+}(k_0)} \lim_{R \to \infty} \prod_{i=R}^{1}  P_{1+}(k_i) \ket{v_{1+}(k_0)} \bra{v_{2-}(k_0)} \prod_{i=R}^{1}  P_{2-}(k_i)  \ket{v_{2-}(k_0)} \\
    \mathcal{W}_{11} &=
    \bra{v_{1-}(k_0)} \lim_{R \to \infty} \prod_{i=R}^{1}  P_{1-}(k_i) \ket{v_{1-}( k_0)} \bra{v_{2+}(k_0)} \prod_{i=R}^{1}  P_{2+}(k_i)  \ket{v_{2+}(k_0)} \\
    \mathcal{W}_{21} &= \mathcal{W}_{12} = 0 \\
    \end{aligned}
\end{equation}

indicating that $\mathcal{W}_{mn}$ is represented by a diagonal matrix that can be written as $\mathcal{W}=\mathcal{W}_1 \mathcal{W}_2$, leading to the same conclusion.\\
\par One can similarly work out the Wannier spectra for the perpendicular MKC, given by the child Hamiltonian in Eqn.~\eqref{MKCperpHamiltonian},
\begin{equation}
H^c_{MKC,\perp}=[-(2t_1\cos k_x+\mu_1)\tau^z+2\Delta_1\sin k_x\tau^y]\otimes [(2t_2\cos k_y+\mu_2)\sigma^z+2\Delta_2\sin k_y\sigma^y].
\end{equation}
Assuming a similar convention to the MKC parallel case described previously while replacing $k$ by the vector $\boldsymbol{k}=(k_x,k_y)$, the definition of $\mathbf{A}(\boldsymbol{k})$ indicates that one should have the following non-Abelian Berry connection vector,
\begin{equation}
\begin{split}
\mathbf{A}(\boldsymbol{k}) =& i
\begin{pmatrix}
\bra{v_{1-}(k_x)}\partial_{k_x}\ket{v_{1-}(k_x)}\boldsymbol{\hat{e}_x}+\bra{v_{2+}(k_y)}\partial_{k_y}\ket{v_{2+}(k_y)}\boldsymbol{\hat{e}_y} & 0\\
0 & \bra{v_{1+}(k_x)}\partial_{k_x}\ket{v_{1+}(k_x)}\boldsymbol{\hat{e}_x}+\bra{v_{2-}(k_y)}\partial_{k_y}\ket{v_{2-}(k_y)}\boldsymbol{\hat{e}_y}
\end{pmatrix}
,\\
=& i
\begin{pmatrix}
\bra{v_{1-}(k_x)}\partial_{k_x}\ket{v_{1-}}\boldsymbol{\hat{e}_x} & 0\\
0 & \bra{v_{1+}(k_x)}\partial_{k_x}\ket{v_{1+}}\boldsymbol{\hat{e}_x}
\end{pmatrix}
+i
\begin{pmatrix}
\bra{v_{2+}(k_y)}\partial_{k_y}\ket{v_{2+}(k_y)}\boldsymbol{\hat{e}_y} & 0\\
0 & \bra{v_{2-}(k_y)}\partial_{k_y}\ket{v_{2-}(k_y)}\boldsymbol{\hat{e}_y}
\end{pmatrix}
,\\
=& A_1\boldsymbol{\hat{e}_x}+A_2\boldsymbol{\hat{e}_y}.
\end{split}
\end{equation}
We first consider the formulation of the Wilson loop in terms of projectors onto occupied states as written in Eqn.~\eqref{WLProj}. At half-filling, the projector onto occupied states for the perpendicular MKC is given as,
\begin{equation}
\begin{split}
P(k_x,k_y) =& \ket{u_1(k_x,k_y)}\bra{u_1(k_x,k_y)}+\ket{u_2(k_x,k_y)}\bra{u_2(k_x,k_y)},\\
=& \ket{v_{1+}(k_x)}\bra{v_{1+}(k_x)}\otimes\ket{v_{2-}(k_y)}\bra{v_{2-}(k_y)}+\ket{v_{1-}(k_x)}\bra{v_{1-}(k_x)}\otimes\ket{v_{2+}(k_y)}\bra{v_{2+}(k_y)}
\end{split}
\end{equation}
One can compute the Wilson loop by integrating over either $k_x$ or $k_y$. Let us assume that the BZ along $k_x$ direction is discretized into $R+1$ segments for a given $k_y$ (assuming sufficient smoothness for the wavefunctions), the other case follows similarly,
\begin{equation}
\begin{split}
\lim_{R\rightarrow\infty}\prod_{i=R}^1P(k_{xi},k_y)=& \lim_{R\rightarrow\infty}\prod_{i=R}^1\bigg{[}\ket{v_{1+}(k_{xi})}\bra{v_{1+}(k_{x1})}\otimes\ket{v_{2-}(k_y)}\bra{v_{2-}(k_y)}+\ket{v_{1-}(k_{xi})}\bra{v_{1-}(k_{xi})}\otimes\ket{v_{2+}(k_y)}\bra{v_{2+}(k_y)}\bigg{]},\\
=& \lim_{R\rightarrow\infty}\bigg{[}\prod_{i=R}^1\ket{v_{1+}(k_{xi})}\bra{v_{1+}(k_{xi})}\bigg{]}\otimes\ket{v_{2-}(k_{y})}\bra{v_{2-}(k_{y})}\\
&+\bigg{[}\prod_{i=R}^1\ket{v_{1-}(k_{xi})}\bra{v_{1-}(k_{xi})}\bigg{]}\otimes\ket{v_{2+}(k_{y})}\bra{v_{2+}(k_{y})},\\
=& \lim_{R\rightarrow\infty}\bigg{[}\prod_{i=R}^1P_{1+}(k_{xi})\bigg{]}\otimes P_{2-}(k_y)+\lim_{R\rightarrow\infty}\bigg{[}\prod_{i=R}^1P_{1-}(k_x)\bigg{]}\otimes P_{2+}(k_y).
\end{split}
\end{equation}
We get the Wilson loop matrix from the definition as follows,
\begin{equation}
\begin{split}
\mathcal{W}_{mn}(k_y)=\bra{u_m(k_{x0},k_y)}\lim_{R\rightarrow\infty}\bigg{[}\prod_{i=R}^1P_{1+}(k_{xi})\bigg{]}\otimes P_{2-}(k_y)+\lim_{R\rightarrow\infty}\bigg{[}\prod_{i=R}^1P_{1-}(k_x)\bigg{]}\otimes P_{2+}(k_y)\ket{u_n(k_{x0},k_y)},
\end{split}
\label{WLMKCperp}
\end{equation}
so that $\mathcal{W}_{11}(k_y) = \mathcal{W}_{11}$ and $\mathcal{W}_{22}(k_y) = \mathcal{W}_{22}$ are expressed as,
\begin{equation}
\begin{split}
\mathcal{W}_{11} =& \bra{v_{1+}(k_{x0})}\lim_{R\rightarrow\infty}\bigg{[}\prod_{i=R}^1P_{1+}(k_{xi})\bigg{]}\ket{v_{1+}(k_{x0})},\\
\mathcal{W}_{22} =& \bra{v_{1-}(k_{x0})}\lim_{R\rightarrow\infty}\bigg{[}\prod_{i=R}^1P_{1-}(k_{xi})\bigg{]}\ket{v_{1-}(k_{x0})},\\
\mathcal{W}_{12} =& \mathcal{W}_{21}=0.
\end{split}
\end{equation}
The projector due to the eigenvectors of the second parent contract with the respective eigenvectors of the second parent in the tensor product eigenvectors for the occupied basis to produce all four matrix elements $\mathcal{W}_{mn}(k_y)=\mathcal{W}_{mn}$ as $k_y$ independent terms. This implies that the Wilson loop computed as an integral over a given momentum component is independent of the other momentum component. Previously we mentioned that the Wilson loop eigenvalues are of the form $e^{i2\pi\nu_i}$ due to the unitary nature of $\mathcal{W}$. Here $\nu_i$ is the $i$\textsuperscript{th} Wannier charge center. Let us refer to the Wannier charge spectra due to the Wilson loop along $k_x$ as $\{\nu_{i}(k_y)\}_x$. The $i$\textsuperscript{th} Wannier charge center for the child Hamiltonian computed as an integral over $k_x$ for each value of $k_y$, $\nu_{i}(k_y)$, is then equal to the Wannier charge center of the parent Hamiltonian that is calculated along a loop across $k_x$ (here parent 1). We similarly find the $i$\textsuperscript{th} Wannier charge center for the child Hamiltonian computed by integrating over $k_y$ for a given $k_x$, $\nu_{i}(k_x)$, is equal to the Wannier charge center of the parent Hamiltonian that is calculated across a loop along $k_y$ (here parent 2), such that
\begin{equation}
\begin{split}
\nu_{i}(k_y)=&\nu^{(1)}\text{mod 1},\\
\nu_{i}(k_x)=&\nu^{(2)}\text{mod 1},
\end{split}
\end{equation}
where $\nu^{(j)}$ is the Wannier charge center due to the $j$\textsuperscript{th} parent. Here the spectrum for both $\{\nu_i(k_y)\}$ and $\{\nu_i(k_x)\}$ are doubly degenerate (up to mod 1) and equal to the respective parent Wannier charge center values.

\section{Calculation for Winding number in the MKC perpendicular case:}\label{WindingPerp}
Consider the Bloch Hamiltonians for the component Hamiltonians in the MKC perpendicular case,
\begin{subequations}
\begin{equation}
\begin{split}
\mc{H}_{\perp,1}(\boldsymbol{k})=&-((\mu_1+2t_1\cos k_x)(\mu_2+2t_2\cos k_y)-4\Delta_1\Delta_2\sin k_x\sin k_y)\sigma^z\\
&+(2\Delta_1\sin k_x(\mu_2+2t_2\cos k_y)+2\Delta_2\sin k_y(\mu_1+2t_1\cos k_x))\sigma^y = (0,d_{1,y},d_{1,z})\cdot\boldsymbol{\sigma},
\end{split}
\end{equation}
\begin{equation}
\begin{split}
\mc{H}_{\perp,2}(\boldsymbol{k})=&-((\mu_1+2t_1\cos k_x)(\mu_2+2t_2\cos k_y)+4\Delta_1\Delta_2\sin k_x\sin k_y)\sigma^z\\
&+(2\Delta_1\sin k_x(\mu_2+2t_2\cos k_y)-2\Delta_2\sin k_y(\mu_1+2t_1\cos k_x))\sigma^y=(0,d_{2,y},d_{2,z})\cdot\boldsymbol{\sigma}.
\end{split}
\end{equation}
\end{subequations}
Since we want to plot the Bloch vectors $(d_{1,y},d_{1,z})$ and $(d_{2,y},d_{2,z})$ for varying $k_x$ at given $k_y$ and vice-versa in PBC on both directions, let us find the locus of the curve for parameter $k_x$ at given $k_y$. Denote, $M_2=\mu_2+2t_2\cos k_y$ and $R_2=2\Delta_2\sin k_y$. Further denote, $\cos\theta=\frac{M_2}{\sqrt{M_2^2+R_2^2}}$ and $\sin\theta=\frac{R_2}{\sqrt{M_2^2+R_2^2}}$. Then the locus of the parametric curve $(d_{1,y},d_{1,z})$ is shown below,
\begin{equation}
\frac{(\cos\theta d_{1,y}+\sin\theta d_{1,z})^2}{4\Delta_1^2(M_2^2+R_2^2)}+\frac{(\cos\theta d_{1,z}-\sin\theta d_{1,y}+\mu_1\sqrt{M_2^2+R_2^2})^2}{4t_1^2(M_2^2+R_2^2)} = 1.
\end{equation}
Since we plot for $t_1=\Delta_1$ in the main text, implementing this condition we further get,
\begin{equation}
(\cos\theta d_{1,y}+\sin\theta d_{1,z})^2+(\cos\theta d_{1,z}-\sin\theta d_{1,y}+\mu_1\sqrt{M_2^2+R_2^2}) = (2t_1\sqrt{M_2^2+R_2^2})^2.
\end{equation}
It is easy to notice that this is a circle whose coordinates have been rotated by and angle $\theta$ and the center, and radii have been modulated by the other parent by $\sqrt{M_2^2+R_2^2}$, which however does not change the range of parameters where the system is topological. For PBC with $L_y$ sites in the y-direction, we obtain $L_y$ number of such circles rotated at $L_y$ angles, which is essentially obtain in the main text.


\section{Dependence of energy with parameter $\mu$ near critical points:}\label{Evsmuexpl}
We will consider two cases for the dispersion of the MKC parallel Hamiltonian, (i)$\mu_1=\mu_2=\mu$, $t_1=t_2=t$ and $\Delta_1=\Delta_2=\Delta$ and (ii)$\frac{\mu_1}{t_1}=-\frac{\mu_2}{t_2}=\frac{\mu}{t}$ and $\Delta_1=\Delta_2=\Delta$. For (i), the child dispersion is of the form,
\begin{equation}
E(k) = (2t\cos k+\mu)^2+4\Delta^2\sin^2k.
\end{equation}
Let $\mu = -2t+\delta\mu$ near the critical point at $k=0$, so that we must have,
\begin{equation}
E(k=0)\approx (2t-2t+\delta\mu)^2, \implies E(k=0)\sim \delta\mu^2.
\label{Evsmuexplplus}
\end{equation}
For (ii), the child dispersion is given as,
\begin{equation}
\begin{split}
E(k) =& \sqrt{((2t\cos k+\mu)^2+4\Delta^2\sin^2k)((2t\cos k-\mu)^2+4\Delta^2\sin^2k)},\\
=& \sqrt{(4t^2\cos^2k+4\Delta^2\sin^2k+\mu^2)^2-16t^2\mu^2\cos^2k}.
\end{split}
\end{equation}
Again, we let $\mu=-2t+\delta\mu$ near the critical point, $k=0$, and we then get,
\begin{equation}
\begin{split}
E(k=0)&\approx \sqrt{(4t^2+(-2t+\delta\mu)^2)^2-16t^2(-2t+\delta\mu)^2},\\
&\approx 4t^2-(-2t+\delta\mu)^2\approx 4t\delta\mu+\delta\mu^2\implies E(k=0)\sim \delta\mu.
\end{split}
\label{Evsmuexplminus}
\end{equation}

\newpage
\section{Robustness of MMZMs in the MKC parallel system:}

Here, we show additional slab spectra for the MKC as a function of chemical potential $\mu_1$ for a variety of disorder terms, for $\mu_1 = -\mu_2$ in Fig.~\ref{mu1eqminusmu2robustMKCpll}, $\mu_2 = 0$ in Fig.~\ref{mu20mu1robustMKCpll}, and $\mu_2 = 3$ in Fig.~\ref{mu23mu1robustMKCpll}. Stability of MMZMs against on-site disorder terms proportional to  $\tau^i \sigma^j$ correspond to presence of zero-energy modes over a finite interval in $\mu_1$.

\begin{minipage}{0.45\textwidth}
    \includegraphics[width=0.9\textwidth]{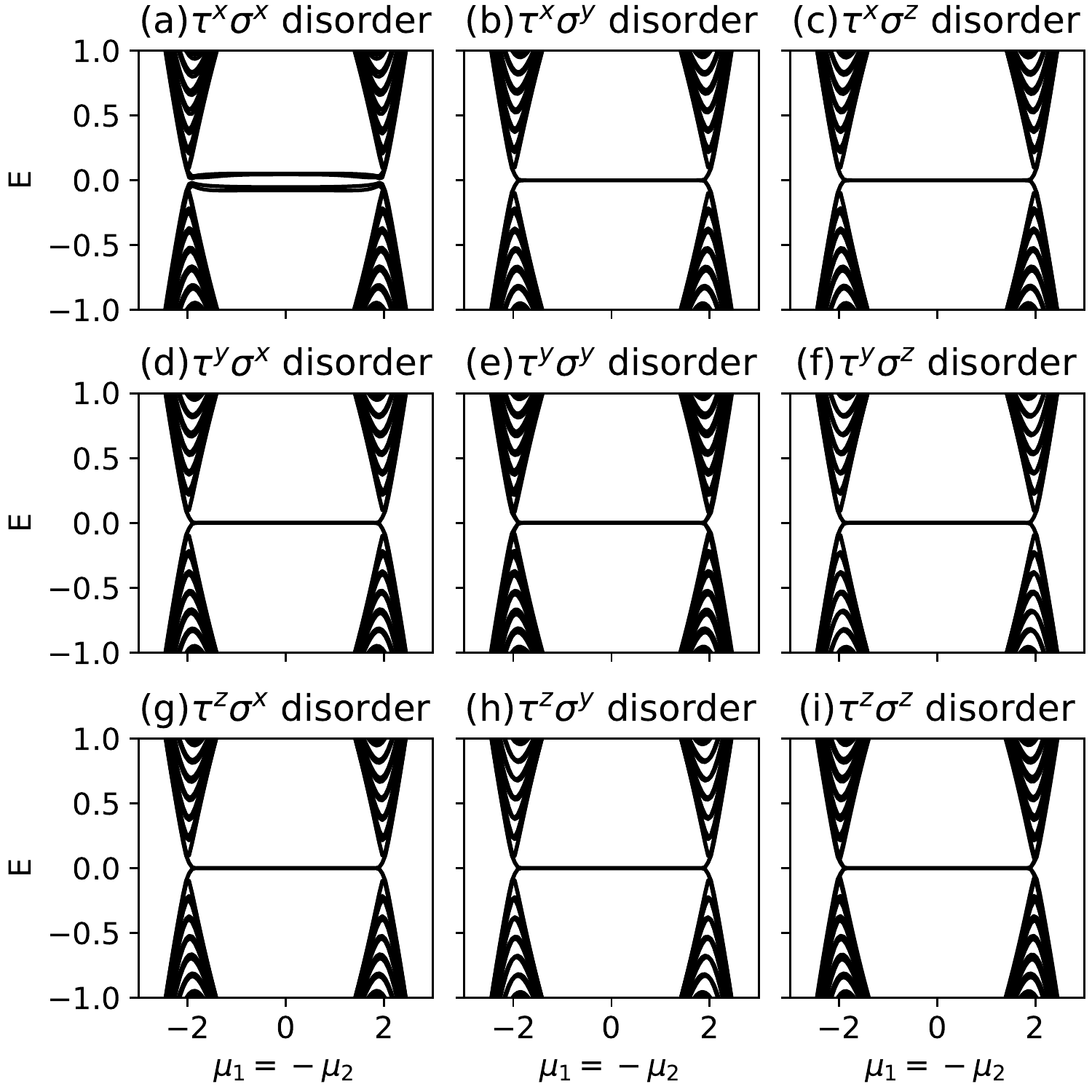}
    \captionof{figure}{For the case $\mu_1=-\mu_2$ again the MMZMs for the MKC parallel system are robust for all $\tau^i\sigma^j$ disorders except $\tau^x\sigma^x$.}
    \label{mu1eqminusmu2robustMKCpll}
\end{minipage}\hfill
\begin{minipage}{0.45\textwidth}
    \includegraphics[width=0.9\textwidth]{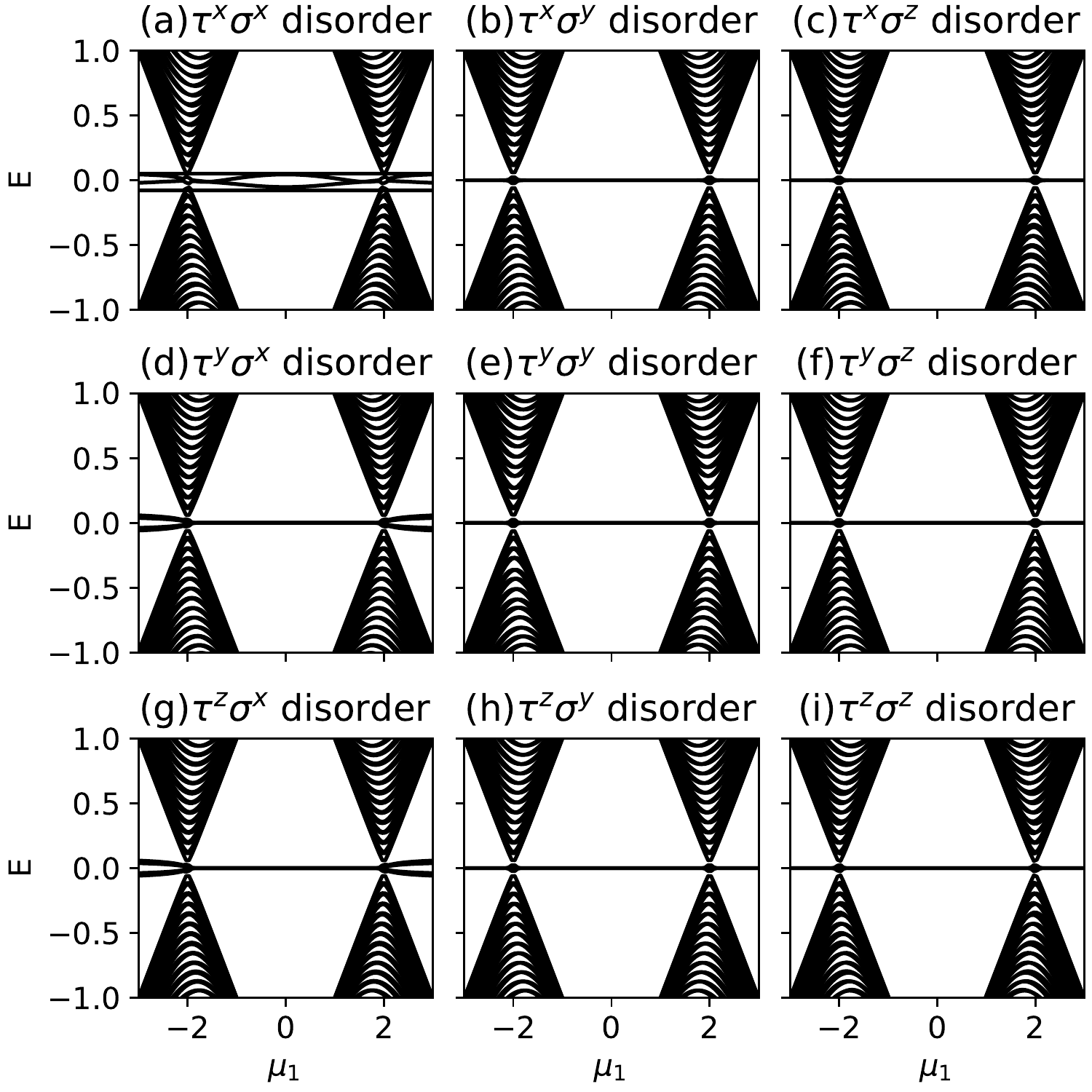}
    \captionof{figure}{Checking for robustness for given $\mu_2=0$ and varying across $\mu_1$ for all disorder $\tau^i\sigma^j$ combinations. The MMZM in the range $(-2t_1, 2t_1)$ is always robust except $\tau^x\sigma^x$, while MZMs beyond that range succumb to disorder for $\tau^i\sigma^x$.}
    \label{mu20mu1robustMKCpll}
\end{minipage}
\begin{minipage}{0.45\textwidth}
    \includegraphics[width=0.9\textwidth]{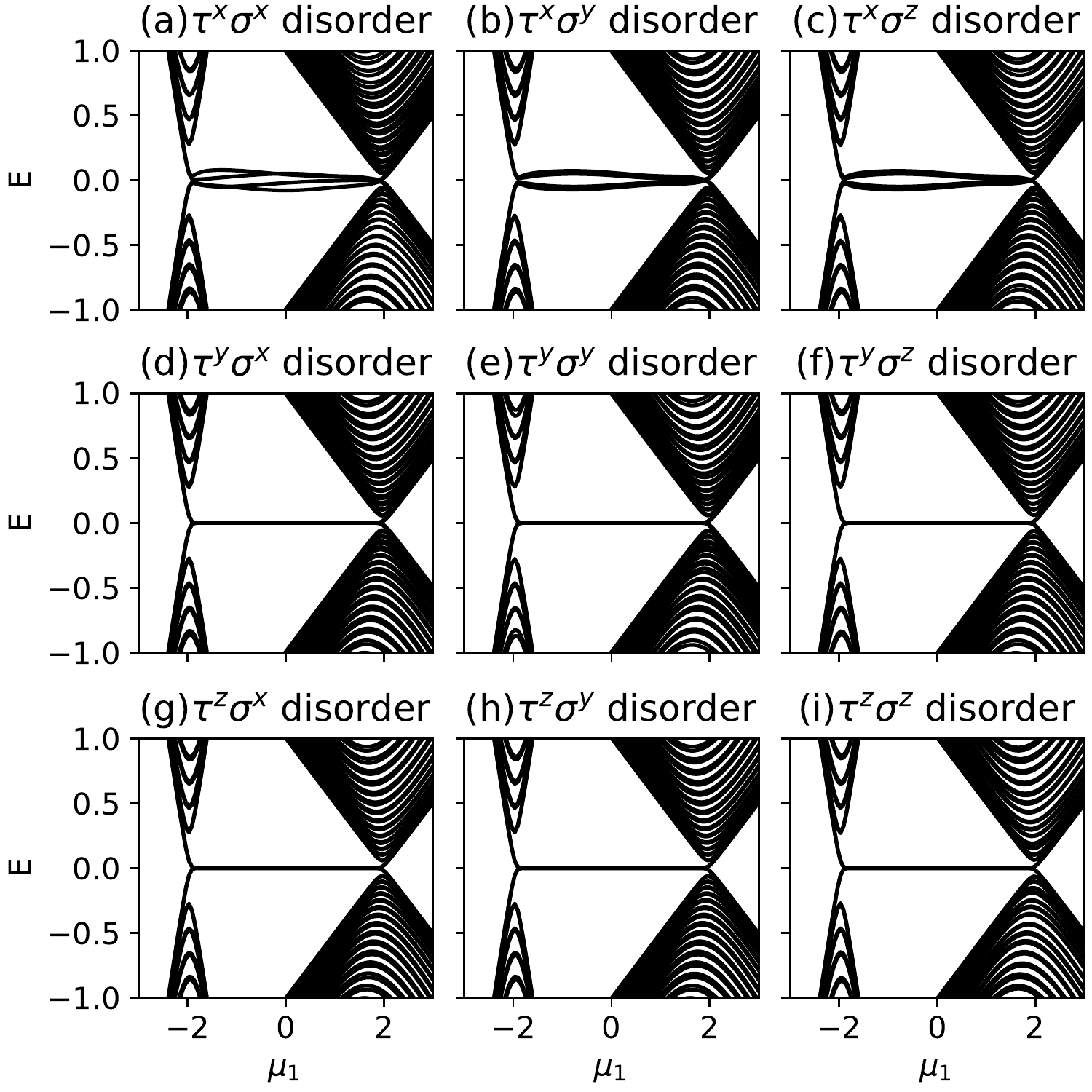}
    \captionof{figure}{Checking for robustness for given $\mu_2=3$ and varying across $\mu_1$ for all disorder $\tau^i\sigma^j$ combinations. The MMZM in the range $(-2t_1, 2t_1)$ is always robust except $\tau^x\sigma^x$, while MZMs beyond that range succumb to disorder for $\tau^x\sigma^j$.}
    \label{mu23mu1robustMKCpll}
\end{minipage}

\end{document}